\documentclass[longauth]{aa}

\usepackage{graphicx}
\usepackage{txfonts}
\usepackage{natbib}
\usepackage[colorlinks=true,citecolor=blue,breaklinks=true]{hyperref}
\usepackage{xspace}
\usepackage{amsmath,amssymb}
\usepackage{gensymb}
\usepackage{enumitem}
\usepackage{upgreek}
\usepackage{subcaption}
\usepackage{tabularx}
\usepackage{float}
\usepackage{todonotes}
\usepackage{afterpage}
\usepackage{placeins}

\bibpunct{(}{)}{;}{a}{}{,}

\newcommand{\MSun}{\ensuremath{M_{\odot}}\xspace}
\newcommand{\MJup}{\ensuremath{M_{\mathrm{Jup}}}\xspace}
\newcommand{\RJup}{\ensuremath{R_{\mathrm{Jup}}}\xspace}
\newcommand{\Teff}{\ensuremath{T_{\mathrm{e\!f\!f}}}\xspace}
\newcommand{\logg}{\ensuremath{\log g}\xspace}
\newcommand{\met}{\ensuremath{\mathrm{[Fe/H]}}\xspace}
\newcommand{\co}{\ensuremath{\mathrm{C/O}}\xspace}
\newcommand{\mic}{\ensuremath{\upmu\mathrm{m}}\xspace}
\newcommand{\as}{\hbox{$^{\prime\prime}$}\xspace}
\newcommand{\vsini}{\hbox{$v \sin i$}\xspace}
\newcommand{\loD}{\hbox{$\lambda/D$}\xspace}
\newcommand{\degre}{\degree\xspace}

\newcommand{\kms}{\ensuremath{\mathrm{km}\,\mathrm{s}^{-1}}\xspace}
\newcommand{\crires}{CRIRES$^{+}$\xspace}
\newcommand{\hirise}{HiRISE\xspace}

\newcommand{\formosa}{\texttt{ForMoSA}\xspace}
\newcommand{\uves}{UVES\xspace}
\newcommand{\jwst}{James Webb Space Telescope\xspace}

\begin{document}

\title{Characterization of AF\,Lep\,b\\at high spectral resolution with VLT/\hirise\footnote{Based on observations made with ESO Telescopes at the La Silla Paranal Observatory under programme ID 112.25FU}}
\titlerunning{AF\,Lep\,b characterization with VLT/\hirise}

\author{
    A.~Denis\inst{\ref{lam}}
    \and
    A.~Vigan\inst{\ref{lam}}
    \and
    J.~Costes\inst{\ref{lam}}
    \and
    G.~Chauvin\inst{\ref{lagrange}}
    \and
    A.~Radcliffe\inst{\ref{lesia}}
    \and
    M.~Ravet\inst{\ref{lagrange}, \ref{ipag}}
    \and
    W.~Balmer\inst{\ref{johnhopkins}, \ref{stsci}}
    \and
    P.~Palma-Bifani\inst{\ref{lagrange}, \ref{lesia}}
    \and
    S.~Petrus\inst{\ref{portales}, \ref{nucleo}}
    \and
    V.~Parmentier\inst{\ref{lagrange}}
    \and
    S.~Martos\inst{\ref{ipag}}
    \and
    A.~Simonnin\inst{\ref{lagrange}}
    \and
    M.~Bonnefoy\inst{\ref{ipag}}
    \and
    R.~Cadet\inst{\ref{lam}}
    \and
    T.~Forveille\inst{\ref{ipag}}
    \and
    B.~Charnay\inst{\ref{lesia}}
    \and
    F.~Kiefer\inst{\ref{lesia}}
    \and
    A.-M.~Lagrange\inst{\ref{lesia}}
    \and
    A.~Chiavassa\inst{\ref{lagrange}}
    \and
    T.~Stolker\inst{\ref{leiden}}
    \and
    A.~Lavail\inst{\ref{irap}}
    \and
    N.~Godoy\inst{\ref{lam}}
    \and
    M.~Janson\inst{\ref{albanova}}
    \and
    R.~Pourcelot\inst{\ref{stsci}}
    \and
    P.~Delorme\inst{\ref{ipag}}
    \and
    E.~Rickman\inst{\ref{esa}}
    \and
    D.~Cont\inst{\ref{munchen},\ref{garching}}
    \and
    A.~Reiners\inst{\ref{gott}}
    \and
    R.~De Rosa\inst{\ref{esoc}}
    \and
    H.~Anwand-Heerwart\inst{\ref{gott}}
    \and
    Y.~Charles\inst{\ref{lam}}
    \and
    A.~Costille\inst{\ref{lam}}
    \and
    M.~El Morsy\inst{\ref{sanantonio}}
    \and
    J.~Garcia\inst{\ref{lam}}
    \and
    M.~Houllé\inst{\ref{lam}}
    \and
    M.~Lopez\inst{\ref{lam}}
    \and
    G.~Murray\inst{\ref{durham}}
    \and
    E.~Muslimov\inst{\ref{lam},\ref{oxf},\ref{kazan}}
    \and
    G.~P.~P.~L.~Otten\inst{\ref{lam},\ref{taiwan}}
    \and
    J.~Paufique\inst{\ref{esog}}
    \and
    M.~Phillips\inst{\ref{exeter},\ref{ifa}}
    \and
    U.~Seemann\inst{\ref{esog}}
    \and
    A.~Viret\inst{\ref{lam}}
    \and
    G.~Zins\inst{\ref{esog}}
    }

\institute{
    Aix Marseille Univ, CNRS, CNES, LAM, Marseille, France \label{lam}
    \\ \email{\href{mailto:allan.denis@lam.fr}{allan.denis@lam.fr}}
    \and
    Laboratoire J.L. Lagrange, Université Côte d'Azur, Observatoire de la côte d'Azur, CNRS, 06304 Nice, France \label{lagrange}
    \and
    LESIA, Observatoire de Paris, Université PSL, Sorbonne Université, Université de Paris, 5 place Jules Janssen, 92195 Meudon, France \label{lesia}
    \and
    Univ. Grenoble Alpes, CNRS, IPAG, F-38000 Grenoble, France \label{ipag}
    \and
    Department of Physics \& Astronomy, John Hopkins University, 3400 N. Charles Street, Baltimore, MD 21218, USA \label{johnhopkins}
    \and
    Space Telescope Science Institute, 3700 San Martin Drive, Baltimore, MD 21218, USA \label{stsci}
    \and
    Instituto de Estudios Astrofísicos, Facultad de Ingeniería y Ciencias, Universidad Diego Portales, Av. Ejército 441, Santiago, Chile \label{portales}
    \and
    Millennium Nucleus on Young Exoplanets and their Moons (YEMS), Santiago, Chile \label{nucleo}
    \and
    Leiden Observatory, Leiden University, Einsteinweg 55, 2333 CC Leiden, The Netherlands \label{leiden}
    \and
    Institut de Recherche en Astrophysique et Planétologie (IRAP), 9 avenue Colonel Roche, BP 44346, 31028 Toulouse \label{irap}
    \and
    Department of Astronomy, Stockholm University, AlbaNova University Center, 10691 Stockholm, Sweden \label{albanova}
    \and
    European Space Agency (ESA), ESA Office, Space Telescope Science Institute, 3700 San Martin Drive, Baltimore, MD 21218, USA \label{esa}
    \and
    Universitäts-Sternwarte, Ludwig-Maximilians-Universität München, Scheinerstraße 1, 81679 München, Germany \label{munchen}
    \and
    Exzellenzcluster Origins, Boltzmannstraße 2, 85748 Garching, Germany \label{garching}
    \and
    Institute for Astrophysics und Geophysik, Georg-August University, Friedrich-Hund-Platz 1, 37077 Göttingen, Germany \label{gott}
    \and
    European Southern Observatory, Alonso de Cordova 3107, Vitacura, Santiago, Chile \label{esoc}
    \and
    Department of Physics and Astronomy, University of Texas-San Antonio, San Antonio, TX, USA \label{sanantonio}
    \and
    Center for Advanced Instrumentation, Durham University, Durham, DH1 3LE, United Kindgom \label{durham}
    \and
    Dept. of Astrophysics, University of Oxford, Keble Road, Oxford, OX1 3RH, UK \label{oxf}
    \and
    Optical and Electronic Systems Department, Kazan National Research Technical University \label{kazan}
    \and
    Academia Sinica, Institute of Astronomy and Astrophysics, 11F Astronomy-Mathematics Building, NTU/AS campus, No. 1, Section 4, Roosevelt Rd., Taipei 10617, Taiwan \label{taiwan}
    \and
    European Southern Observatory (ESO), Karl-Schwarzschild-Str. 2, 85748 Garching, Germany \label{esog}
    \and
    Physics \& Astronomy Dpt, University of Exeter, Exeter, EX4 4QL, UK \label{exeter}
    \and
    Institute for Astronomy, University of Hawaii at Manoa, Honolulu, HI 96822, USA \label{ifa}
}

\date{Received 21 November 2024; accepted 26 February 2025}

\abstract
{Since the recent discovery of the directly imaged super-Jovian planet AF\,Lep\,b, several studies have been conducted to characterize its atmosphere and constrain its orbital parameters. AF\,Lep\,b has a measured dynamical mass of $3.68 \pm 0.48$\,\MJup, a radius of $1.3 \pm 0.15$\,\RJup, a nearly circular orbit in spin-orbit alignment with the host star, a relatively high metallicity, and a near-solar to super-solar \co ratio. However, key parameters such as the rotational velocity and radial velocity could not be estimated as they require high-resolution spectroscopic data that is impossible to obtain with classical spectrographs.}
{AF\,Lep\,b was recently observed with the new \hirise visitor instrument at the VLT, with the goal of obtaining high-resolution (R $\approx$ 140,000) spectroscopic observations to better constrain the orbital and atmospheric parameters of the young giant exoplanet.}
{We compare the extracted spectrum of AF\,Lep\,b to self-consistent atmospheric models using \formosa, a forward modeling tool based on Bayesian inference methods. We then use our measurements of the radial velocity of the planet to provide new constraints on the orbit of the planet.}
{From the forward modeling, we find a \co ratio that aligns with previous low-resolution analyses, and we confirm the super-solar metallicity. We also confirm unambiguously the presence of methane in the atmosphere of the companion. Based on all available relative astrometry and radial velocity measurements of the host star, we show that two distinct orbital populations are possible for the companion. We derive the radial velocity of AF\,Lep\,b to be $10.51 \pm 1.03$\,\kms, and show that this value agrees well with one of the two orbital solutions, allowing us to rule out an entire family of orbits. Additionally, assuming that the rotation and orbit are coplanar, the derived planet's rotation rate is consistent with the observed trend of increasing spin velocity with higher planet mass.}
{}


\maketitle

\section{Introduction}
\label{sec:introduction}

In the wake of the large-scale surveys conducted with ground-based extreme adaptive optics planet imagers like GPI at Gemini-South \citep{Macintosh2014} and SPHERE at the VLT \citep{Chauvin2017sf2a}, direct imaging has recently passed a fundamental milestone in the selection and follow-up of sources using the astrometric acceleration measurements information provided by the \textit{Gaia} telescope \citep{Kervella2019,Brandt2021HipGaia,Kervella2022}. This astrometric information considerably increases the success rate of surveys based on this prior knowledge compared to fully blind searches \citep[][]{Nielsen2019,Vigan2021}, as illustrated by several imaging discoveries of substellar companions over the recent years \citep{Bowler2021,Bonavita2022,Currie2023,rickman_discovery_2024}. It also offers the unique possibility of inferring the dynamical masses of the companions, which in turn can be used to directly constrain the formation and evolution models of giant planets \citep{Brandt2021,Zhang2024}.

An emblematic discovery in that context concerns the accelerating star AF\,Lep (HD\,35850, HR\,1817, HIP\,25486) around which the discovery of a young giant planet (hereafter AF\,Lep\,b) was announced almost simultaneously by three independent teams \citep{DeRosa2023,Mesa2023,franson_astrometric_2023}. The detection was obtained in direct imaging, but based on a selection of the target using the long time baseline astrometric acceleration, or proper-motion anomaly, measured between the \textit{Hipparcos} and \textit{Gaia} missions \citep{Brandt2021HipGaia}.

AF Lep is a $1.09\pm0.06$\,\MSun star, of spectral type F8 \citep{gray_contributions_2006}, located at a distance of $26.825\pm0.014$\,pc \citep{bailer-jones_estimating_2021} with a super-solar metallicity \citep[\met = $0.29 \pm 0.03$, estimated from a high S/N HARPS spectrum,][]{perdelwitz_analysis_2024}. As member of the $\beta$ Pictoris moving group, the isochronal age of the star is estimated at $24\pm3$\,Myr \citep{Bell2015}. The mass and orbital elements of the young giant planet AF\,Lep\,b have been recently estimated by combining all available direct imaging data, including an early detection from 2011 obtained from archival VLT/NaCo data \citep{Bonse2024} and VLTI/GRAVITY data \citep{balmer_vltigravity_2024}: it has an estimated dynamical mass of $3.68^{+0.47}_{-0.48}$\,\MJup, and is orbiting at a semi-major axis of $8.98^{+0.15}_{-0.08}$\,au with an eccentricity $e = 0.013^{+0.010}_{-0.024}$ and an inclination of $i = 57.5^{+0.6}_{-0.7}$\,\degree.

Recent chemically-consistent retrievals for AF\,Lep\,b using \texttt{petitRADTRANS} \citep{molliere_petitradtrans_2019} of all previously published emission spectra and photometry spanning 0.9--4.2\,\mic with a maximum resolution of $R_{\lambda} = 30$ confirmed a cool temperature of $\Teff \simeq 800$\,K and low surface gravity of $\rm{log(g)} \simeq 3.7$\,dex, with the presence of silicate clouds and disequilibrium chemistry in the atmosphere of AF\,Lep\,b, as expected for a young, cold, early-T super-Jovian planet \citep{zhang_elemental_2023}. This analysis also revealed a metal-enriched atmosphere ($\met > 1.0$\,dex) compared to the host star's metallicity, supported by follow-up study including \textit{JWST/NIRCam} observations \citep{Franson2024}. These results were independently confirmed by \cite{palma-bifani_atmospheric_2024} who used a forward-modelling approach with the Exo-REM atmospheric radiative-convective equilibrium model, which includes the effects of non-equilibrium processes and clouds \citep{charnay_self-consistent_2018}. The results of the latter study point towards a lower metallicity of 0.5--0.7 and a \co ratio solution ranging between 0.4 and 0.8, compatible with a solar to super-solar value ($\co_{\odot} = 0.55$, \citealt{Asplund2009}).

We report here the results of much higher spectroscopic observations of AF\,Lep\,b made with the new High-Resolution Imaging and Spectroscopy of Exoplanets \citep[\hirise,][]{Vigan2024} instrument at the Very Large Telescope (VLT), which combines the exoplanet imager SPHERE \citep{Beuzit2019} with the recently upgraded high-resolution spectrograph \crires \citep{Dorn2023}. \hirise operates in the $H$-band at a spectral resolution of the order of $R_{\lambda} = 140\,000$. In Sect.~\ref{sec:obs} we present the observations and calibration data related to AF Lep. In Sect.~\ref{sec:data_reduction}, we describe the \hirise data reduction and signal extraction steps applied to the data, in particular the corrections for the tellurics and instrumental response and the wavelength calibration. The strategy used to model the data is presented in Sect.~\ref{sec:data_modeling}. In Sect.~\ref{sec:atmospheric_characterization} we report the results of our forward modelling atmospheric analysis. In Sect.~\ref{sec:orbit_fitting}, we present new constraints coming from the radial velocity (RV) measurements of the planet itself and their implications for the orbital solutions derived in combination with previous studies. Finally, we discuss our results in Sect.~\ref{sec:discussion} and conclude in Sect.~\ref{sec:conclusions}.

\section{Observations}
\label{sec:obs}

\begin{table*}
  \caption[]{AF Lep observations.}
  \renewcommand{\arraystretch}{1.2}
  \label{tab:observations}
  \centering
  \begin{tabular}{cccccccccc}
    \hline\hline
    UT date    & Object     & Setting & Offset & DIT   & Integration time & Airmass    & Seeing        & Coherence time & Transmission \\
               &            &         &        & [min] & [min]            &            & [\as]         & [ms]     & [\%] \\
    \hline
    2023-11-20 & AF\,Lep\,A & H1567   & 1     & 2     & 2                & 1.20       & 0.6            &          & 4.0\% \\
    2023-11-20 & AF\,Lep\,b & H1567   & 1     & 20    & 100              & 1.19--1.05 & 0.6            & 4.7      & \\
    2023-11-20 & Background & H1567   & 1     & 2     & 2  \\
    2023-11-20 & Background & H1567   & 1     & 20    & 100 \\
    \hline
    2023-11-23 & AF\,Lep\,A & H1567   & 1     & 2     & 4                & 1.25       & 0.8--1.0       & 4.6       & 3.8\% \\
    2023-11-23 & AF\,Lep\,A & H1567   & 2     & 2     & 4                & 1.25       &    0.8--1.0       & 4.6       & 3.8\% \\
    2023-11-23 & AF\,Lep\,b & H1567   & 1     & 20    & 80               & 1.19--1.03 & 0.7--1.0       & 5.1       & \\
    2023-11-23 & AF\,Lep\,b & H1567   & 2     & 20    & 60               & 1.19--1.03 &    0.7--1.0       & 5.1       & \\
    2023-11-23 & Background & H1567   & 1     & 2     & 4  \\
    2023-11-23 & Background & H1567   & 2     & 2     & 4  \\
    2023-11-23 & Background & H1567   & 1     & 20    & 100 \\
    2023-11-23 & Background & H1567   & 2     & 20    & 20 \\
    \hline
  \end{tabular}
  \tablefoot{During the second night we used an offset procedure in which the science fiber is shifted by approximately 10 pixels along the slit (see Sect. \ref{sec:obs}).}
\end{table*}

The AF Lep system was observed on UT 2023 November 20 and 23 with VLT/\hirise. These observations are summarised in Table~\ref{tab:observations}. At each epoch, we started the observations by placing the host star on the science fiber to acquire a reference spectrum. We did not use a coronagraph for the observations to maximize the end-to-end transmission of the instrument \citep{Vigan2024}. The centering of the star on the single-mode fiber was optimized using a dedicated procedure performed on the internal source of the SPHERE instrument \citep{ElMorsy2022,Vigan2024}, which provides a typical centering accuracy better than 8\,mas (0.2\loD in the $H$ band). The \crires spectrograph was set up in the \texttt{H1567} spectral setting and the detector integration time (DIT) of the science detector was set to 120\,sec, resulting in $\sim$20\,000\,ADU per exposure, the saturation limit of the \crires detector beeing 37\,000\,ADU (see Table 2 of the \href{https://www.eso.org/sci/facilities/paranal/instruments/crires/doc/CRIRES_User_Manual_P108_Phase2.pdf}{\crires user manual}). Then, the centering procedure was used again to place the planetary companion AF\,Lep\,b on the science fiber. For this, we relied on the astrometric calibration of the tracking camera obtained during the \hirise commissioning \citep{Vigan2024} and on the accurate astrometry of the companion measured a few days earlier by VLTI/GRAVITY \citep[][private communication]{balmer_vltigravity_2024}. For the observations of the companion, the DIT of the science detector was set to 1\,200\,s, resulting in $\sim$60\,ADU per exposure. This DIT was chosen as a compromise between the number of detector readouts over a long exposure, which will increase the overall noise level, and the risk of losing an exposure in case of unstable observing conditions. During all science observations, the internal metrology of \crires is enabled to provide an improved precision in the wavelength solution.

After the science exposures, we also acquired sky backgrounds on an empty region of the sky about 30\as away from AF\,Lep. The goal of these exposures is mainly to help subtract the leakage term associated to the \hirise MACAO guide fiber described in \citet{Vigan2024}. Although this leakage term has been decreased by a factor $\sim$10 since commissioning by the addition of an optical attenuator on the guide fiber, it is still visible in long exposures and needs to be properly subtracted before the star and companion's spectra can be extracted. We note that this strategy is not optimal in very low signal regime because subtracting the same background to all science exposures tends to increase the noise level. Fortunately the leakage term is stable, and the 1\,200\,s DIT was used for all targets during the \hirise observing run, so we were able to combine the different backgrounds to decrease the noise in the backgrounds. The impact of the number of backgrounds on the derived parameters of the planet is discussed in Appendix~\ref{sec:S/N_ratio_impact}.

The two epochs on AF\,Lep are mostly identical, except that on the second night we used an offset procedure where the science fiber is moved by $\sim$10\,pixels along the slit axis to help reducing the impact of the bad pixels on the \crires science detectors. This procedure intends to move the science signal with respect to static bad pixels, but the two offsets are not subtracted to each other like in classical nodding. This is why an offset of a few pixels is enough since the full width at half-maximum (FWHM) of the science signal is 2\,pixels. On both nights the companion was observed at very low airmass, but the observing conditions during the first night were more stable: the DIMM seeing on the first night remained at 0.6\as with very little variations, while on the second night it varied between 0.7\as and 1.0\as. The coherence time, $\tau_0$, was very similar during the two nights, with a value between 4.6 and 5.1\,ms. The end-to-end transmission of the system was measured on the star, with 95$^{th}$ percentile values of 4.0\% and 3.8\% on the first and second nights, respectively. These values are consistent with instrumental predictions, which confirm that the centering of the stellar PSF on the science fiber was within the specifications of 0.2\,\loD in H-band.

Standard \crires calibrations were acquired automatically the next morning based on the science observations of each night. This includes dark, flat fields and wavelength calibration files, as detailed in the \crires calibration plan. \hirise observations do not require any specific internal calibrations in addition to standard daily calibrations.

\section{Data reduction}
\label{sec:data_reduction}

\begin{figure*}
  \centering
  \includegraphics[width=\textwidth]{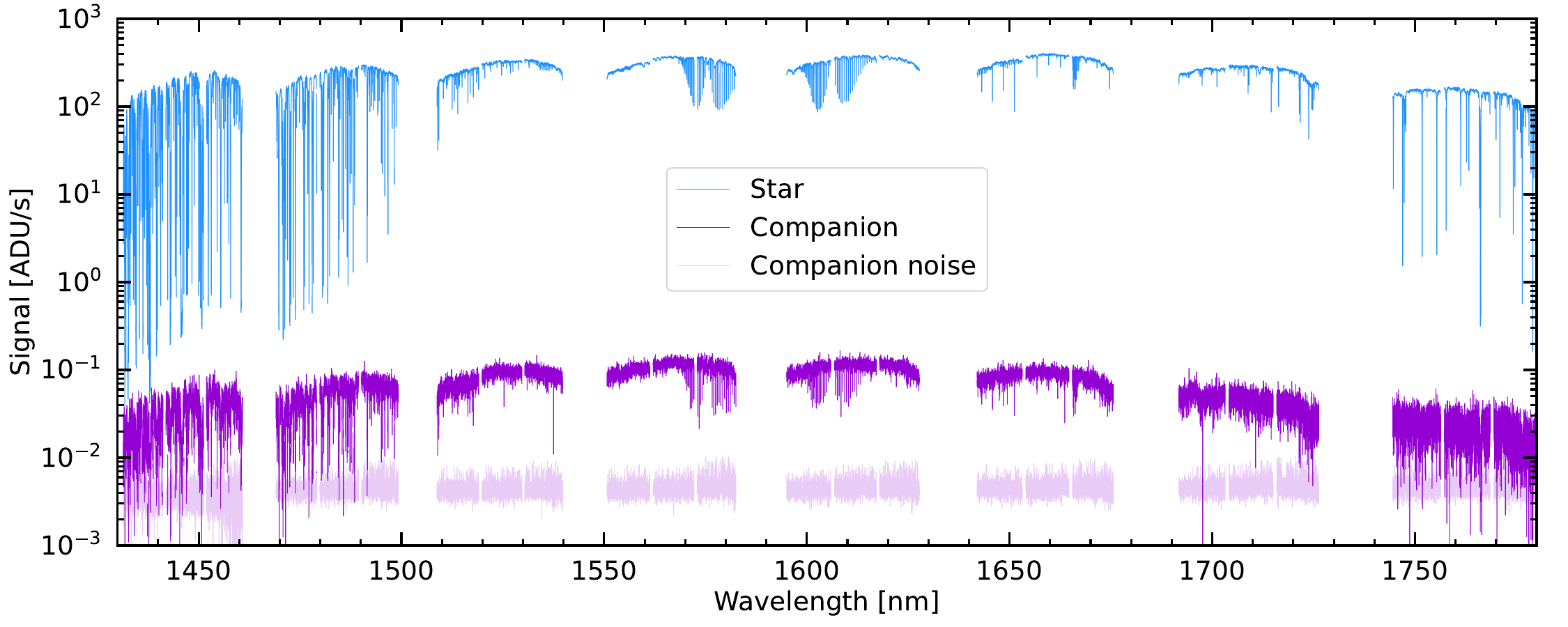}
  \caption{Data for the star and companion obtained on 2023-11-23 and extracted using our custom pipeline. The read-out noise of the detector is also plotted to demonstrate that the data obtained on the companion has a mean S/N of $\sim$25 per spectral channel at the center of the $H$ band. The effect of the telluric lines is clearly visible at the start and end of the band, but also close to the center where many telluric CO bandheads are visible. We remind that data for the companion refers to the spectrum at the location of the companion which contains primarily stellar flux.}
  \label{fig:hirise_data}
\end{figure*}

\subsection{Raw data reduction and signal extraction}
\label{sec:reduction_extraction}

A dedicated pipeline was developed to reduce and calibrate \hirise data \citep{hipipe2024}\footnote{\url{https://gitlab.lam.fr/hirise/pipeline}}. This pipeline relies on the official \crires pipeline provided by ESO\footnote{\url{https://www.eso.org/sci/software/pipelines/cr2res/cr2res-pipe-recipes.html}} to produce the flat-field and wavelength calibrations, and on custom python routines for data combination, spectral extraction, filtering and wavelength recalibration. We briefly describe below the main reduction steps performed with our python tools.

The first step is to clean the raw data, independently for the star and the companion. The appropriate backgrounds are median-combined and subtracted to the science frames, and the science frames are divided by the detector flat field generated by the \crires pipeline. The bad pixels, flagged by the \crires pipeline, are all replaced by NaN values in the images. Finally, if multiple science frames are available, they are also mean-combined to increase the signal-to-noise ratio (S/N) of the data.

The second step is to locate the trace position of the eight orders dispersed on the three science detectors of \crires. For this step we use the science data acquired on the star, which is usually at high S/N at the location of the science fiber. For each of the 24 orders segment, we fit a 1d Gaussian function in each of the 2048 spectral channels to provide an accurate position of the trace on the detector and of its FWHM. Some regions of the spectrum are strongly affected by the absorption of telluric lines, which may result in a poor fit. We remove these channels using an iterative sigma-clipping that detects the outliers with respect to a parabolic fit to the position of the trace. The algorithm converges in 3 iterations and the final trace position is defined as the result of the parabolic fit to the trace with all the outliers removed.

The third step is the extraction of the stellar and companion signal in each spectral channel. In each spectral channel, the signal is summed in a 6-pixel window for the companion and the star, centered around the trace position measured at the previous step. We refer to Fig.~\ref{fig:trace} of Appendix~\ref{sec:detector traces} for more details. The noise is estimated as the standard deviation in a 20-pixel window located 60 pixels away from the location of the science fiber. In this spectral extraction step, the impact of bad pixels is estimated by computing a weight parameter in each spectral channel. The signal coming through the science fiber is Gaussian, with a standard deviation in the spatial direction of $\sim$0.85\,pix that is stable over all segments of orders. In each spectral channel, the weight is computed by the integrated value of the normalized Gaussian of standard deviation 0.8\,pix, centered at the calibrated position of the trace, with a value of zero attributed to the pixels flagged as bad. The weight is equal to 1 when there are no bad pixels in the extracted window, 0 when there are only bad pixels, and intermediate values between 0 and 1 when there are some bad pixels in the extracted window. This method only estimates the impact of bad pixels but cannot be used to compensate for them. For now, in the data analysis, we simply remove all data that result in an extracted window containing at least one bad pixel.

A fourth optional step can be executed when the observations have been performed with an offset procedure, e.g. for our observations of AF\,Lep on the second night. In this case, the spectra obtained at the two offset position will be affected by different bad pixels and can be combined. The weight vectors computed at the previous step are used to help with the combination: channels where weights equal to 1 in both channels will be averaged together, while channels where weight of one spectrum is equal to 1 and the other is not will use the value of the spectrum not affected by bad pixels. Channels where both spectra have a value $< 1$ will be flagged as bad and will be discarded in the analysis. We measure that typically $< 1\%$ of channels are bad in both spectra in the observation of AF\,Lep on the second night.

Finally, a filtering step is applied to the extracted signals for the star and the companion to remove strong outliers that may be the result of bad pixels that were not properly flagged in the initial calibrations. This filtering step usually removes less than 0.05\% of the data.

Figure~\ref{fig:hirise_data} illustrates the data extracted on the star and companion from the 2023-11-23 observations. The two offsets have been combined to remove the impact of bad pixels. The spectra are mostly shaped by (i) the overall transmission of the system, which gives an overall parabolic shape, (ii) the blaze function of the spectrograph that adds an additional parabolic shape to each of the eight individual orders, and (iii) the deep telluric lines that affect the data below 1500\,nm, above 1750\,nm and in some specific regions in between. As the scientific data is located at specific positions of the fibers on the detector, in background exposures without strong starlight diffraction, the main part of the detector contains only the detector's read-out noise, allowing us to estimate this noise. We estimate a S/N of $\sim$25 per spectral channel at 1600\,nm for the data acquired at the location of the companion. Although the spectrum obtained at the location of the companion is still dominated by the stellar PSF coupling into the science fiber, we refer to the spectrum obtained at that location as the ``spectrum of the companion'' for practicality.

\subsection{Sky and instrumental response}
\label{sec:sky_response}

The sky and instrumental transmissions are an important parameter for the modelling of the data and the search for the planet's signal (see Sect.~\ref{sec:data_modeling}). At high spectral resolution, the sky imprints many telluric lines over the spectra with varying width and depth. Then, the telescope and instrument also have an impact in shaping the signal, although this instrumental contribution is mostly smooth with wavelength. The main contributors that do not have a flat or almost flat contribution are the SPHERE dichroic filter and the \crires blaze function \citep[see Fig.~10 of][]{Vigan2022spie}. Finally, the \crires science detectors have numerous bad pixels that have an impact on the extracted signal that needs to be taken into account or modelled.

We use the observations of dedicated early-type stars to compute the telluric and instrumental response, in that case  $\beta$\,Pic\,A (A6) on 2023-11-20 and AF\,Lep\,A itself (F8) on 2023-11-23. The spectra measured for the calibrators are divided by a PHOENIX stellar model from \citet{Husser2013} at the appropriate \Teff and \logg, rotationally broadened using the measured \vsini for the stars, and velocity shifted by the known RV of the stars. This division effectively removes the stellar effect, leaving mostly the telluric and instrument effects. The response is then normalized to have a median value of one.

\subsection{Wavelength recalibration}
\label{sec:wave_recal}

\begin{figure*}
  \centering
  \includegraphics[width=\textwidth]{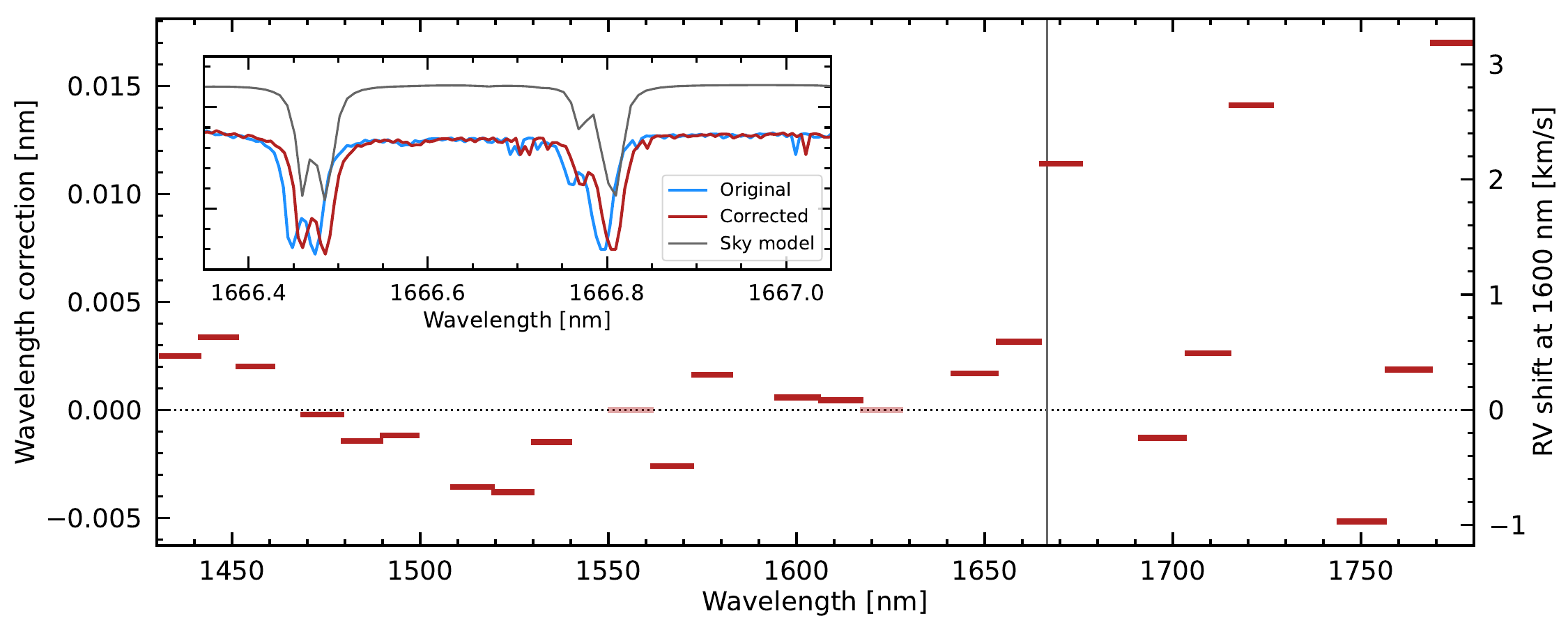}
  \caption{Corrections to the wavelength solution for each segments of orders based on the analysis of the telluric lines for the AF\,Lep data acquired on 2023-11-23. For the two segments centered at 1556 and 1623\,nm (light red), there are not enough telluric lines to compute an accurate correction and the default wavelength solution is therefore adopted. The right axis of the main plot shows the RV shift corresponding to the level of correction of the left axis, computed at 1600\,nm. The top-left inset shows the effect of the correction at 1666.5\,nm (grey vertical line in the main plot), which is a segment that requires one of the largest corrections. The telluric model obtained from \texttt{SkyCalc} is overplotted.}
  \label{fig:wavelength_recalibration}
\end{figure*}

Another critical aspect of high-resolution spectroscopy is the wavelength calibration of the data. Previous authors have already demonstrated the importance of adopting an additional level of correction to the wavelength solution provided by the \crires pipeline \citep[e.g.,][]{landman_beta_2024,Nortmann2024}.

For the \hirise observations of AF\,Lep, we use a correction based on the telluric lines imprinted into the spectrum of the star. The main limitation of this approach is that telluric lines are not evenly distributed over the $H$ band and there are even small parts of the band where no telluric lines are detectable. This prevents using a quadratic wavelength correction in most segments of orders \citep[as in][]{landman_beta_2024}, so we finally settled on a simple constant correction for each of the segments. The resulting corrections are showed in Fig.~\ref{fig:wavelength_recalibration}. They are all below 0.02\,nm, and even below 0.005\,nm for 20 of the 24 segments. The associated RVs of the correction are up to 3 \kms as shown in Fig.~\ref{fig:wavelength_recalibration}. This demonstrates the importance of the wavelength recalibration. To compute the wavelength correction coefficients for each of the segments, we use a nested sampling algorithm \citep{Skilling2006}. For the loglikelihood function, we use the CCF mapping defined in \cite{zucker_cross-correlation_2003}. More details on the nested sampling algorithm are given in Sect.~\ref{sec:formosa}. In conjunction with the wavelength constant corrections, the nested sampling algorithm estimates the uncertainties of these corrections. In the telluric dominated segments, where we have high confidence in our wavelength recalibration, we estimate uncertainties in the RV correction on the order of 100 m/s.

In the future, a more evolved recalibration is foreseen using dedicated observations of an M star calibrator. This approach is used for the KPIC instrument \citep[e.g.,][]{Wang2021} and has demonstrated good accuracy \citep{Morris2020, Ruffio2023, horstman_fringing_2024}. For \hirise, an M star calibrator was observed every night, but we detected some shifts in wavelength between the calibrator and the science target during the first night, which are not fully understood yet. For the present work, we consider the self-calibration of the data using the telluric lines safer since the same deep telluric lines are observed at identical positions in both the AF\,Lep\,A and AF\,Lep\,b data.

Finally, the recalibrated wavelength is corrected for the mean barycentric component computed for the time of the science exposures. The value is computed using the \texttt{helcorr} function of the \texttt{PyAstronomy} package \citep{PyAstronomy}.

\section{Data modeling}
\label{sec:data_modeling}

As seen in Sect.~\ref{sec:data_reduction}, the extracted 1D spectrum at the location of the companion is dominated by starlight diffraction and speckles coupling into the science fiber. Since we do not have an independent measurement of this stellar contamination (e.g., from a measurement at the same separation but with a different position angle), we must rely on a joint estimation of the stellar contamination and the planetary model. This approach is feasible because the two components have very difference spectral shapes. To achieve this, we adopt a method similar to \cite{landman_beta_2024}. The signal extracted from the science fiber at the location of the companion can be decomposed into a planetary signal term, $d_p$, a starlight contamination term, $d_s$, and an additional noise term, $\eta$:

\begin{equation}
    d(\lambda) = d_p(\lambda) + d_s(\lambda) + \eta,
\label{eq1}
\end{equation}
where d is the reduced 1D spectrum at the location of the companion. Following \cite{landman_beta_2024}, we can modulate the stellar contamination at the location of the companion with the equation:
\begin{equation} \label{eq:star_contamination}
    d_s(\lambda) = c_s\alpha(\lambda)f_s(\lambda)
\end{equation}
where $c_s$ is a scaling factor, $\alpha$ is a low-order function and $f_s$ is the stellar master spectrum, that is the observation of the star when the science fiber is centered on the star. Following \cite{landman_beta_2024}, we directly estimate this modulation from the data, which corresponds to the following final model for the starlight contamination:
\begin{equation}
    d_s(\lambda) = c_s \frac{\mathcal{L}\left[d(\lambda)\right]}{\mathcal{L}\left[f_s(\lambda)\right]}f_s(\lambda),
\end{equation}
where $\mathcal{L}$ is a lowpass filtering operation. For this filtering operation, we used a Savitzky-Golay filter of order 2 with a kernel size of 301 pixels. This value was empirically determined by maximizing a cross-correlation function (CCF) with template models for several HiRISE targets (see e.g. Sect~\ref{sec:molecules_detection} for AF\,Lep\,b). We note that a similar value was also determined and used by \citet{landman_beta_2024} on \crires data. The final results are not very sensitive to the value adopted for this parameter as long as the continuum is removed.

Similarly to \citet{landman_beta_2024}, the planetary contribution $d_p$ can be written as:
\begin{equation} \label{eq:planet_model}
    d_p(\lambda) = c_p \left(M_{p,LSF}(\lambda) T(\lambda) - f_s\frac{\mathcal{L}\left[T(\lambda)M_{p,LSF}(\lambda)\right]}{\mathcal{L}\left[f_s(\lambda)\right]}\right),
\end{equation}
where $c_p$ is a linear scaling factor accounting for the brightness of the companion and M$_{p,LSF}$ is the model of the planet convolved at the spectral resolution of the instrument. The terms $f_s\frac{\mathcal{L}\left[T(\lambda)M_{p,LSF}(\lambda)\right]}{\mathcal{L}\left[f_s(\lambda)\right]}$ comes from the leaking of the planet model continuum into the estimate of $\alpha$ \citep[see][]{landman_beta_2024}. This effectively removes the continuum of the planet model.

The total transmission can be estimated using our stellar master spectrum, $f_s$. We simply divide the stellar master spectrum by a model of the star:
\begin{equation}
    T(\lambda) = \frac{f_s(\lambda)}{M_s(\lambda)},
\end{equation}
where $M_s(\lambda)$ is a PHOENIX model selected at the known effective temperature and surface gravity, and rotationally broadened (see Sect. \ref{sec:sky_response}).

For simplicity, we will drop the ($\lambda$) notation in the rest of the paper, but it is important to note that all the parameters used in this model depend on the wavelength. Our final model can be written as:
\begin{equation}
    d =  c_s\frac{\mathcal{L}\left[d\right]}{\mathcal{L}\left[f_s\right]} f_s + c_p \left(M_{p,LSF} T - \frac{f_s \mathcal{L}\left[T M_{p,LSF}\right]}{\mathcal{L}\left[f_s\right]}\right) + \eta
    \label{eq:model}
\end{equation}

We can forward model our spectrum in order to jointly estimate the stellar and planetary contributions to the data. Equation~\ref{eq:model} can be written in matrix form using the same conventions as in \citealt{Wang2021}:
\begin{equation}
    \left(\begin{matrix}
        \vdots \\
        d_i\\
        \vdots
    \end{matrix}\right) =
    \left(\begin{matrix}
        \vdots & \vdots \\
        \frac{\mathcal{L}\left[d\right]}{\mathcal{L}\left[f_s\right]}f_s & M_{p,LSF}T-\frac{\mathcal{L}\left[M_{p,LSF}T\right]}{\mathcal{L}\left[f_s\right]}f_s \\
        \vdots & \vdots
    \end{matrix}\right)
    \left(\begin{matrix}
    c_s \\
    c_p
    \end{matrix}\right) + \eta,
    \label{eq:model_matrix}
\end{equation}

where the left-hand side of the equation corresponds to a column vector with length equal to the number of spectral channels, $N_{\lambda}$. The first matrix in the right-hand side of the equations has an $N_{\lambda}\times 2$ dimension. The last term in the right-hand side equations is just a vector of length 2 corresponding to the linear scaling coefficients $c_s$ and $c_p$ that we want to determine. Finally, we can write this model as:
\begin{equation}
    d = M_{\psi}*c + \eta,
    \label{eq:model_LSQ}
\end{equation}
where $\psi$ represents the planetary parameters that will determine the shape of the planetary spectrum model, $M_{p,LSF}$. The parameters include the atmospheric parameters \Teff, \logg, \met and \co (see Sect. \ref{sec:atmospheric_characterization}), but also the RV shift and the projected rotational velocity \vsini. For the projected rotational velocity, we used the \texttt{fastRotBroad} function from PyAstronomy \citep{PyAstronomy}.

\section{Atmospheric characterization}
\label{sec:atmospheric_characterization}

To characterize the atmosphere of AF\,Lep\,b, we use forward modeling analysis, consisting in using pre-computed grids of self-consistent models that can cover a range of parameters, including \Teff, \logg, \met, \co, or more parameters (see Sect.~\ref{sec:atmosheric_models}). We use the \formosa python package \citep{petrus_medium-resolution_2021}\footnote{\url{https://formosa.readthedocs.io/en/latest/}}, which we upgraded to work efficiently with high-spectral resolution data and in which we implemented the model described in Sect.~\ref{sec:data_modeling}. We present our results in Sects.~\ref{sec:atmosheric_models} and \ref{sec:molecules_detection}.

\subsection{\formosa: A forward modeling analysis tool}
\label{sec:formosa}

\formosa has already been extensively described and used in previous works to characterize exoplanets such as HIP\,65426\,b \citep{petrus_medium-resolution_2021}, VHS\,1256\,AB\,b \citep{petrus_x-shyne_2023,petrus_jwst_2024}, AB\,Pic\,b \citep{palma-bifani_peering_2023} and AF\,Lep\,b \citep{palma-bifani_atmospheric_2024}. These works were all based on low and medium resolution data.

\formosa relies on a nested sampling algorithm involving Bayesian inference \citep{Skilling2006}. The nested sampling method was built to naturally estimate the marginal likelihood, using a multi-surface approach of space parameters exploration during the Bayesian inversion. The marginal likelihood $z$ is defined as:
\begin{equation}
    z = \int \pi (\theta )L(\theta )d\theta,
\end{equation}
with $L(\theta)$ the likelihood function and $\pi (\theta)$ the prior distribution. The prior distribution and the likelihood function are defined by the users, given assumptions on the model. Calculating the value of z has the advantage of different model assumptions to be compared through the ratio of their evidence values. This ratio is known as the Bayes factor. Let $z_1$ and $z_2$ be the marginal likelihoods of two set of model assumptions. Then the Bayes factor is defined as:
\begin{equation}
    B = \frac{z_1}{z_2}
\end{equation}
However, as \formosa returns the logarithm of the evidence ($\log z$), we can make use of the logarithm of the Bayes factor. Two model assumptions can then be compared through the difference of the logarithm of the Bayes factor:
\begin{equation}
    \log B = \log z_1 - \log z_2
\end{equation}
A positive value of $\log B$ denotes a statistical preference of the data to the first model assumption. More quantitatively, Bayes factors can be interpreted against the Jeffrey scale \citep[see Table 2 of][]{benneke_how_2013}, in which case a value of $\log B$ above 5 is considered a strong statistical preference for the first model assumption compared to the second model assumption.

For this work, we chose the following log-likelihood function:
\begin{equation}
    \chi _0^2 = (d-M_{\psi}\hat c)^{T}\Sigma_0^{-1}(d-M_{\psi}\hat c),
\label{Chi2}
\end{equation}
where $\Sigma_0$ is the covariance matrix of the data and $\hat c$ is the linear least squares solution to Eq.~\ref{eq:model_LSQ} such that:
\begin{equation}
    M_{\psi}^{T}\Sigma_0^{-1}M_{\psi}c = M_{\psi}^{T}\Sigma_0^{-1}d
    \label{eq:LSQ}
\end{equation}
For the covariance matrix, we assume a simple model where the photon noise, mainly coming from the stellar contamination, is considered as the dominant source of noise in the data. Table \ref{tab:observations} shows that the read-out noise is very low, which justifies this assumption. For this paper, we also assume uncorrelated pixels, which means a diagonal covariance matrix. Practically, this matrix is calculated as the continuum of the extracted 1D spectrum at the location of the companion:
\begin{equation}
    \Sigma_0 = \mathcal{L}\left[d\right]
\end{equation}

To account for flux calibration offsets between spectral segments, which can impact the final results, we apply Eq.~\ref{eq:model} to each order separately in order to properly rescale the stellar contamination to the level of the flux of the data. This also allows us to better estimate the continuum of the data as it can be problematic to evaluate the continuum with such offsets between segments.

Since telluric lines dominate the spectrum up to 1500\,nm and from 1740\,nm onward (see Fig. \ref{fig:hirise_data}), we exclude the spectral segments corresponding to these parts of the spectrum from our analysis. This represents a total of nine spectral segments out of the 24 available. We thus end up using 15 spectral segments, covering a wavelength range approximately between 1500 and 1730\,nm.

\subsection{Atmospheric models}
\label{sec:atmosheric_models}

We run \formosa with a model referred to as Exo-REM/Exo\_k. This model is a result of using Exo-REM volume mixing ratio profiles in tandem with Exo\_k to produce high-resolution spectra (Radcliffe et al., in prep). On one hand, Exo-REM is a radiative-convective model developed to simulate atmospheres of young giant exoplanets \citep{baudino_interpreting_2015,charnay_self-consistent_2018,charnay_formation_2021}. The chemical abundances of each element is defined according to \cite{lodders_solar_2010}. It implements disequilibrium chemistry. The flux is solved iteratively assuming 64 pressure levels over the grid, with a minimum pressure of $10^{-6}$\,bar and a maximum pressure of $10^{2}$\,bar. This model considers sources of opacities from H$_2$-H$_2$ and H$_2$-H$_e$ collision induced absorption, ro-vibrational bands from nine molecules (H$_2$O, CH$_4$, CO, CO$_2$, NH$_3$, TiO, VO and FeH) and resonant lines from Na and K.

However, the highest resolution Exo-REM models stand at $R = 20\,000$. Thus, we use Exo\_k, a library constructed to handle radiative opacities from various sources for the subsequent computation emission spectra for 1D planetary atmospheres \citep{leconte_spectral_2021}, to recompute the radiative transfer at a higher resolution for a fixed atmospheric structure. We use the cross section data listed in \texttt{petitRADTRANS} \citep{molliere_petitradtrans_2019} and the collision-induced absorption data from HiTRAN \citep{karman_update_2019}. The resulting model which is used in this work has as free parameters \Teff, \logg, \met and \co. \Teff extends from 500 to 1000 K, \logg from 3 to 5\,dex, \met from -0.5 to 1.0\,dex, and \co from 0.1 to 0.8. The models are initially computed from 1 to 5\,\mic at a spectral resolution of $R = 1\,000\,000$. For the analysis in \formosa, we downgrade the spectral resolution of the model to 200\,000 but keep the bin sampling at 600\,000. The Exo-REM/Exo\_k model we use does not include the effects of clouds. This is discussed in section ~\ref{sec:clouds_sensitivity}.

To compare the model to the data in \formosa, we first apply the RV and \vsini correction, then we downgrade the resolution of the model at the spectral resolution of our data, which is estimated at 140\,000. The resolution of the \crires spectrograph is constant to better than 2.5\% in the H band \citep{Dorn2023} and the value of 140\,000 can be estimated from instrumental parameters. This value is also in perfect agreement with the value derived by \citet{Nortmann2024} using on-sky data obtained in very good observing conditions, where the stellar PSF has the same 2-pixel FWHM as the PSF of the HiRISE science fiber. The degradation of the resolution is performed using a Gaussian convolution adapted to the resolution at each wavelength.

\begin{table}[!ht]
    \centering
    \renewcommand{\arraystretch}{1.3}
    \caption{Priors used in the forward modeling.}
    \begin{tabular}{cc} \hline\hline
        Parameter & Prior \\
        \hline
        \Teff & $\mathcal{N}(800,50)$ \\
        \logg & $\mathcal{N}(3.7,0.2)$ \\
        \met  & $\mathcal{N}(0.75,0.25)$ \\
        \co   & $\mathcal{N}(0.55,0.10)$ \\
        \hline
    \end{tabular}
    \tablefoot{$\mathcal{N}(\mu,\sigma)$ means a normal distribution of mean $\mu$ and standard deviation $\sigma$.}
    \label{tab:priors}
\end{table}

\begin{table*}[!ht]
    \centering
    \caption{\formosa results on AF\,Lep\,b for two nights of observation.}
    \renewcommand{\arraystretch}{1.5}
    \resizebox{\textwidth}{!}{
    \begin{tabular}{l c c c c c c c c}
    \hline \hline
    Priors\tablefoottext{a} & \Teff & \logg & \met & \co & RV & \vsini & R\tablefoottext{b} & $\Delta \log z$\tablefoottext{c} \\
    & [K] & [dex] & [dex] & & [\kms] & [\kms] & [\RJup] & \\
    \hline \hline
    \multicolumn{9}{c}{2023-11-20} \\
    \hline
    --- & $930_{-83}^{+51}$ & $3.22_{-0.15}^{+0.23}$ & $-0.13_{-0.22}^{+0.26}$ & $0.47_{-0.15}^{+0.12}$ & $31.68_{-1.14}^{+1.20}$ & $10.09_{-2.42}^{+2.74}$ & $2.37_{-0.60}^{+0.46}$ & 0 \\
    \Teff, \logg  & $812_{-26}^{+23}$ & $3.68_{-0.10}^{+0.09}$ & $0.07_{-0.27}^{+0.24}$ & $0.39_{-0.14}^{+0.15}$ & $32.04_{-1.26}^{+1.28}$ & $11.60_{-2.97}^{+3.14}$ & $1.40_{-0.16}^{+0.20}$ & -1 \\
    \Teff, \logg, \met & $808_{-25}^{+55}$ & $3.68_{-0.10}^{+0.09}$ & $0.50_{-0.19}^{+0.16}$ & $0.57_{-0.08}^{+0.07}$ & $31.77_{-1.25}^{+1.20}$ & $12.44_{-2.83}^{+2.69}$ & $1.39_{-0.16}^{+0.19}$ & -2 \\
    \Teff, \logg, \met, \co & $811_{-25}^{+22}$ & $3.68_{-0.10}^{+0.08}$ & $0.47_{-0.16}^{+0.15}$ & $0.57_{-0.06}^{+0.05}$ & $31.77_{-1.16}^{+1.16}$ & $12.27_{-2.63}^{+2.46}$ & $1.39_{-0.16}^{+0.19}$ & -1 \\
    \hline \hline
    \multicolumn{9}{c}{2023-11-23 offset 1} \\
    \hline
    --- & $880_{-116}^{+86}$ & $3.51_{-0.33}^{+0.37}$ & $-0.06_{-0.2}^{+0.32}$ & $0.28_{-0.12}^{+0.17}$ & $32.10_{-1.33}^{+1.35}$ & $11.09_{-4.08}^{+4.07}$ & $1.72_{-0.60}^{+0.80}$ & 0 \\
    \Teff, \logg & $814_{-25}^{+23}$ & $3.66_{-0.09}^{+0.09}$ & $-0.01_{-0.26}^{+0.32}$ & $0.21_{-0.08}^{+0.15}$ & $31.95_{-1.34}^{+1.36}$ & $11.16_{-4.35}^{+3.85}$ & $1.44_{-0.16}^{+0.19}$ & -1 \\
    \Teff, \logg, \met & $805_{-26}^{+24}$ & $3.67_{-0.09}^{+0.09}$ & $0.54_{-0.22}^{+0.23}$ & $0.43_{-0.15}^{+0.11}$ & $32.17_{-1.32}^{+1.18}$ & $10.99_{-4.42}^{+3.72}$ & $1.42_{-0.16}^{+0.18}$ & -1 \\
    \Teff, \logg, \met, \co & $809_{-25}^{+24}$ & $3.66_{-0.09}^{+0.09}$ & $0.63_{-0.18}^{+0.19}$ & $0.51_{-0.07}^{+0.06}$ & $32.19_{-1.33}^{+1.21}$ & $11.45_{-4.23}^{+3.42}$ & $1.43_{-0.17}^{+0.19}$ & -0.5 \\
    \hline \hline
    \multicolumn{9}{c}{2023-11-23 offset 2} \\
    \hline
    --- & $897_{-107}^{+71}$ & $3.50_{-0.33}^{+0.29}$ & $-0.12_{-0.18}^{+0.21}$ & $0.15_{-0.04}^{+0.07}$ & $32.92_{-1.56}^{+1.46}$ & $17.36_{-2.51}^{+3.29}$ & $1.74_{-0.51}^{+0.81}$ & 0  \\
    \Teff, \logg & $841_{-40}^{+38}$ & $3.56_{-0.15}^{+0.15}$ & $-0.03_{-0.17}^{+0.22}$ & $0.15_{-0.03}^{+0.06}$ & $32.87_{-1.57}^{+1.47}$ & $17.23_{-2.55}^{+3.49}$ & $1.62_{-0.28}^{+0.35}$ & -2 \\
    \Teff, \logg, \met & $805_{-24}^{+24}$ & $3.68_{-0.09}^{+0.09}$ & $0.36_{-0.21}^{+0.23}$ & $0.19_{-0.06}^{+0.12}$ & $32.55_{-1.59}^{+1.50}$ & $17.20_{-3.15}^{+6.55}$ & $1.41_{-0.16}^{+0.19}$ & -3 \\
    \Teff, \logg, \met, \co & $814_{-25}^{+24}$ & $3.65_{-0.09}^{+0.08}$ & $0.63_{-0.19}^{+0.21}$ & $0.43_{-0.08}^{+0.07}$ & $32.10_{-1.64}^{+1.54}$ & $16.76_{-3.40}^{+7.08}$ & $1.46_{-0.16}^{+0.18}$ & -4.5 \\
    \hline
    \end{tabular}
    } 
    \label{tab:formosa_results}
    \tablefoot{
        \tablefoottext{a}{The definition of the priors is provided in Table~\ref{tab:priors}.}
        \tablefoottext{b}{R represents the self consistent radius using Newton's law with the corresponding value of \logg and estimated value of mass ($3.68 \pm 0.48$\,\MJup) from \cite{balmer_vltigravity_2024}. When \logg is fixed, we consider an uncertainty on \logg of 0.}
        \tablefoottext{c}{For each night, $\Delta \log z$ indicates the difference in $\log z$ between the specific case being analysed and the cas where all parameters are free.}
    }
\end{table*}

\begin{figure*}[!ht]
    \centering
    \includegraphics[width=0.9\linewidth]{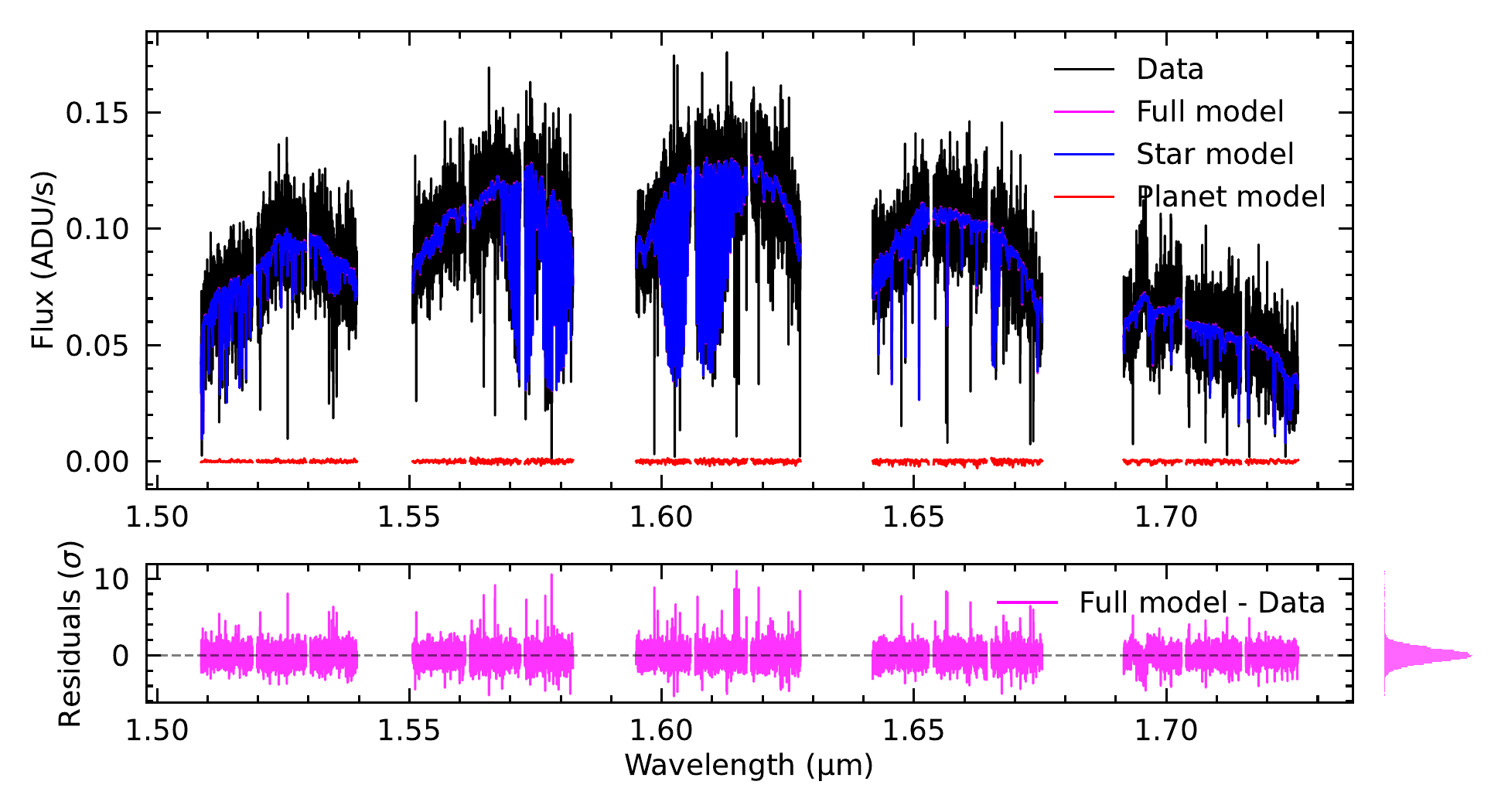}
    \caption{Best fit model for the 2023-11-23 data with the Exo-REM/Exo\_k model. The top panel shows the data, the full model (magenta), the stellar model (blue) and the planet model (red). The last two are scaled to their respective amplitude in the total signal. The stellar component in the data completely dominates the planetary component, so the full model (magenta) is mostly hidden by the star model (blue). The bottom panel shows the residuals, i.e., the data subtracted by the full model. For this panel, the scale is the standard deviation of the residuals. The top bottom of the figure depicts the distribution of the residuals which is well centered around 0.}
    \label{fig:full_model}
\end{figure*}

\begin{figure}[!ht]
    \centering
    \includegraphics[width=\linewidth]{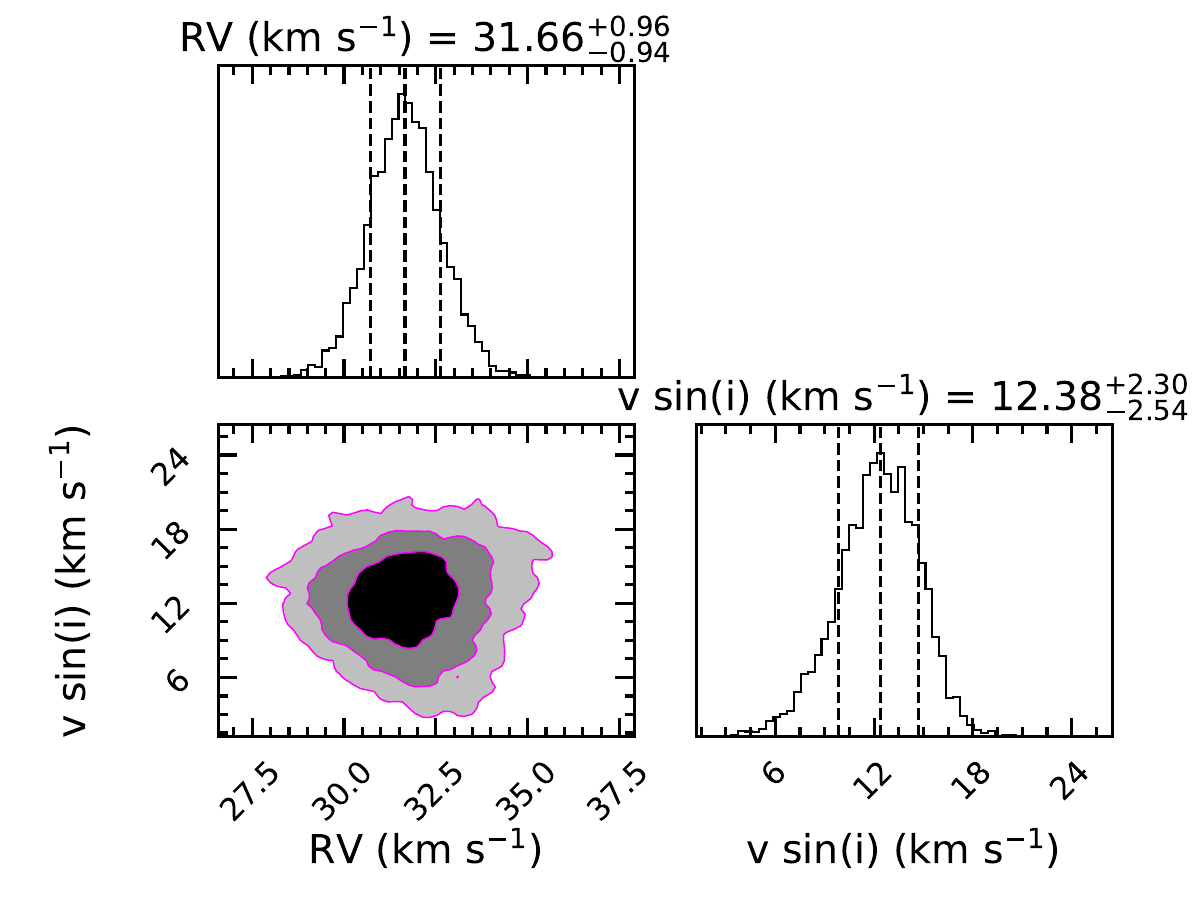}
    \caption{Posterior distributions of RV and \vsini for the two nights combined with the Exo-REM/Exo\_k model. These results are the final adopted values for these parameters.}
    \label{fig:corner_combined}
\end{figure}

For the present work, we use the nested sampling in \formosa with 500 living points. We consider two types of priors for the parameters: either uninformative uniform priors, or informative priors for \Teff, \logg, \met and \co. In all cases, since we have no prior information on RV and \vsini, we consistently apply uninformative priors for these parameters, that is $\mathcal{U}(-100,100)$ for RV and $\mathcal{U}(0,100)$ for \vsini. We derive our priors from the final atmospheric parameter values reported in \cite{balmer_vltigravity_2024}: $\Teff = 800 \pm 50$\,K, $\logg = 3.7 \pm 0.2$\,dex, $\met = 0.75 \pm 0.25$\,dex, and $\co = 0.55 \pm 0.1$. Table~\ref{tab:priors} presents the priors we use in the forward modeling.

The results are presented in Table \ref{tab:formosa_results}. In the present section, we present only our results on RV and \vsini and discuss our results and their implications further in Sect. \ref{sec:discussion}. From the results we can estimate the total RV of the system, which includes both the orbital velocity of the companion and the systemic velocity of the system. For the first night and the first offset of the second night we find RV and \vsini values that agree well. For the second offset of the second night we find RV values slightly higher, but mostly \vsini significantly higher than for the other data sets. We note that for the second offset we have a lower integration time for the science and an even lower integration time for the backgrounds, resulting in much noisier data (see Table~\ref{tab:observations}). This could explain the discrepancy between the results of the second offsets and the results of the other datasets. Therefore, we decided to exclude this dataset for the rest of the analysis. We refer the reader to Appendix~\ref{sec:S/N_ratio_impact} for a more detailed analysis on the impact of the number of backgrounds to the derived parameters of the planet. We also note that observing conditions were different between the two nights (see Table~\ref{tab:observations}), with a more variable weather on the second night.

To derive the final values adopted for RV and \vsini, we combined the datasets of the first night and the first offset of the second night. To do so, we co-added the log-likelihoods associated to the two datasets on \formosa. The result of this combination is presented on Fig.~\ref{fig:corner_combined}. We estimate a final RV of $31.61_{-0.94}^{+0.96}$\,\kms and a final \vsini of $12.38_{-2.54}^{+2.30}$\,\kms. This value must be corrected from the systemic velocity, which is estimated to $21.1 \pm 0.37$\,\kms by \citealt{gaia_collaboration_2023}. This value is perfectly consistent with ground-based high-resolution \uves data \citep[][$20.90 \pm 1.11$\,\kms]{zuniga-fernandez_search_2021}. Using the value of \cite{gaia_collaboration_2023}, and propagating our $100 m.s^{-1}$ uncertainty on the wavelength recalibration (see Sect.~\ref{sec:wave_recal}), we finally infer a relative radial velocity between the companion and the star of $10.51_{-1.02}^{+1.03}$\,\kms. This value is consistent at 2 $\sigma$ with the value estimated with the orbital solution derived from previous astrometric measurements (see Sect. \ref{sec:orbit_fitting}). Another team of researchers \citep{Hayoz2025} have derived the radial velocity of the planet around the same epoch as our measurement using ERIS/SPIFFIER observations. They derived a value of $7.8 \pm 1.7$\,\kms, which confirms our measurement.

\begin{figure*}[!ht]
    \centering
    \includegraphics[width=\textwidth]{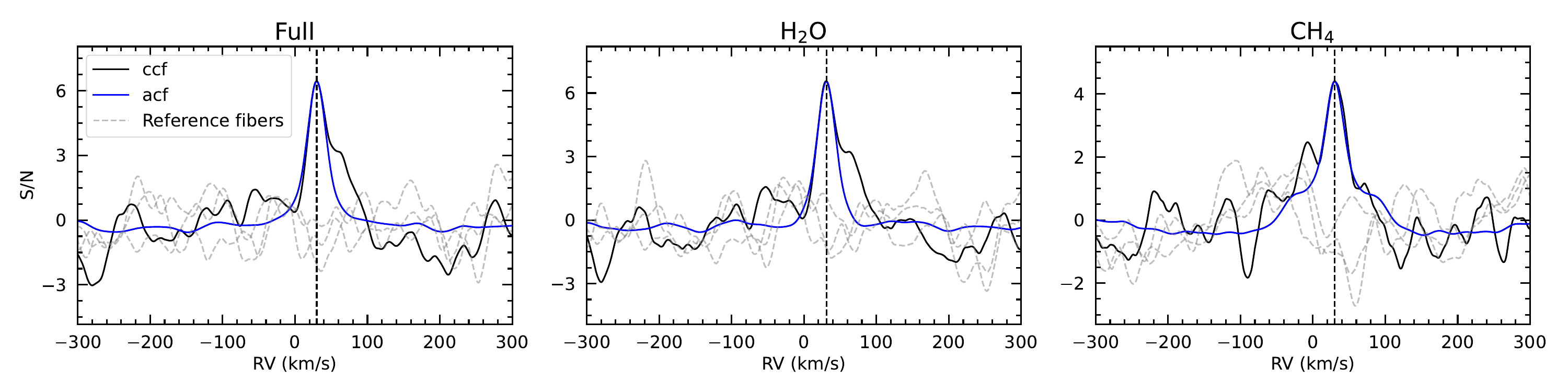}
    \caption{Cross-correlation functions for the first night with the Exo-REM/Exo\_k model. In grey, the cross-correlation between the signal of the 3 reference fibers and the best model. In blue, the auto-correlation between the best model and itself. The model has been broadened at the \vsini given by the fit.}
    \label{fig:molecules_detection}
\end{figure*}

\begin{figure}[!ht]
    \centering
    \includegraphics[width=\linewidth]{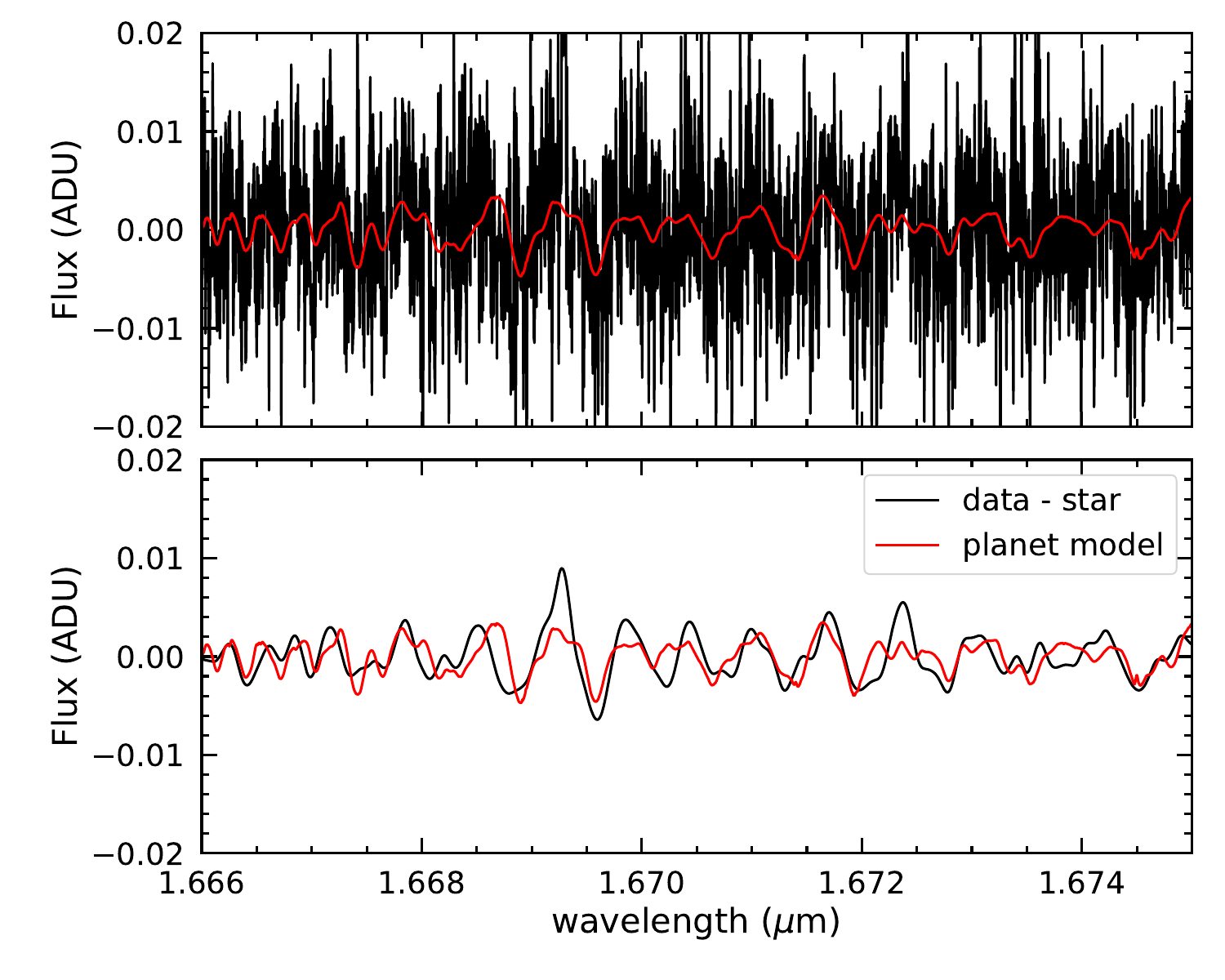}
    \caption{Comparison between the 2023-11-23 data after subtraction of the stellar contribution and the best fit Exo-REM/Exo\_k model in a segment of order dominated by CH$_4$ absorption lines. The top panel shows the raw data subtracted by the estimated stellar contribution at the full spectral resolution, while in the bottom panel, the data has been smoothed to a spectral resolution of $\sim$30\,000 that takes into account the decrease of resolution induced by the \vsini.}
    \label{fig:full_model_zoomed}
\end{figure}

To assess the quality of the fit, we plot the best Exo-REM/Exo\_k \formosa fit with the data in Fig.~\ref{fig:full_model}. We see that the data is highly dominated by the stellar contribution. We note that for this plot, the continuum of the planet model was removed from the planet model (see Eq.~\ref{eq:planet_model}). This is because the continuum of the planet model leaks into the estimate of $\alpha$ (see Eq.~\ref{eq:star_contamination} and \citealt{landman_beta_2024}). Using the estimate of $c_p$ defined in Eq.~\ref{eq:model_matrix}, we can give an estimate of the total flux of the companion model we estimate (Table~\ref{tab:delta_H}). In the H band, \cite{DeRosa2023} derives $\Delta H = 12.48 \pm 0.12$\,mag. Our estimate is fully consistent with this value for the first night and the first offset of the second night. This confirms that, for these data, we are neither over- nor underestimating the planetary model. For the second offset of the second night, however, the flux of the planet model is slightly overestimated by $\sim$0.3\,mag.

In Fig.~\ref{fig:full_model_zoomed}, we focus on one of the best fitted segments and compare the data after subtraction of the stellar model with the planet model. In this wavelength band, the absorption lines are mainly coming from CH$_4$. Some features of the planetary model can be identified in the raw residuals but become obvious when smoothing the data at the intrinsic spectral resolution of the model, which has been rotationally broadened by the fitted value of \vsini.

\begin{table}[!ht]
    \centering
    \small
    \renewcommand{\arraystretch}{1.3}
    \caption{Estimated fluxes and differential magnitudes for each of the dataset.}
    \begin{tabular}{p{2.5cm}ccc} \hline\hline
        Dataset & Companion & Star & $\Delta H$ \\
        & [ADU/s] & [ADU/s] & [dex] \\ \hline
        2023-11-20 & $91 \pm 8$ & $8769275$ & $12.67 \pm 0.14$ \\
        2023-11-23 offset 1 & $65 \pm 8$ & $8423726$ & $12.75 \pm 0.15$ \\
        2023-11-23 offset 2 & $113 \pm 13$ & $8406735$ & $12.18 \pm 0.13$ \\ \hline
    \end{tabular}
    \tablefoot{The estimated $\Delta H$ corresponds to the estimated differential magnitude for the best planet model only. It does not take into account the distribution of planet models explored by the nested sampling.}
    \label{tab:delta_H}
\end{table}

\subsection{Detection of molecules}
\label{sec:molecules_detection}

Another approach to the analysis high-resolution data is to use CCF analysis, for example to assess the presence of individual molecules. In this analysis we cross-correlate template models with the data subtracted by the estimated stellar contribution (see Sect. \ref{sec:data_modeling}):
\begin{equation}
    \hat{d} = d - \hat{c_s} \frac{\mathcal{L}\left[d\right]}{\mathcal{L}\left[f_s\right]}f_s,
\end{equation}
where $\hat{c_s}$ is the first component of the solution $\hat{c}$ to the linear least square Eq.~\ref{eq:LSQ}. We calculate the CCF as:
\begin{equation}
    \mathrm{CCF}(RV) = \sum_{\lambda_{i}}\hat{d}(\lambda_{i})m_{RV}(\lambda_{i}),
\end{equation}
where $m_{RV}$ is the template model Doppler-shifted at the radial velocity RV and broadened at the \vsini given by our results (see table \ref{tab:formosa_results}). For this analysis we compute the CCF using a grid of RV ranging from $-1000$ to $+1000$\,\kms in steps of 0.5\,\kms. We perform the analysis both for best Exo-REM/Exo\_k model inferred from the \formosa analysis with priors on all the bulk parameters and for model templates of individual molecules. We consider only H$_2$O and CH$_4$ as they are the species expected to be dominating in the $H$ band at the \Teff of AF\,Lep\,b. The templates of individual molecules have been generated using Exo-REM volume mixing ratio profiles in tandem with Exo\_k.

The results are presented in Fig.~\ref{fig:molecules_detection} for the night 2023-11-20. For the second night, the results are presented in Appendix~\ref{sec:ccf_second_night} and yield similar conclusions to the ones discussed below. To improve visibility, we plot the RV between $-300$ and $+300$\,\kms. The S/N is estimated by normalizing the CCF by its standard deviation computed over two windows 100\,\kms away from peak of the CCF (30\,\kms), which in this case is approximately between $-70$ and $-1000$\,\kms and between $+130$ and $+1000$\,\kms. We overplot the autocorrelation function (ACF) of the model, which we shift to the estimated RV and normalize at the peak S/N of the cross-correlation function.

To accurately estimate the RV associated to the CCF, we fit a Gaussian function to the CCF. We find this method of estimating the RV to be more robust compared to estimating the RV with the maximum of the CCF. With the full model, we estimate a S/N of 6.6 at RV = 31.0\,\kms. For H$_2$O, we find a detection with a S/N of 6.6 at 31.4\,\kms, consistent with the full model. And finally for CH$_4$, the detection is at a S/N of 4.4 with a RV of 31.5\,\kms, again consistent with the full model. We note that these values are within the error bars of the value inferred from our \formosa analysis of the first night with the priors on all the bulk parameters. Although the CH$_4$ CCF is noisier than for the ones for the full model and for H$_2$O, the location of the peak in RV makes us confident that the detection of CH$_4$ in the atmosphere of AF\,Lep\,b is real, confirming the recent findings of \citet{balmer_vltigravity_2024}. This also supports the possibility of inferring a \co ratio in the H band.

To enhance the confidence in the detections, we overplot in Fig.~\ref{fig:molecules_detection} the CCF between the model and the other three reference fibers of the instrument, which contains signal from the star speckles \citep[see][for details on these fibers]{Vigan2024}. The CCF with the reference fibers shows similar features to the planet's CCF. We interpret this by the fact that all fibers see some common signal coming from thermal background noise, which is expected to be at the same level for all fibers, and from star speckles and systematic effects, which are imperfectly removed from the science signal.

\subsection{Is our data sensitive to clouds?} \label{sec:clouds_sensitivity}

At the low temperature of AF\,Lep\,b, condensate clouds should sink below the photosphere, reducing their effect in the emission spectrum \citep{lodders_chemistry_2006}. However, \cite{zhang_elemental_2023} noted the possible presence of silicate clouds in the atmosphere of AF\,Lep\,b \citep[see Fig. 15 of][]{zhang_elemental_2023}. As illustrated in Fig.~6 of \cite{xuan_clear_2022}, high-resolution data are sensitive to lower pressures compared to low-resolution data. This is because high-resolution data resolve the cores of absorption lines (from H$_2$O and CH$_4$ molecules in this case) which, compared to the continuum, comes from regions of high molecular opacity. The optical depth in a layer of the atmosphere is proportional to the integral of opacity and abundance, integrated over the path length. Since the opacity at line cores is high, the path length must be small, so these regions come from lower pressures. As a consequence, our very high-resolution data may be sensitive to pressures lower than the silicate clouds base pressure inferred in \cite{zhang_elemental_2023}.

\begin{figure}
    \centering
    \includegraphics[width=\linewidth]{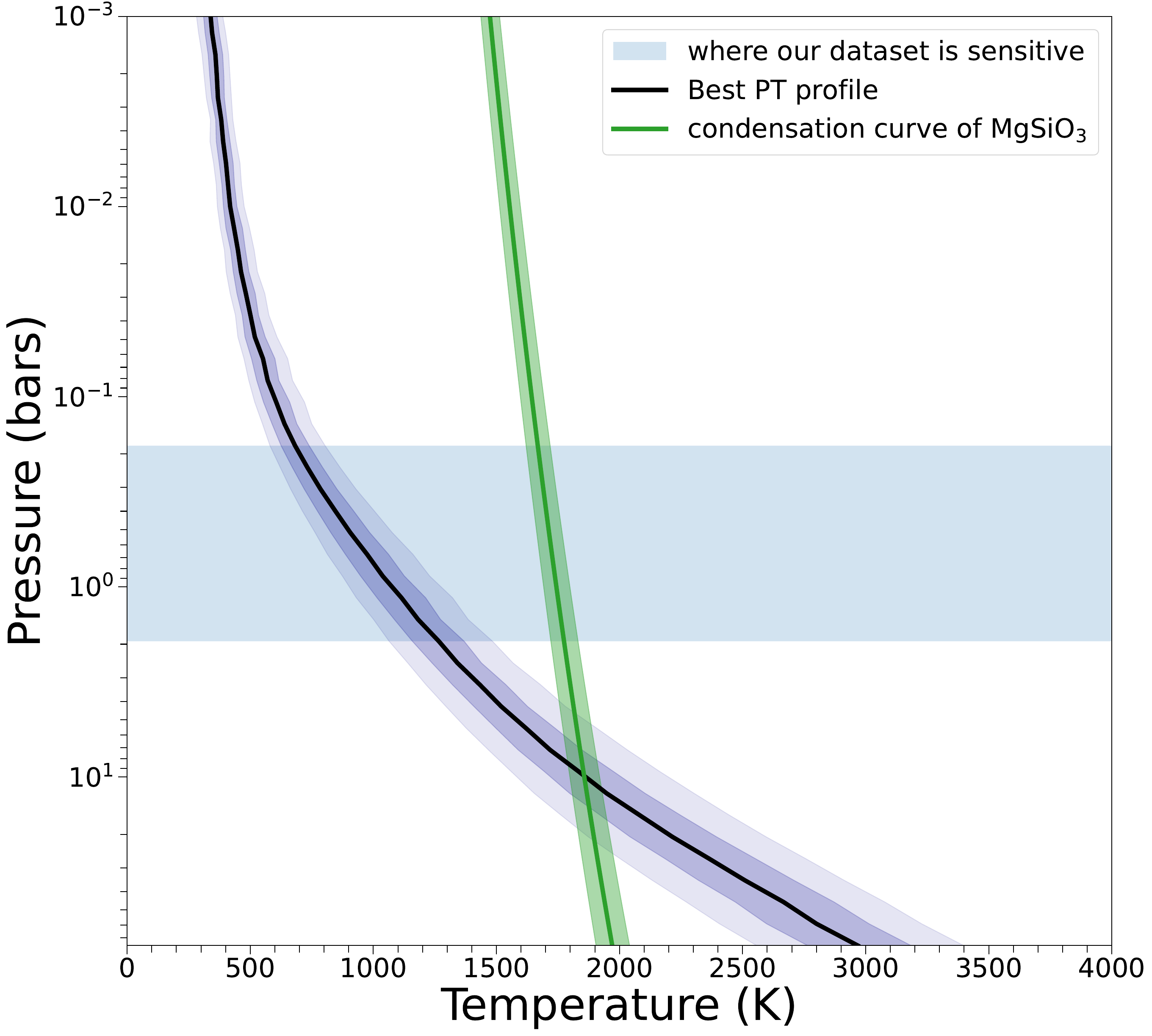}
    \caption{P-T profile with 1$\sigma$ and 2$\sigma$ uncertainties derived from the results of combining the 2 nights. In green we show the 1$\sigma$ condensation curve of MgSiO$_3$. The shaded blue region corresponds to the pressure levels to which our data is most sensitive.}
    \label{fig:PT_clouds}
\end{figure}

To verify this hypothesis, we plot the best P-T profile inferred from our results of the combination of the two nights in Fig.~\ref{fig:PT_clouds}. We derive the condensation curve of MgSiO$_3$ from Eq.~20 of \cite{visscher_atmospheric_2010}, which depends on the metallicity. To plot the 1$\sigma$ condensation curve of MgSiO$_3$, we use the 1$\sigma$ uncertainties on metallicity derived in \cite{balmer_vltigravity_2024} (\met = $0.75 \pm 0.25$\,dex). We also estimate the pressure levels to which our data is most sensitive by calculating the brightness temperature of our best model at each wavelength and match it to the pressure in the PT profile. The relatively narrow region to which our data is most sensitive is a result of the relatively small wavelength range covered by our data (1.43--1.77\,\mic). As we see on Fig.~\ref{fig:PT_clouds}, our data is sensitive to layers in the atmosphere significantly above the silicate cloud base pressure level, justifying our assumption that our data should not be sensitive to clouds and that we can safely use cloudless models.

\section{Orbit fitting}
\label{sec:orbit_fitting}

We use \texttt{Orvara} \citep{brandt_orvara_2021} to constrain the orbit of AF\,Lep\,b. We run the parallel-tempering Markov Chain Monte Carlo (MCMC) sampler (\texttt{ptemcee}) of \texttt{Orvara} \citep{foreman-mackey_emcee_2013,vousden_dynamic_2016} with 10 temperatures, 100 walkers, 500\,000 steps per walker. The chains are saved every 50 steps and in each chain we discard the first 3000 saved steps as burn-in. We fit for the mass of the star (M$_{\star}$), the mass of the companion (M$_{p}$), the orbit semi-major axis ($a$), the inclination angle ($i$), eccentricity ($e$) and argument of periastron ($\omega$) in parametrized forms ($\sqrt{e}\sin\omega$ and $\sqrt{e}\cos\omega$), the longitude of the ascending node ($\Omega$), and the mean longitude at reference epoch J2010.0 ($\lambda$).

In Sect.~\ref{sec:datasets}, we describe the datasets used in the analysis. To assess the contribution of our RV measurements to the orbit of AF\,Lep\,b, we first analyze the orbit excluding our RV measurement (Sect. \ref{subsec:orbit_analysis}) and then we compare with the results obtained using our \hirise measurement (Sect. \ref{subsec:RV_measurement}).

\subsection{Datasets}
\label{sec:datasets}

In our orbit analysis we use relative astrometry spanning a baseline of 11 years. This includes the SPHERE astrometry \citep{Mesa2023,DeRosa2023}, the Keck/NIRC2 astrometry \citep{franson_astrometric_2023}, the archival VLT/NaCo astrometry \citep{Bonse2024}, and the recent VLTI/GRAVITY observations \citep{balmer_vltigravity_2024}. We also include 20 RV measurements of the star obtained with Keck/HIRES \citep{butler_lces_2017}, spanning a baseline of 12 years.

A recent analysis of these datasets lead by \citet{balmer_vltigravity_2024} measured a mass of $3.68_{-0.48}^{+0.47}$\,\MJup for the companion, a semi-major axis of $8.98_{-0.08}^{+0.15}$\,au and a quasi-circular orbit (e = $0.013_{-0.010}^{+0.024}$). It also concluded in a spin-orbit alignment with an inclination angle $i = 57.12_{-0.71}^{+0.67}$\degre. This value is consistent with the stellar inclination angle $i_{\star} = 54_{-9}^{+11}$ \degree derived in \cite{franson_astrometric_2023}.

\subsection{Orbit analysis}
\label{subsec:orbit_analysis}

These datasets provide good constraints on the main orbital parameters, but they leave some ambiguity regarding the direction of motion of the companion in the line of sight. Two families of orbit co-exist depending on the sign of the RV of the companion relative to the host star at epoch J2024.0. This was first demonstrated by \cite{zhang_elemental_2023}. This leaves two degenerate solutions for the mean longitude of the star at epoch J2010.0 ($\lambda$) and the longitude of ascending node ($\Omega$) (see Fig. \ref{fig:corner_plot_orvara} of Appendix~\ref{sec:orvara_results}).

From the orbit, we can derive the RV of the star ($\mathrm{RV}_{\star}$) at any reference epoch $t_{ref}$ using the following equation:
\begin{align}
    & \mathrm{RV}_{\star, ref} = K_{\star}(e\cos(\omega) + \cos(\nu_{ref} + \omega)) \\
    & K_{\star} = \left(\frac{2\pi G}{P}\right)^{1/3} \frac{M_{p}}{\left(M_{p} + M_{\star}\right)^{2/3}}\frac{sin(i)}{\sqrt{1-e^2}},
\end{align}
where $\nu_{ref}$ is the true anomaly of the companion at epoch $t_{ref}$. From this equation, we can infer the RV of the companion ($\mathrm{RV}_{p}$) with the following relation:
\begin{equation}
    \label{eq:RV_comp}
    M_{p}\mathrm{RV}_{p} = -M_{\star}\mathrm{RV}_{\star}
\end{equation}
This is coming from the conservation of angular momentum of the system and the fact that the companion moves in opposite direction of the star. The RV of the companion itself is:
\begin{align}
    & \mathrm{RV}_{p, ref} = K_{p}(e\cos(\omega) + cos(\nu_{ref} + \omega) \\
    & K_{p} = -\left(\frac{2\pi G}{P}\right)^{1/3}\frac{M_{\star}}{\left(M_{p} + M_{\star}\right)^{2/3}}\frac{sin(i)}{\sqrt{1-e^2}}
\end{align}

Finally, the relative RV between the planet and the host star can be expressed as:
\begin{equation} \label{eq:delta_RV}
\begin{split}
    \Delta \mathrm{RV}_{ref} & = -\left(\frac{2\pi G(M_{\star} + M_{p})}{P}\right)^{1/3}\frac{sin(i)}{\sqrt{1-e^2}} \\
    & \times (e\cos(\omega) + \cos(\nu_{ref} + \omega))
\end{split}
\end{equation}

At epoch J2023.88, corresponding to 2023-11-20, we find, using the dataset described on Sect.~\ref{sec:datasets} and Eq.~\ref{eq:delta_RV}, one population of orbit with RV = $8.76_{-0.35}^{+0.22}$\,\kms and another population of orbit with RV = $-8.78_{-0.21}^{+0.33}$\,\kms. The first population is consistent within 2 $\sigma$ with our \hirise measurement of $10.51_{-1.01}^{+1.03}$\,\kms for AF\,Lep\,b (see Sect.~\ref{sec:atmosheric_models}).

\begin{figure*}[!ht]
\centering
    \includegraphics[width=\linewidth]{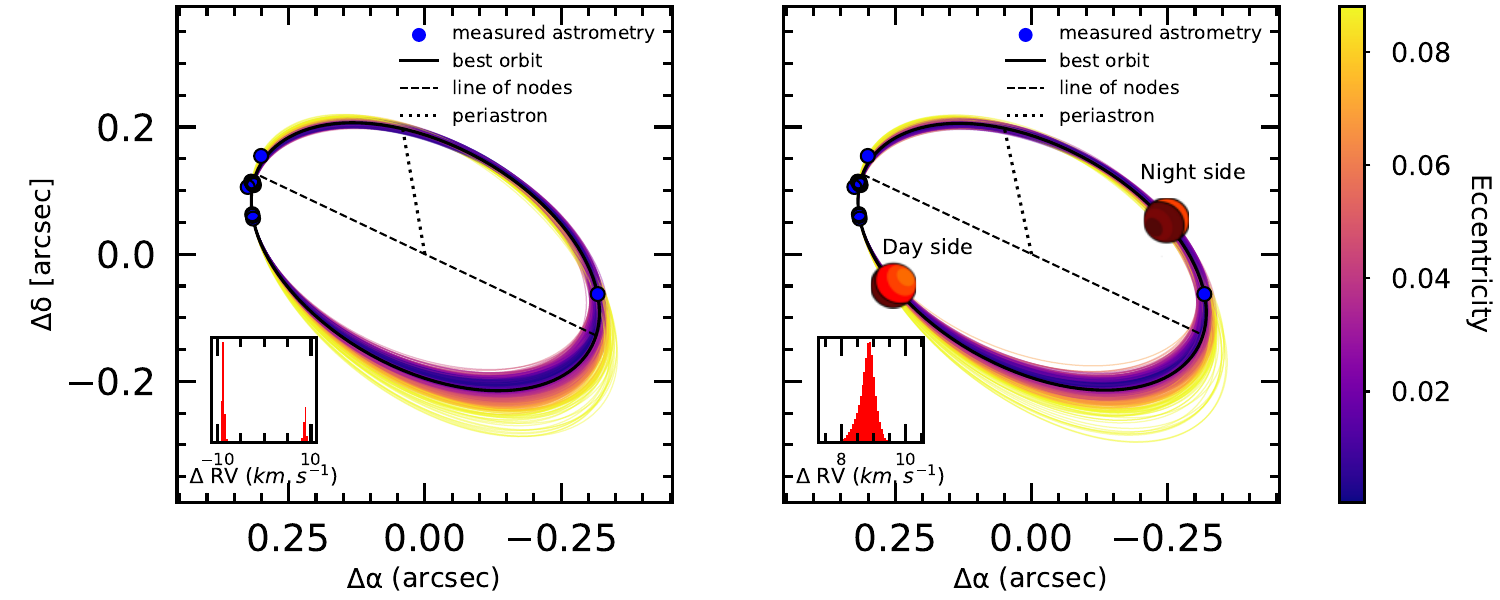}
    \caption{Astrometric orbit of AF\,Lep\,b with 1\,000 orbits randomly drawn from the posterior. The left panel shows the solutions obtained without the \hirise RV measurement and the right panel shows the solutions including our measurement. The inset plot in each panel shows the distributions of the relative RV between the planet and the star. As illustrated on the plot, the determination of the RV of the planet allows us to have information on the phase of the planet.}
    \label{fig:astrometric_orbit}
\end{figure*}

\begin{figure*}[!ht]
    \centering
    \includegraphics[width=\linewidth]{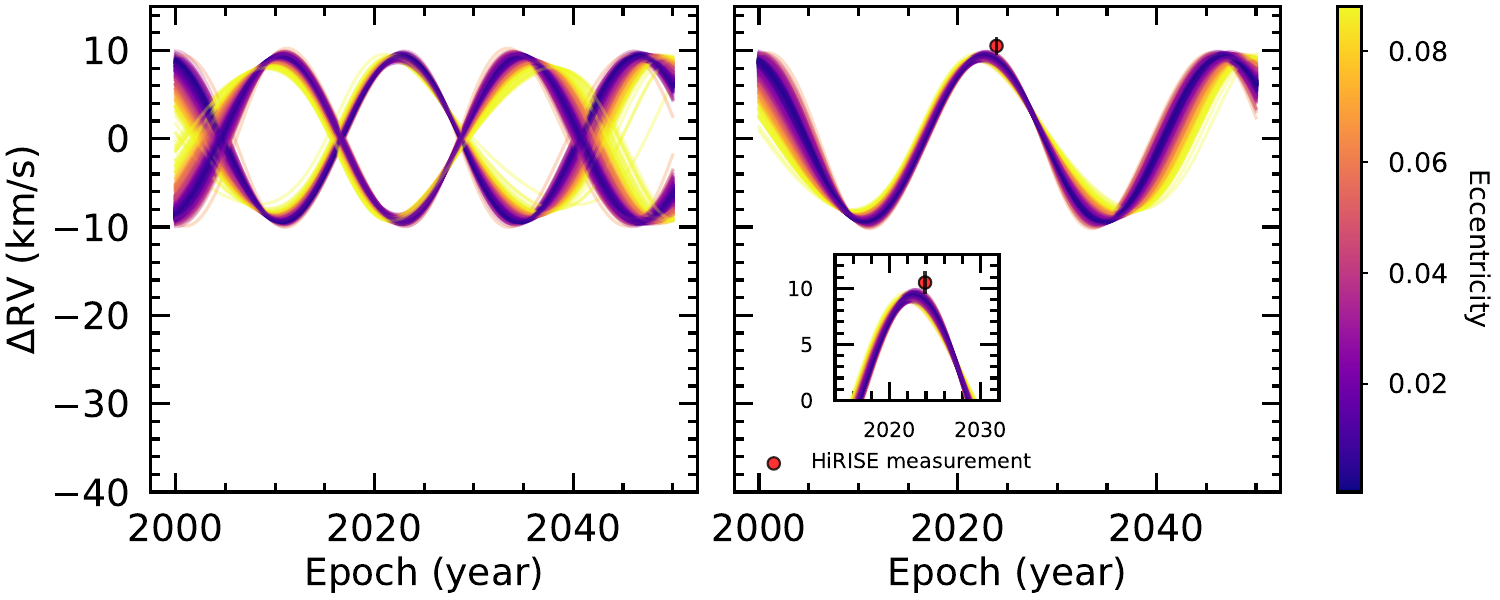}
    \caption{Predictions of the relative RV between the companion and the star obtained with 1\,000 orbits randomly drawn from the posterior. As in Fig.~\ref{fig:astrometric_orbit}, the left panel shows the solutions without the \hirise RV measurement and the right panels shows the solutions including our measurement. The bottom left of the right panel, shows a zoom around the \hirise measurement.}
    \label{fig:RV_comp}
\end{figure*}

\subsection{\hirise RV measurement}
\label{subsec:RV_measurement}

In practice, \texttt{Orvara} only uses the relative RV measurements between the secondary and the primary. However, the ratio of the RVs between the primary and the secondary is scaled by a factor of $\frac{M_{p}}{M_{\star}}$, which in most cases is significantly lower than 1. It is the case here as $\frac{M_{p}}{M_{\star}} = 0.00289_{-0.00036}^{+0.00036}$ (see Table~\ref{tab:orvara_results}). As a result, the RV of the star is entirely dominated by uncertainty in the RV of the companion. We thus directly included the $10.51_{-1.01}^{1.03}$\,\kms RV measurement of the companion on \texttt{Orvara}.

\begin{figure}
    \centering
    \includegraphics[width=1.0\linewidth]{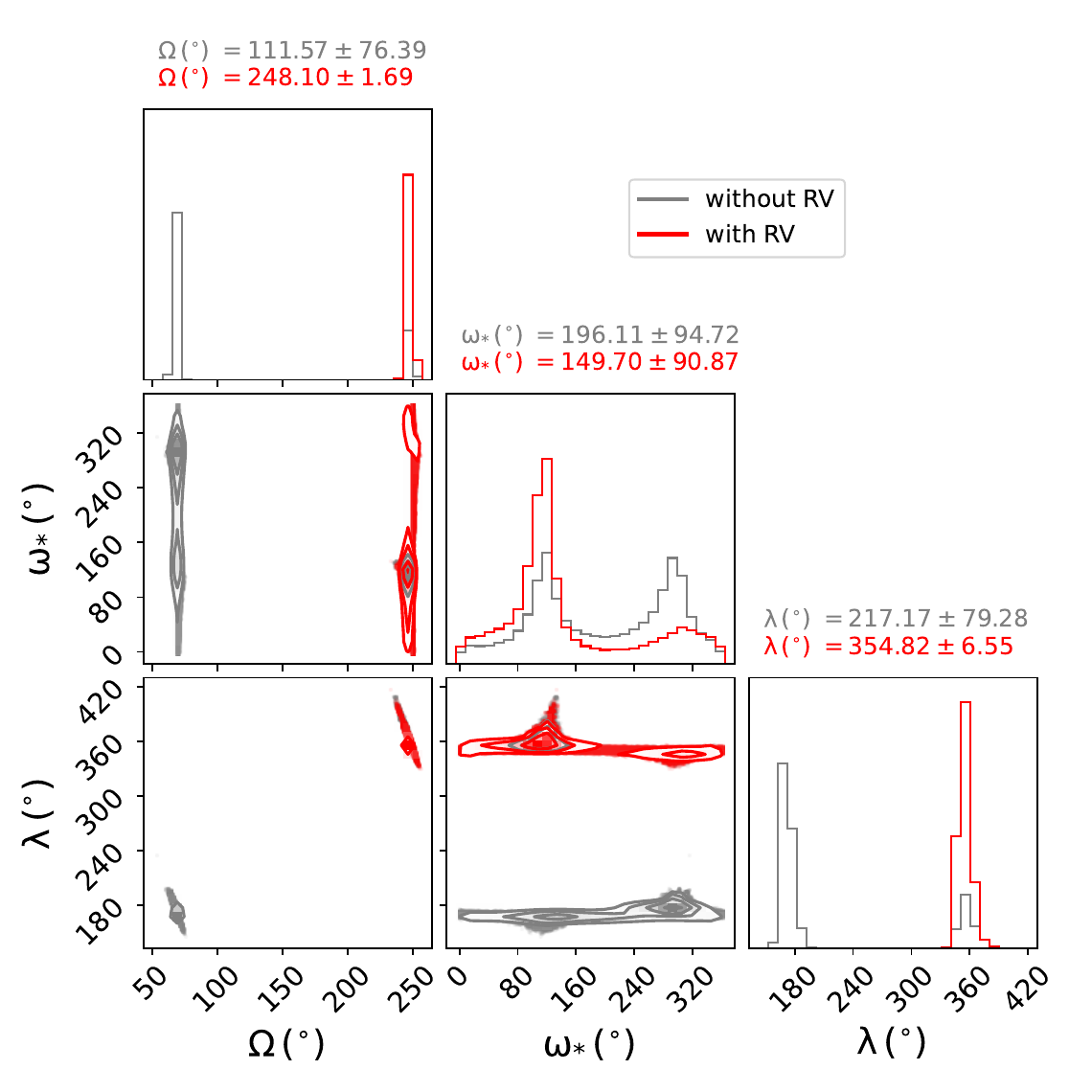}
    \caption{Posterior distribution for the ascending node ($\Omega$), the argument of periastron ($\omega$) and the mean longitude at reference epoch J2010.0 ($\lambda$). Grey : without our RV measurement. Red : with our RV measurement.}
    \label{fig:conrner_plot_orvara_Omega_omega_lam}
\end{figure}

The results of adding our RV measurement are presented on Fig.~\ref{fig:astrometric_orbit} (right) and Table \ref{tab:orvara_results}. The astrometric orbit with our RV measurement appears visually similar to the one without it. But, with the RV measurement included, we rule out an entire family of orbital solutions. The bottom left of each panel in Fig.~\ref{fig:astrometric_orbit} presents the distribution of the relative RV between the companion and the star. We clearly see two widely separated populations of orbits when our RV measurement is not included (left panel), which vanish when we include our measurement (right panel). Also, our measurement sets the value of the argument of periastron $\omega$, albeit still with large uncertainties, the ascending node $\Omega$ and the mean longitude at reference epoch J2010.0 $\lambda$ (see Fig. \ref{fig:conrner_plot_orvara_Omega_omega_lam}). In addition, the very precise astrometric orbit, together with the RV information, allows us to have precise information about the phase of the planet at a given epoch. This will be of particular interest for the next generation of instruments probing the reflected light of exoplanets.

Figure~\ref{fig:RV_comp} presents the prediction of the relative RV between the companion and the star as a function of time. Similarly, our RV measurement resolves the ambiguity in the sign of the RV. However, our measurement does not provide additional constraints on its exact value. This is illustrated in the bottom left inset of the right panel, which presents a zoom around our \hirise measurement.

\begin{table*}[!ht]
    \centering
    \caption{Fitted and derived parameters of AF\,Lep\,b with orvara.}
    \begin{tabular}{lll}
        \hline
        \hline
        Parameter                                & Prior & Median and 1$\sigma$ errors \\
        \hline
        \multicolumn{3}{c}{Fitted parameters}  \\
        \hline
        Primary mass $M_{\star}$ (\MSun)         & $\mathcal{N}(1.2,0.06)$ & $1.218_{-0.039}^{+0.029}$ \\
        Secondary mass $M_{p}$ (\MJup)           & $1/M$ (log-flat)        & $3.61_{-0.48}^{+0.49}$ \\
        Semi-major axis $a$ (au)                 & $1/a$ (log-flat)        & $8.996_{-0.077}^{+0.19}$ \\
        $\sqrt{e}\sin\omega$                     & $\mathcal{U}(0,1)$      & $0.08_{-0.12}^{+0.11}$ \\
        $\sqrt{e}\cos\omega$                     & $\mathcal{U}(0,1)$      & $-0.028_{-0.078}^{+0.089}$ \\
        Inclination $i$ (\degre)                 & $\sin i$ with $i \in [0\degre,180\degre]$ (Uniform on $\sin i$) & $57.50_{-0.65}^{+0.61}$ \\
        Mean longitude\tablefootmark{a} $\lambda_{\mathrm{ref}}$ (\degre)  & $\mathcal{U}(0,360)$ & $353.2_{-3.6}^{+8.6}$ \\
        Ascending node $\Omega$ (\degre)         & $\mathcal{U}(0,360)$    & $248.54_{-2.2}^{+0.91}$ \\
        Parallax $\varpi$ (mas)                  & $\mathcal{N}$(37.254, 0.019) & $37.254_{-0.019}^{+0.019}$ \\
        \hline
        \multicolumn{3}{c}{Derived parameters} \\
        \hline
        Period $P$ (years)                       & ---                     & $24.38_{-0.41}^{+1.1}$ \\
        Argument of periastron $\omega$ (\degre) & ---                     & $121_{-43}^{+158}$ \\
        Eccentricity $e$                         & ---                     & $0.015_{-0.011}^{+0.029}$ \\
        Time of periastron\tablefootmark{b} $t_0$ (JD) & ---               & $2458231_{-1095}^{+3683}$ \\
        Mass ratio $\frac{M_{p}}{M_{\star}}$     & ---                     & $0.00284_{-0.00036}^{+0.00036}$ \\
        \hline  \\
    \end{tabular}
    \label{tab:orvara_results}
    \tablefoot{
        \tablefoottext{a}{The mean longitude $\lambda_{\mathrm{ref}}^{*}$ is computed at reference epoch $t_{\mathrm{ref}} = 2455197.4$ JD (J2010.0).}
        \tablefoottext{b}{The time of periastron is computed as $t_0 = t_{ref} - P\frac{\lambda_{\mathrm{ref}}-\omega}{360\degre}$.}
        }
\end{table*}

\section{Discussion}

\label{sec:discussion}

\subsection{Effective temperature and surface gravity}
\label{sec:Teff_logg}

Previous studies suggest a lower \Teff, of the order of 800\,K \citep{palma-bifani_atmospheric_2024, zhang_elemental_2023}. This parameter, as well as \logg, impact mostly the continuum of the spectra as well as the depth of molecular lines. Our \hirise data tends to favour low \logg values and high \Teff values, compared to what is expected (see Table~\ref{tab:formosa_results}.) We also find strong correlations between \Teff and \logg in our results (see Figs.~\ref{fig:corner_first_night_free} and ~\ref{fig:corner_second_night_free}), indicative that a more consistent lower \Teff would also decrease the estimated value of \logg. We note that low values for \logg are physically inconsistent with the estimated radius of AF\,Lep\,b \citep[$1.3 \pm 0.15$\,\RJup,][]{balmer_vltigravity_2024} as the value of \logg strongly affects the radius estimated from Newton's law (see Table \ref{tab:formosa_results}). For a \logg of 3.22\,dex, corresponding to the value obtained for the first night with uninformative priors on all parameters, the self consistent radius using Newton's law would be 2.37\,\RJup, which is highly inconsistent with the results of \cite{zhang_elemental_2023}, \cite{palma-bifani_atmospheric_2024} and \cite{balmer_vltigravity_2024}. This demonstrates the importance of using an accurate value on \logg when fitting a spectra. To address this issue, we use Gaussian priors on \Teff and \logg (\Teff $\sim \mathcal{N}(800,50)$, \logg $\sim \mathcal{N}(3.7,0.2)$). However, even with these Gaussian priors, our results tend to favour \Teff values slightly higher and \logg slightly lower than the expected values.

As a result, it is difficult to constrain these parameters using only our \hirise data. These two parameters are generally best estimated with high S/N low-resolution data over a large wavelength range, where we have accurate information on the continuum, which is not the case since we have removed the continuum (see Sect.~\ref{sec:data_modeling}), and on the depth of the spectral lines. It would therefore be interesting to combine low-resolution data with \hirise data in order to have a robust estimate of these parameters without the need to apply strong priors to these parameters.

\subsection{Carbon-to-oxygen ratio and metallicity}
\label{sec:CO_met}

\begin{figure}
    \centering
    \includegraphics[width=\linewidth]{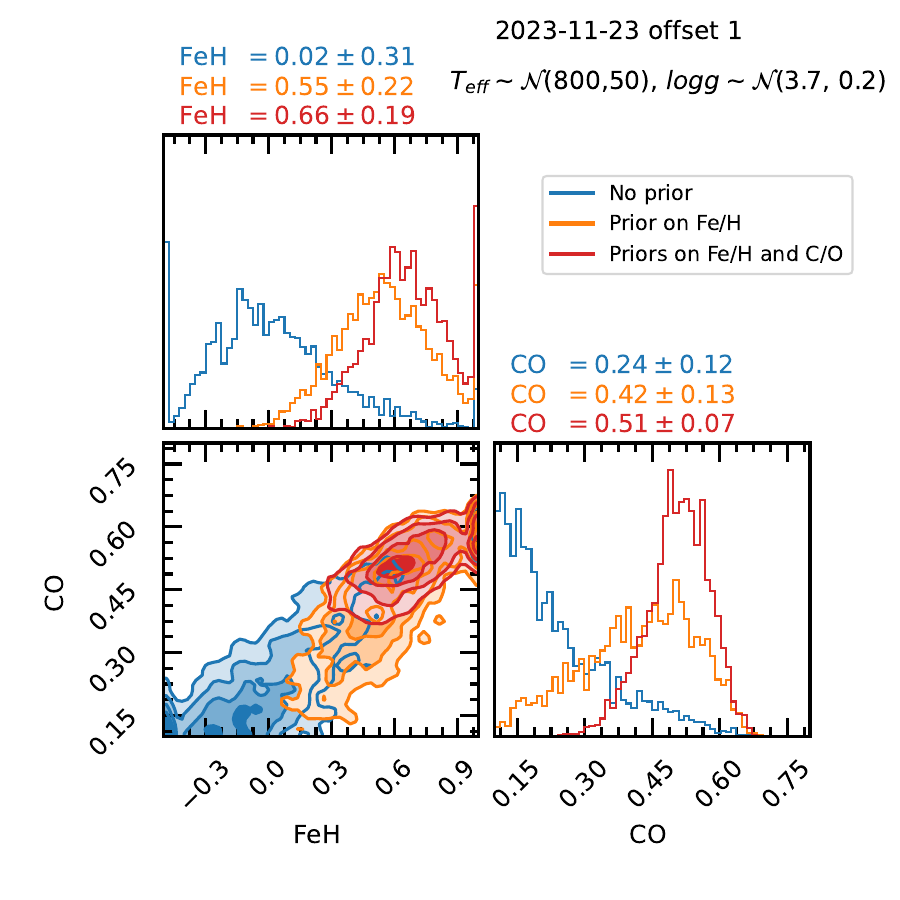}
    \caption{First offset of the second night posterior distribution of \met and \co for the Exo-REM/Exo\_k model with $\Teff = 800$\,K and $\logg = 3.7$\,dex. Three cases are considered : No prior (blue), Prior on \met (orange), Priors on \met and \co (red).}
    \label{fig:corner_second_night_FeH_CO}
\end{figure}

The literature reports enriched metallicities for AF\,Lep\,b, higher than 0.4\,dex for \cite{palma-bifani_atmospheric_2024} and around 0.75\,dex for \citet{balmer_vltigravity_2024}, but based on low-resolution data. When we leave all parameters free, we find values of \met inconsistent with these results at 3 $\sigma$ for the first night and 2 $\sigma$ for the second night. The difficulty to constrain \met with high-resolution data alone has been previously noted by \cite{landman_beta_2024}. This is mainly the results of the low S/N of our data. Indeed, \cite{Xuan2024} and \cite{costes_fresh_2024} show that at high S/N, high-resolution data can give very accurate \met estimates. Again, combining \hirise data with other datasets at lower resolution with high S/N and over a larger wavelength range would certainly help provide better constraints on this parameter.

For the \co ratio, the detection of CH$_4$ in the atmosphere of AF\,Lep\,b (see Sect.~\ref{sec:molecules_detection}), as also supported by \cite{balmer_vltigravity_2024}, reinforces the possibility of deriving a \co ratio in the H band. The \co ratio estimated using \hirise data alone for the second night appears inconsistent with the near-solar values reported in the literature when all parameters are free. However, this parameter tends to vary when we apply a prior on \met (see Table~\ref{tab:formosa_results}).  For both nights, the \co ratio converges to near-solar or slightly super-solar values without the need for a constrained prior on this parameter as depicted on Fig.~\ref{fig:corner_second_night_FeH_CO}. This demonstrates that combining high-resolution with low-resolution data may help break degeneracies that can exist between these parameters.

The \co ratio is often assumed as a tracer of the formation location of the planet in the protoplanetary disk \citep{oberg_effects_2011,mordasini_imprint_2016,molyarova_gas_2017,madhusudhan_atmospheric_2017,booth_chemical_2017}. This is because gaseous H$_2$O, CO$_2$ and CO molecules condensate into solid icy grains at various snowlines \citep{oberg_effects_2011}, therefore affecting the \co ratio in the gas and solid phases. However, caution should be taken when interpreting the \co ratio as it is particularly sensitive to planet formation and evolution assumptions \citep[see e.g.,][]{molliere_interpreting_2022,hoch_assessing_2023}. As shown in \citet{balmer_vltigravity_2024}, the \co ratio and \met for AF\,Lep\,b derived from GRAVITY data could either lead to a core formed beyond the CO iceline, which would have migrated inwards, or to an in-situ formation, depending on the assumptions made on the chemical evolution of the disk.

\subsection{Orbit, radial velocity and projected rotational velocity}

Previous orbital analyses of AF\,Lep\,b have been conducted by \citet{DeRosa2023}, \citet{franson_astrometric_2023}, and \citet{Mesa2023} who independently concluded in very different mass estimates: $4.3_{-1.2}^{+2.9}$\,\MJup, $3.2_{-0.6}^{+0.7}$\,\MJup, and $5.24_{-0.10}^{+0.09}$\,\MJup, respectively. These discrepencies come from the fact that their relative astrometry was measured at different epochs and over different baselines. By combining all the astrometry measurements used in these analyses, \citep{zhang_elemental_2023} gave an update on the orbital parameters of AF\,Lep\,b, but they were still unable to capture the circular nature of the planet's orbit because of the short time-baseline covered by the relative astrometry data. The addition of the recovered archival VLT/NaCo astrometry of 2011 relieves this issue and confirms the quasi-circular nature of the planet's orbit \citep{Bonse2024}. This is because the NaCo astrometric data point is almost opposite to the more recent astrometry data points with respect to the star (Fig. \ref{fig:astrometric_orbit}). However, for the same reason, it does not help better constrain the inclination of the orbit. \citet{balmer_vltigravity_2024} further confirms the circular nature of the orbit with the addition of the GRAVITY measurements, and make a significant improvement in the constraints on the planet's orbital parameters. With this dataset, the amplitudes of RVs of the star and the companion are well constrained, but still leave some ambiguity regarding the sign of the RV. Our \hirise RV measurement on the companion itself allows to resolve this ambiguity.

The estimated radial velocity and projected rotational velocity do not vary when we change the priors configuration, as shown on Table \ref{tab:formosa_results} (see also Figs. \ref{fig:corner_first_night_priors} and \ref{fig:corner_second_night_priors}). This comforts us in the derived value for these parameters. The measured projected rotational velocity gives us an idea of the period of rotation of the companion. Rotation rates of exoplanets are an important parameter since they determine the magnitude of their Coriolis force, which is a key parameter in understanding the climate of exoplanets. It is also linked to the formation of exoplanets since it is related to the accretion of angular momentum during the formation phase in the proto-planetary disk \citep{lissauer_origin_1991,dones_origin_1993,johansen_prograde_2010, batygin_terminal_2018}. From the combination of the orbit fitting and forward modeling, we infer a rotational velocity of $\frac{\vsini}{sin(i)} = 14.63_{-2.87}^{+2.63}$\,\kms for AF\,Lep\,b, assuming the orbit and rotation of the planet are coplanar ($i = 57.78_{-0.58}^{+0.57}$\degre). From this rotational velocity, we can compute the rotation period with :
\begin{equation}
    P_{\mathrm{rot}} = \frac{2\pi R}{v_{\mathrm{rot}}}
\end{equation}
with R the radius. We find a period of rotation of $10.86_{-2.08}^{+3.00}$\,hours, given a radius of $1.3 \pm 0.15$\,\RJup \citep{balmer_vltigravity_2024}. The rotation period of Jupiter is ($\sim$10\,hours), corresponding to a rotational velocity of 12.20 \kms. This result is in line with the tentative trend in spin velocity with planet mass \citep{hughes_planetary_2003,Wang2021,morris_kappa_2024,Xuan2024,Hsu2024}. However, caution should be made when inferring the rotation period from \vsini and $\sin i$, or equivalently, when inferring $\sin i$ from the rotation period and \vsini. Indeed, because \vsini and $\sin i$ are correlated, combining these estimates is somewhat more complex \citep{masuda_inference_2020, bryan_obliquity_2021, bowler_rotation_2023, zhang__testbed_2024}.

\subsection{Limitations and possible improvements}

Despite the analysis conducted above, it is important to keep in mind that high-resolution models still perform poorly in predicting observations. Discrepancies such as continuum mismatches \citep{lim_atmospheric_2023,jahandar_comprehensive_2024}, unidentified spectral features in observations \citep{jahandar_comprehensive_2024}, and shifts in spectral features \citep{tannock_146248_2022,jahandar_comprehensive_2024} can strongly bias parameter estimations. Systematics between models \citep[Ravet et al. 2024, in preparation;][]{petrus_jwst_2024} also play a significant role. These issues could be mitigated by fitting only the spectral lines or parts of the spectrum we can trust \citep{jahandar_comprehensive_2024,petrus_jwst_2024}. Additionally, the effects of systematics could be incorporated into a refined modeling of the covariance matrix. At present, our covariance matrix modeling assumes uncorrelated noise. However, correlated residuals across adjacent pixels may arise due to rotational broadening, imperfections in the template model, or systematics. This issue is generally addressed through the use of Gaussian Processes \citep{czekala_constructing_2015,kawahara_autodifferentiable_2022,iyer_sphinx_2023,regt_eso_2024}.

A second limitation lies in the extraction of the signal itself. Currently, we extract the spectrum by simply summing the signal within a window centered around the trace, extending over 6 pixels (see Sect. \ref{sec:data_reduction}). This approach has been shown to be suboptimal and could be improved by applying weighting to the sum based on the shape of the fitted trace \citep[i.e., using an ``optimal extraction'' algorithm;][]{horne_optimal_1986,marsh_extraction_1989,zechmeister_flat-relative_2014,piskunov_optimal_2021,holmberg_first_2022}, typically offering a 70 \% improvement in effective exposure time. This would probably result in more robust estimates (lower error bars) of the parameters of the planet.

A final limitation lies in the observation strategy. The current wavelength calibration correction is likely not fully optimal. Future observations could leverage M-dwarf or M-giant stars, which have deep and stable spectral lines, to achieve robust wavelength calibration across all order segments (see Sect. \ref{sec:wave_recal}).

\section{Conclusions}
\label{sec:conclusions}

We obtained new $H$-band high-resolution ($R \approx 140\,000$, 1.4--1.8\,\mic) spectroscopic data of AF\,Lep\,b with the new VLT/\hirise instrument. These observations result in a detection of AF\,Lep\,b with the first direct measurement of its projected rotational velocity and RV. We also detect methane absorption in the planet's atmosphere at a 3$\sigma$ significance level. This detection, a major finding of this paper, is consistent with the observations of \citet{balmer_vltigravity_2024}. It also supports our derivation of the \co ratio, as the only significant carbon-bearing species in the H-band is CH$_4$.

We give new constraints on the orbit of AF\,Lep\,b, ruling out a whole family of orbital solutions. Combining our results with the low-resolution results from \citet{balmer_vltigravity_2024}, we can give an update on the atmospheric and orbital parameters of the planet: $\Teff = 800 \pm 50$\,K, $\logg = 3.7 \pm 0.2$\,dex, $R = 1.3 \pm 0.15$\,\RJup, $\met = 0.75 \pm 0.25$\,dex, $\co = 0.55 \pm 0.10$ and $\mathrm{RV} = 8.76_{-0.35}^{+0.22}$\,\kms at epoch J2023.88 (2023-11-20).

Our data uniquely provide an estimation of the planet's projected rotational velocity, $\vsini = 12.58_{-2.54}^{+2.30}$\,\kms. We predict the rotation period of AF\,Lep\,b to be $10.9 \pm 3$\,hours, assuming that the orbit and rotation of the planet are coplanar. This issue could be resolved by acquiring time-resolved high-resolution photometric measurements of AF\,Lep\,b in order to measure variations the in brightness of its atmosphere. This has been done for the first time on the 4\,\MJup directly-imaged companion of 2M1207b \citep{zhou_discovery_2016}, but also for VHS 1256b \citep{zhou_spectral_2020}. However, for high-contrast planets, variability monitoring from the ground is limited by varying speckle noise \citep{biller_high-contrast_2021, wang_atmospheric_2022}. From space, the stability and high-contrast imaging capabilities of the \jwst could be promising for variability monitoring on high-contrast planets such as AF\,Lep\,b. \cite{Franson2024} presents a detection along with the first attempt at variability monitoring of AF\,Lep\,b with the \jwst, but shows no evidence of variability. However, the authors used 5 epochs of observations spread over 2 years. Given the \vsini estimated for the planet, it would be interesting to obtain photometric data points distributed over a few hours. Still, the value derived for the projected rotational velocity of AF\,Lep\,b, in line with the known trend in spin velocity with planet mass, suggests that the planetary orbit and rotation axes should be aligned as is the case for Jupiter, the main gas giant in our system.

With the characterization of AF Lep b with \hirise, we are getting closer to characterizing Jupiter-mass companions in direct imaging at high spectral resolution. \hirise demonstrates a promising instrument for the characterization of such faint, low-mass companions; moreover, it excels at low-separation detection in direct imaging standards. This is particularly interesting since \citet{santos_observational_2017} and \citet{schlaufman_evidence_2018} have reported the existence of two populations of exoplanets which split at 4\,\MJup. Planets with masses above 4\,\MJup are found orbiting less metal rich stars compared to planets with masses below 4\,\MJup, where the host stars are more metal rich. The latter is associated to accretion formation pathway. Similarly, \citet{hoch_assessing_2023} determined there were two populations of companion when looking at \co ratio and mass. Interestingly, they also concluded that the split is around 4\,\MJup. Hopefully, the ESA/Gaia DR4 results expected in the coming years will provide additional Jupiter-mass planetary candidates that could help confirm such a trend.

The orbit of AF\,Lep\,b is now well constrained, both in astrometry and in RV. \citet{balmer_vltigravity_2024} proposed that astrometric monitoring of AF\,Lep\,b could detect deviations from the well-established astrometry, which might be caused by an inner planet or a massive exomoon in the system. This method has proven to be effective, even allowing constraints to be placed on the orbit and mass of the inner planet in the $\beta$\,Pictoris system \citep{lacour_mass_2021}. Another way to do this would be through the acquisition of a high S/N spectrum of the star to study the variations of RV of the star. The instrument VLT/ESPRESSO would be a good candidate in such a study \citep{pepe_espresso_2021}.

Compared to transiting planets, directly imaged planets orbit farther away, which results in larger Hill spheres and makes these planets more favorable for moon formation and retention. In this context, it has been proposed that studying RV variations could be a promising way to detect exomoons around directly imaged planets \citep{vanderburg_detecting_2018,vanderburg_first_2021,Ruffio2023}. \hirise, as well as future generations of instruments, could be promising, not necessarily for detection, but for setting initial constraints on the parameters of binary planets and exomoons around AF\,Lep\,b.

Altogether, it is clear that AF\,Lep\,b will remain a benchmark for exoplanet characterization, providing valuable insights into planet formation and evolution both within and beyond our Solar System.

\begin{acknowledgements}
    This project has received funding from \emph{Agence Nationale de la Recherche} (ANR) under grant ANR-23-CE31-0006-01 (MIRAGES). The \hirise instrument has been developed with funding from the European Research Council (ERC) under the European Union's Horizon 2020 research and innovation programme, grant agreements No. 757561 (\hirise) and 678777 (ICARUS), from the \emph{Commission Spécialisée Astronomie-Astrophysique} (CSAA) of CNRS/INSU, and from the \emph{Action Spécifique Haute Résolution Angulaire} (ASHRA) of CNRS/INSU co-funded by CNES, and from Région Provence-Alpes-Côte d'Azur under grant agreement 2014-0276 (ASOREX).
    This research has made use of computing facilities operated by CeSAM data center at LAM, Marseille, France.
    S.P. is supported by the ANID FONDECYT postdoctoral program No 3240145.
    N.G. acknowledges funding by the European Union (ERC, ESCAPE, projet No 101044152). Views and opinions expressed are however those of the author(s) only and do not reflect those of the European Union or the European Research Council Executive Agency. Neither the European Union nor the granting authority can be held responsible for them.
\end{acknowledgements}

\bibliographystyle{aa}
\bibliography{AFLep}

\begin{thebibliography}{106}
\expandafter\ifx\csname natexlab\endcsname\relax\def\natexlab#1{#1}\fi

\bibitem[{{Asplund} {et~al.}(2009){Asplund}, {Grevesse}, {Sauval}, \&
  {Scott}}]{Asplund2009}
{Asplund}, M., {Grevesse}, N., {Sauval}, A.~J., \& {Scott}, P. 2009, \araa, 47,
  481

\bibitem[{{Bailer-Jones} {et~al.}(2021){Bailer-Jones}, {Rybizki}, {Fouesneau},
  {Demleitner}, \& {Andrae}}]{bailer-jones_estimating_2021}
{Bailer-Jones}, C.~A.~L., {Rybizki}, J., {Fouesneau}, M., {Demleitner}, M., \&
  {Andrae}, R. 2021, \aj, 161, 147

\bibitem[{{Balmer} {et~al.}(2025){Balmer}, {Franson}, {Chomez}, {Pueyo},
  {Stolker}, {Lacour}, {Nowak}, {Nasedkin}, {Bonse}, {Thorngren},
  {Palma-Bifani}, {Molli{\`e}re}, {Wang}, {Zhang}, {Chavez}, {Kammerer},
  {Blunt}, {Bowler}, {Bonnefoy}, {Brandner}, {Charnay}, {Chauvin}, {Henning},
  {Lagrange}, {Pourr{\'e}}, {Rickman}, {De Rosa}, {Vigan}, \&
  {Winterhalder}}]{balmer_vltigravity_2024}
{Balmer}, W.~O., {Franson}, K., {Chomez}, A., {et~al.} 2025, {VLTI/GRAVITY
  Observations of AF Lep b: Preference for Circular Orbits, Cloudy Atmospheres,
  and a Moderately Enhanced Metallicity}

\bibitem[{{Batygin}(2018)}]{batygin_terminal_2018}
{Batygin}, K. 2018, \aj, 155, 178

\bibitem[{{Baudino} {et~al.}(2015){Baudino}, {B{\'e}zard}, {Boccaletti},
  {Bonnefoy}, {Lagrange}, \& {Galicher}}]{baudino_interpreting_2015}
{Baudino}, J.~L., {B{\'e}zard}, B., {Boccaletti}, A., {et~al.} 2015, \aap, 582,
  A83

\bibitem[{{Bell} {et~al.}(2015){Bell}, {Mamajek}, \& {Naylor}}]{Bell2015}
{Bell}, C. P.~M., {Mamajek}, E.~E., \& {Naylor}, T. 2015, \mnras, 454, 593

\bibitem[{{Benneke} \& {Seager}(2013)}]{benneke_how_2013}
{Benneke}, B. \& {Seager}, S. 2013, \apj, 778, 153

\bibitem[{{Beuzit} {et~al.}(2019){Beuzit}, {Vigan}, {Mouillet}, {Dohlen},
  {Gratton}, {Boccaletti}, {Sauvage}, {Schmid}, {Langlois}, {Petit},
  {Baruffolo}, {Feldt}, {Milli}, {Wahhaj}, {Abe}, {Anselmi}, {Antichi},
  {Barette}, {Baudrand}, {Baudoz}, {Bazzon}, {Bernardi}, {Blanchard}, {Brast},
  {Bruno}, {Buey}, {Carbillet}, {Carle}, {Cascone}, {Chapron}, {Charton},
  {Chauvin}, {Claudi}, {Costille}, {De Caprio}, {de Boer}, {Delboulb{\'e}},
  {Desidera}, {Dominik}, {Downing}, {Dupuis}, {Fabron}, {Fantinel}, {Farisato},
  {Feautrier}, {Fedrigo}, {Fusco}, {Gigan}, {Ginski}, {Girard}, {Giro},
  {Gisler}, {Gluck}, {Gry}, {Henning}, {Hubin}, {Hugot}, {Incorvaia}, {Jaquet},
  {Kasper}, {Lagadec}, {Lagrange}, {Le Coroller}, {Le Mignant}, {Le Ruyet},
  {Lessio}, {Lizon}, {Llored}, {Lundin}, {Madec}, {Magnard}, {Marteaud},
  {Martinez}, {Maurel}, {M{\'e}nard}, {Mesa}, {M{\"o}ller-Nilsson}, {Moulin},
  {Moutou}, {Orign{\'e}}, {Parisot}, {Pavlov}, {Perret}, {Pragt}, {Puget},
  {Rabou}, {Ramos}, {Reess}, {Rigal}, {Rochat}, {Roelfsema}, {Rousset}, {Roux},
  {Saisse}, {Salasnich}, {Santambrogio}, {Scuderi}, {Segransan}, {Sevin},
  {Siebenmorgen}, {Soenke}, {Stadler}, {Suarez}, {Tiph{\`e}ne}, {Turatto},
  {Udry}, {Vakili}, {Waters}, {Weber}, {Wildi}, {Zins}, \&
  {Zurlo}}]{Beuzit2019}
{Beuzit}, J.~L., {Vigan}, A., {Mouillet}, D., {et~al.} 2019, \aap, 631, A155

\bibitem[{{Biller} {et~al.}(2021){Biller}, {Apai}, {Bonnefoy}, {Desidera},
  {Gratton}, {Kasper}, {Kenworthy}, {Lagrange}, {Lazzoni}, {Mesa}, {Vigan},
  {Wagner}, {Vos}, \& {Zurlo}}]{biller_high-contrast_2021}
{Biller}, B.~A., {Apai}, D., {Bonnefoy}, M., {et~al.} 2021, \mnras, 503, 743

\bibitem[{{Bonavita} {et~al.}(2022){Bonavita}, {Fontanive}, {Gratton},
  {Mu{\v{z}}i{\'c}}, {Desidera}, {Mesa}, {Biller}, {Scholz}, {Sozzetti}, \&
  {Squicciarini}}]{Bonavita2022}
{Bonavita}, M., {Fontanive}, C., {Gratton}, R., {et~al.} 2022, \mnras, 513,
  5588

\bibitem[{{Bonse} {et~al.}(2024){Bonse}, {Gebhard}, {Dannert}, {Absil},
  {Cantalloube}, {Christiaens}, {Cugno}, {Garvin}, {Hayoz}, {Kasper},
  {Matthews}, {Sch{\"o}lkopf}, \& {Quanz}}]{Bonse2024}
{Bonse}, M.~J., {Gebhard}, T.~D., {Dannert}, F.~A., {et~al.} 2024, arXiv
  e-prints, arXiv:2406.01809

\bibitem[{{Booth} {et~al.}(2017){Booth}, {Clarke}, {Madhusudhan}, \&
  {Ilee}}]{booth_chemical_2017}
{Booth}, R.~A., {Clarke}, C.~J., {Madhusudhan}, N., \& {Ilee}, J.~D. 2017,
  \mnras, 469, 3994

\bibitem[{{Bowler} {et~al.}(2021){Bowler}, {Endl}, {Cochran}, {MacQueen},
  {Crepp}, {Doppmann}, {Dulz}, {Brandt}, {Mirek Brandt}, {Li}, {Dupuy},
  {Franson}, {Kratter}, {Morley}, \& {Zhou}}]{Bowler2021}
{Bowler}, B.~P., {Endl}, M., {Cochran}, W.~D., {et~al.} 2021, \apjl, 913, L26

\bibitem[{{Bowler} {et~al.}(2023){Bowler}, {Tran}, {Zhang}, {Morgan}, {Ashok},
  {Blunt}, {Bryan}, {Evans}, {Franson}, {Huber}, {Nagpal}, {Wu}, \&
  {Zhou}}]{bowler_rotation_2023}
{Bowler}, B.~P., {Tran}, Q.~H., {Zhang}, Z., {et~al.} 2023, \aj, 165, 164

\bibitem[{{Brandt} {et~al.}(2021{\natexlab{a}}){Brandt}, {Brandt}, {Dupuy},
  {Michalik}, \& {Marleau}}]{Brandt2021}
{Brandt}, G.~M., {Brandt}, T.~D., {Dupuy}, T.~J., {Michalik}, D., \& {Marleau},
  G.-D. 2021{\natexlab{a}}, \apjl, 915, L16

\bibitem[{{Brandt}(2021)}]{Brandt2021HipGaia}
{Brandt}, T.~D. 2021, \apjs, 254, 42

\bibitem[{{Brandt} {et~al.}(2021{\natexlab{b}}){Brandt}, {Dupuy}, {Li},
  {Brandt}, {Zeng}, {Michalik}, {Bardalez Gagliuffi}, \&
  {Raposo-Pulido}}]{brandt_orvara_2021}
{Brandt}, T.~D., {Dupuy}, T.~J., {Li}, Y., {et~al.} 2021{\natexlab{b}}, \aj,
  162, 186

\bibitem[{{Bryan} {et~al.}(2021){Bryan}, {Chiang}, {Morley}, {Mace}, \&
  {Bowler}}]{bryan_obliquity_2021}
{Bryan}, M.~L., {Chiang}, E., {Morley}, C.~V., {Mace}, G.~N., \& {Bowler},
  B.~P. 2021, \aj, 162, 217

\bibitem[{{Butler} {et~al.}(2017){Butler}, {Vogt}, {Laughlin}, {Burt},
  {Rivera}, {Tuomi}, {Teske}, {Arriagada}, {Diaz}, {Holden}, \&
  {Keiser}}]{butler_lces_2017}
{Butler}, R.~P., {Vogt}, S.~S., {Laughlin}, G., {et~al.} 2017, \aj, 153, 208

\bibitem[{{Charnay} {et~al.}(2018){Charnay}, {B{\'e}zard}, {Baudino},
  {Bonnefoy}, {Boccaletti}, \& {Galicher}}]{charnay_self-consistent_2018}
{Charnay}, B., {B{\'e}zard}, B., {Baudino}, J.~L., {et~al.} 2018, \apj, 854,
  172

\bibitem[{{Charnay} {et~al.}(2021){Charnay}, {Blain}, {B{\'e}zard}, {Leconte},
  {Turbet}, \& {Falco}}]{charnay_formation_2021}
{Charnay}, B., {Blain}, D., {B{\'e}zard}, B., {et~al.} 2021, \aap, 646, A171

\bibitem[{{Chauvin} {et~al.}(2017){Chauvin}, {Desidera}, {Lagrange}, {Vigan},
  {Feldt}, {Gratton}, {Langlois}, {Cheetham}, {Bonnefoy}, \&
  {Meyer}}]{Chauvin2017sf2a}
{Chauvin}, G., {Desidera}, S., {Lagrange}, A.~M., {et~al.} 2017, in SF2A-2017:
  Proceedings of the Annual meeting of the French Society of Astronomy and
  Astrophysics, ed. C.~{Reyl{\'e}}, P.~{Di Matteo}, F.~{Herpin}, E.~{Lagadec},
  A.~{Lan{\c{c}}on}, Z.~{Meliani}, \& F.~{Royer}, Di

\bibitem[{{Costes} {et~al.}(2024{\natexlab{a}}){Costes}, {Denis}, \&
  {Vigan}}]{hipipe2024}
{Costes}, J., {Denis}, A., \& {Vigan}, A. 2024{\natexlab{a}}, {hipipe:
  VLT/HiRISE reduction pipeline}, Astrophysics Source Code Library, record
  ascl:2407.019

\bibitem[{{Costes} {et~al.}(2024{\natexlab{b}}){Costes}, {Xuan}, {Vigan},
  {Wang}, {D'Orazi}, {Molli{\`e}re}, {Baker}, {Bartos}, {Blake}, {Calvin},
  {Cetre}, {Delorme}, {Doppmann}, {Echeveri}, {Finnerty}, {Fitzgerald}, {Hsu},
  {Jovanovic}, {Lopez}, {Mawet}, {Morris}, {Pezzato}, {Phillips}, {Ruffio},
  {Sappey}, {Schneeberger}, {Schofield}, {Skemer}, {Wallace}, \&
  {Wang}}]{costes_fresh_2024}
{Costes}, J.~C., {Xuan}, J.~W., {Vigan}, A., {et~al.} 2024{\natexlab{b}}, \aap,
  686, A294

\bibitem[{{Currie} {et~al.}(2023){Currie}, {Brandt}, {Brandt}, {Lacy},
  {Burrows}, {Guyon}, {Tamura}, {Liu}, {Sagynbayeva}, {Tobin}, {Chilcote},
  {Groff}, {Marois}, {Thompson}, {Murphy}, {Kuzuhara}, {Lawson}, {Lozi}, {Deo},
  {Vievard}, {Skaf}, {Uyama}, {Jovanovic}, {Martinache}, {Kasdin}, {Kudo},
  {McElwain}, {Janson}, {Wisniewski}, {Hodapp}, {Nishikawa}, {He{\l}miniak},
  {Kwon}, \& {Hayashi}}]{Currie2023}
{Currie}, T., {Brandt}, G.~M., {Brandt}, T.~D., {et~al.} 2023, Science, 380,
  198

\bibitem[{{Czekala} {et~al.}(2015){Czekala}, {Andrews}, {Mandel}, {Hogg}, \&
  {Green}}]{czekala_constructing_2015}
{Czekala}, I., {Andrews}, S.~M., {Mandel}, K.~S., {Hogg}, D.~W., \& {Green},
  G.~M. 2015, \apj, 812, 128

\bibitem[{{Czesla} {et~al.}(2019){Czesla}, {Schr{\"o}ter}, {Schneider},
  {Huber}, {Pfeifer}, {Andreasen}, \& {Zechmeister}}]{PyAstronomy}
{Czesla}, S., {Schr{\"o}ter}, S., {Schneider}, C.~P., {et~al.} 2019, {PyA:
  Python astronomy-related packages}

\bibitem[{{de Regt} {et~al.}(2024){de Regt}, {Gandhi}, {Snellen}, {Zhang},
  {Ginski}, {Gonz{\'a}lez Picos}, {Kesseli}, {Landman}, {Molli{\`e}re},
  {Nasedkin}, {S{\'a}nchez-L{\'o}pez}, \& {Stolker}}]{regt_eso_2024}
{de Regt}, S., {Gandhi}, S., {Snellen}, I.~A.~G., {et~al.} 2024, \aap, 688,
  A116

\bibitem[{{De Rosa} {et~al.}(2023){De Rosa}, {Nielsen}, {Wahhaj}, {Ruffio},
  {Kalas}, {Peck}, {Hirsch}, \& {Roberson}}]{DeRosa2023}
{De Rosa}, R.~J., {Nielsen}, E.~L., {Wahhaj}, Z., {et~al.} 2023, \aap, 672, A94

\bibitem[{{Dones} \& {Tremaine}(1993)}]{dones_origin_1993}
{Dones}, L. \& {Tremaine}, S. 1993, \icarus, 103, 67

\bibitem[{{Dorn} {et~al.}(2023){Dorn}, {Bristow}, {Smoker}, {Rodler}, {Lavail},
  {Accardo}, {van den Ancker}, {Baade}, {Baruffolo}, {Courtney-Barrer},
  {Blanco}, {Brucalassi}, {Cumani}, {Follert}, {Haimerl}, {Hatzes}, {Haug},
  {Heiter}, {Hinterschuster}, {Hubin}, {Ives}, {Jung}, {Jones}, {Kaeufl},
  {Kirchbauer}, {Klein}, {Kochukhov}, {Korhonen}, {K{\"o}hler}, {Lizon},
  {Moins}, {Molina-Conde}, {Marquart}, {Neeser}, {Oliva}, {Pallanca},
  {Pasquini}, {Paufique}, {Piskunov}, {Reiners}, {Schneller}, {Schmutzer},
  {Seemann}, {Slumstrup}, {Smette}, {Stegmeier}, {Stempels}, {Tordo},
  {Valenti}, {Valenzuela}, {Vernet}, {Vinther}, \& {Wehrhahn}}]{Dorn2023}
{Dorn}, R.~J., {Bristow}, P., {Smoker}, J.~V., {et~al.} 2023, \aap, 671, A24

\bibitem[{{El Morsy} {et~al.}(2022){El Morsy}, {Vigan}, {Lopez}, {Otten},
  {Choquet}, {Madec}, {Costille}, {Sauvage}, {Dohlen}, {Muslimov}, {Pourcelot},
  {Floriot}, {Benedetti}, {Blanchard}, {Balard}, \& {Murray}}]{ElMorsy2022}
{El Morsy}, M., {Vigan}, A., {Lopez}, M., {et~al.} 2022, \aap, 667, A171

\bibitem[{{Foreman-Mackey} {et~al.}(2013){Foreman-Mackey}, {Hogg}, {Lang}, \&
  {Goodman}}]{foreman-mackey_emcee_2013}
{Foreman-Mackey}, D., {Hogg}, D.~W., {Lang}, D., \& {Goodman}, J. 2013, \pasp,
  125, 306

\bibitem[{{Franson} {et~al.}(2024){Franson}, {Balmer}, {Bowler}, {Pueyo},
  {Zhou}, {Rickman}, {Zhang}, {Mukherjee}, {Pearce}, {Bardalez Gagliuffi},
  {Biddle}, {Brandt}, {Bowens-Rubin}, {Crepp}, {Davidson}, {Faherty}, {Ginski},
  {Horch}, {Morgan}, {Morley}, {Perrin}, {Sanghi}, {Salama}, {Theissen},
  {Tran}, \& {Wolf}}]{Franson2024}
{Franson}, K., {Balmer}, W.~O., {Bowler}, B.~P., {et~al.} 2024, \apjl, 974, L11

\bibitem[{{Franson} {et~al.}(2023){Franson}, {Bowler}, {Zhou}, {Pearce},
  {Bardalez Gagliuffi}, {Biddle}, {Brandt}, {Crepp}, {Dupuy}, {Faherty},
  {Jensen-Clem}, {Morgan}, {Sanghi}, {Theissen}, {Tran}, \&
  {Wolf}}]{franson_astrometric_2023}
{Franson}, K., {Bowler}, B.~P., {Zhou}, Y., {et~al.} 2023, \apjl, 950, L19

\bibitem[{{Gaia Collaboration} {et~al.}(2023){Gaia Collaboration}, {Vallenari},
  {Brown}, {Prusti}, {de Bruijne}, {Arenou}, {Babusiaux}, {Biermann},
  {Creevey}, {Ducourant}, \& et~al.}]{gaia_collaboration_2023}
{Gaia Collaboration}, {Vallenari}, A., {Brown}, A.~G.~A., {et~al.} 2023, \aap,
  674, A1

\bibitem[{{Gray} {et~al.}(2006){Gray}, {Corbally}, {Garrison}, {McFadden},
  {Bubar}, {McGahee}, {O'Donoghue}, \& {Knox}}]{gray_contributions_2006}
{Gray}, R.~O., {Corbally}, C.~J., {Garrison}, R.~F., {et~al.} 2006, \aj, 132,
  161

\bibitem[{{Hayoz} {et~al.}(2025){Hayoz}, {Quanz}, \& {et al.}}]{Hayoz2025}
{Hayoz}, J., {Quanz}, S., \& {et al.} 2025, \aap, submitted

\bibitem[{{Hoch} {et~al.}(2023){Hoch}, {Konopacky}, {Theissen}, {Ruffio},
  {Barman}, {Rickman}, {Perrin}, {Macintosh}, \&
  {Marois}}]{hoch_assessing_2023}
{Hoch}, K. K.~W., {Konopacky}, Q.~M., {Theissen}, C.~A., {et~al.} 2023, \aj,
  166, 85

\bibitem[{{Holmberg} \& {Madhusudhan}(2022)}]{holmberg_first_2022}
{Holmberg}, M. \& {Madhusudhan}, N. 2022, \aj, 164, 79

\bibitem[{{Horne}(1986)}]{horne_optimal_1986}
{Horne}, K. 1986, \pasp, 98, 609

\bibitem[{{Horstman} {et~al.}(2024){Horstman}, {Ruffio}, {Wang}, {Hsu},
  {Baker}, {Finnerty}, {Xuan}, {Echeverri}, {Mawet}, {Blake}, {Bartos}, {Bond},
  {Calvin}, {Cetre}, {Delorme}, {Doppmann}, {Fitzgerald}, {Jovanovic}, {Lopez},
  {Martin}, {Morris}, {Pezzato}, {Ruane}, {Sappey}, {Schofield}, {Skemer},
  {Venenciano}, {Wallace}, {Wang}, \& {Wizinowich}}]{horstman_fringing_2024}
{Horstman}, K.~A., {Ruffio}, J.-B., {Wang}, J.~J., {et~al.} 2024, in Society of
  Photo-Optical Instrumentation Engineers (SPIE) Conference Series, Vol. 13096,
  Ground-based and Airborne Instrumentation for Astronomy X, ed. J.~J.
  {Bryant}, K.~{Motohara}, \& J.~R.~D. {Vernet}, 130962E

\bibitem[{{Hsu} {et~al.}(2024){Hsu}, {Burgasser}, {Theissen}, {Birky},
  {Aganze}, {Gerasimov}, {Schmidt}, {Blake}, {Covey}, {Moreno-Hilario},
  {Gelino}, {Serna}, {Brownstein}, \& {Cunha}}]{Hsu2024}
{Hsu}, C.-C., {Burgasser}, A.~J., {Theissen}, C.~A., {et~al.} 2024, \apjs, 274,
  40

\bibitem[{{Hughes}(2003)}]{hughes_planetary_2003}
{Hughes}, D.~W. 2003, \planss, 51, 517

\bibitem[{{Husser} {et~al.}(2013){Husser}, {Wende-von Berg}, {Dreizler},
  {Homeier}, {Reiners}, {Barman}, \& {Hauschildt}}]{Husser2013}
{Husser}, T.~O., {Wende-von Berg}, S., {Dreizler}, S., {et~al.} 2013, \aap,
  553, A6

\bibitem[{{Iyer} {et~al.}(2023){Iyer}, {Line}, {Muirhead}, {Fortney}, \&
  {Gharib-Nezhad}}]{iyer_sphinx_2023}
{Iyer}, A.~R., {Line}, M.~R., {Muirhead}, P.~S., {Fortney}, J.~J., \&
  {Gharib-Nezhad}, E. 2023, \apj, 944, 41

\bibitem[{{Jahandar} {et~al.}(2024){Jahandar}, {Doyon}, {Artigau}, {Cook},
  {Cadieux}, {Lafreni{\`e}re}, {Forveille}, {Donati}, {Fouqu{\'e}}, {Carmona},
  {Cloutier}, {Cristofari}, {Gaidos}, {Gomes da Silva}, {Malo}, {Martioli}, {do
  Nascimento}, {Pelletier}, {Vandal}, \& {Venn}}]{jahandar_comprehensive_2024}
{Jahandar}, F., {Doyon}, R., {Artigau}, {\'E}., {et~al.} 2024, \apj, 966, 56

\bibitem[{{Johansen} \& {Lacerda}(2010)}]{johansen_prograde_2010}
{Johansen}, A. \& {Lacerda}, P. 2010, \mnras, 404, 475

\bibitem[{{Karman} {et~al.}(2019){Karman}, {Gordon}, {van der Avoird},
  {Baranov}, {Boulet}, {Drouin}, {Groenenboom}, {Gustafsson}, {Hartmann},
  {Kurucz}, {Rothman}, {Sun}, {Sung}, {Thalman}, {Tran}, {Wishnow},
  {Wordsworth}, {Vigasin}, {Volkamer}, \& {van der Zande}}]{karman_update_2019}
{Karman}, T., {Gordon}, I.~E., {van der Avoird}, A., {et~al.} 2019, \icarus,
  328, 160

\bibitem[{{Kawahara} {et~al.}(2022){Kawahara}, {Kawashima}, {Masuda},
  {Crossfield}, {Pannier}, \& {van den
  Bekerom}}]{kawahara_autodifferentiable_2022}
{Kawahara}, H., {Kawashima}, Y., {Masuda}, K., {et~al.} 2022, \apjs, 258, 31

\bibitem[{{Kervella} {et~al.}(2019){Kervella}, {Arenou}, {Mignard}, \&
  {Th{\'e}venin}}]{Kervella2019}
{Kervella}, P., {Arenou}, F., {Mignard}, F., \& {Th{\'e}venin}, F. 2019, \aap,
  623, A72

\bibitem[{{Kervella} {et~al.}(2022){Kervella}, {Arenou}, \&
  {Th{\'e}venin}}]{Kervella2022}
{Kervella}, P., {Arenou}, F., \& {Th{\'e}venin}, F. 2022, \aap, 657, A7

\bibitem[{{Lacour} {et~al.}(2021){Lacour}, {Wang}, {Rodet}, {Nowak},
  {Shangguan}, {Beust}, {Lagrange}, {Abuter}, {Amorim}, {Asensio-Torres},
  {Benisty}, {Berger}, {Blunt}, {Boccaletti}, {Bohn}, {Bolzer}, {Bonnefoy},
  {Bonnet}, {Bourdarot}, {Brandner}, {Cantalloube}, {Caselli}, {Charnay},
  {Chauvin}, {Choquet}, {Christiaens}, {Cl{\'e}net}, {Coud{\'e} Du Foresto},
  {Cridland}, {Dembet}, {Dexter}, {de Zeeuw}, {Drescher}, {Duvert}, {Eckart},
  {Eisenhauer}, {Gao}, {Garcia}, {Garcia Lopez}, {Gendron}, {Genzel},
  {Gillessen}, {Girard}, {Haubois}, {Hei{\ss}el}, {Henning}, {Hinkley},
  {Hippler}, {Horrobin}, {Houll{\'e}}, {Hubert}, {Jocou}, {Kammerer},
  {Keppler}, {Kervella}, {Kreidberg}, {Lapeyr{\`e}re}, {Le Bouquin},
  {L{\'e}na}, {Lutz}, {Maire}, {M{\'e}rand}, {Molli{\`e}re}, {Monnier},
  {Mouillet}, {Nasedkin}, {Ott}, {Otten}, {Paladini}, {Paumard}, {Perraut},
  {Perrin}, {Pfuhl}, {Rickman}, {Pueyo}, {Rameau}, {Rousset}, {Rustamkulov},
  {Samland}, {Shimizu}, {Sing}, {Stadler}, {Stolker}, {Straub}, {Straubmeier},
  {Sturm}, {Tacconi}, {van Dishoeck}, {Vigan}, {Vincent}, {von Fellenberg},
  {Ward-Duong}, {Widmann}, {Wieprecht}, {Wiezorrek}, {Woillez}, {Yazici},
  {Young}, \& {Gravity Collaboration}}]{lacour_mass_2021}
{Lacour}, S., {Wang}, J.~J., {Rodet}, L., {et~al.} 2021, \aap, 654, L2

\bibitem[{{Landman} {et~al.}(2024){Landman}, {Stolker}, {Snellen}, {Costes},
  {de Regt}, {Zhang}, {Gandhi}, {Molliere}, {Kesseli}, {Vigan}, \&
  {Sanchez-L{\'o}pez}}]{landman_beta_2024}
{Landman}, R., {Stolker}, T., {Snellen}, I.~A.~G., {et~al.} 2024, \aap, 682,
  A48

\bibitem[{{Leconte}(2021)}]{leconte_spectral_2021}
{Leconte}, J. 2021, \aap, 645, A20

\bibitem[{{Lim} {et~al.}(2023){Lim}, {Benneke}, {Doyon}, {MacDonald},
  {Piaulet}, {Artigau}, {Coulombe}, {Radica}, {L'Heureux}, {Albert}, {Rackham},
  {de Wit}, {Salhi}, {Roy}, {Flagg}, {Fournier-Tondreau}, {Taylor}, {Cook},
  {Lafreni{\`e}re}, {Cowan}, {Kaltenegger}, {Rowe}, {Espinoza}, {Dang}, \&
  {Darveau-Bernier}}]{lim_atmospheric_2023}
{Lim}, O., {Benneke}, B., {Doyon}, R., {et~al.} 2023, \apjl, 955, L22

\bibitem[{{Lissauer} \& {Kary}(1991)}]{lissauer_origin_1991}
{Lissauer}, J.~J. \& {Kary}, D.~M. 1991, \icarus, 94, 126

\bibitem[{Lodders(2010)}]{lodders_solar_2010}
Lodders, K. 2010, in Principles and Perspectives in Cosmochemistry, ed.
  A.~Goswami \& B.~E. Reddy (Springer Berlin Heidelberg), 379--417

\bibitem[{{Lodders} \& {Fegley}(2006)}]{lodders_chemistry_2006}
{Lodders}, K. \& {Fegley}, Jr., B. 2006, in Astrophysics Update 2, ed. J.~W.
  {Mason}, 1

\bibitem[{{Macintosh} {et~al.}(2014){Macintosh}, {Anthony}, {Atwood}, {Bauman},
  {Cardwell}, {Caputa}, {Chilcote}, {De Rosa}, {Dillon}, {Doyon}, {Dunn},
  {Erickson}, {Fitzgerald}, {Gavel}, {Galvez}, {Goodsell}, {Graham},
  {Greenbaum}, {Hartung}, {Hibon}, {Ingraham}, {Kerley}, {Konopacky}, {Labrie},
  {Larkin}, {Maire}, {Marchis}, {Marois}, {Millar-Blanchaer}, {Morzinski},
  {Nunez}, {Oppenheimer}, {Palmer}, {Pazder}, {Perrin}, {Poyneer}, {Pueyo},
  {Quiroz}, {Rantakyro}, {Reshetov}, {Saddlemyer}, {Sadakuni}, {Savransky},
  {Serio}, {Sivaramakrishnan}, {Smith}, {Soummer}, {Thomas}, {Wallace}, {Wang},
  {Weiss}, {Wiktorowicz}, \& {Wolff}}]{Macintosh2014}
{Macintosh}, B.~A., {Anthony}, A., {Atwood}, J., {et~al.} 2014, in Society of
  Photo-Optical Instrumentation Engineers (SPIE) Conference Series, Vol. 9148,
  Adaptive Optics Systems IV, ed. E.~{Marchetti}, L.~M. {Close}, \& J.-P.
  {Vran}, 91480J

\bibitem[{{Madhusudhan} {et~al.}(2017){Madhusudhan}, {Bitsch}, {Johansen}, \&
  {Eriksson}}]{madhusudhan_atmospheric_2017}
{Madhusudhan}, N., {Bitsch}, B., {Johansen}, A., \& {Eriksson}, L. 2017,
  \mnras, 469, 4102

\bibitem[{{Marsh}(1989)}]{marsh_extraction_1989}
{Marsh}, T.~R. 1989, \pasp, 101, 1032

\bibitem[{{Masuda} \& {Winn}(2020)}]{masuda_inference_2020}
{Masuda}, K. \& {Winn}, J.~N. 2020, \aj, 159, 81

\bibitem[{{Mesa} {et~al.}(2023){Mesa}, {Gratton}, {Kervella}, {Bonavita},
  {Desidera}, {D'Orazi}, {Marino}, {Zurlo}, \& {Rigliaco}}]{Mesa2023}
{Mesa}, D., {Gratton}, R., {Kervella}, P., {et~al.} 2023, \aap, 672, A93

\bibitem[{{Molli{\`e}re} {et~al.}(2022){Molli{\`e}re}, {Molyarova}, {Bitsch},
  {Henning}, {Schneider}, {Kreidberg}, {Eistrup}, {Burn}, {Nasedkin},
  {Semenov}, {Mordasini}, {Schlecker}, {Schwarz}, {Lacour}, {Nowak}, \&
  {Schulik}}]{molliere_interpreting_2022}
{Molli{\`e}re}, P., {Molyarova}, T., {Bitsch}, B., {et~al.} 2022, \apj, 934, 74

\bibitem[{{Molli{\`e}re} {et~al.}(2019){Molli{\`e}re}, {Wardenier}, {van
  Boekel}, {Henning}, {Molaverdikhani}, \&
  {Snellen}}]{molliere_petitradtrans_2019}
{Molli{\`e}re}, P., {Wardenier}, J.~P., {van Boekel}, R., {et~al.} 2019, \aap,
  627, A67

\bibitem[{{Molyarova} {et~al.}(2017){Molyarova}, {Akimkin}, {Semenov},
  {Henning}, {Vasyunin}, \& {Wiebe}}]{molyarova_gas_2017}
{Molyarova}, T., {Akimkin}, V., {Semenov}, D., {et~al.} 2017, \apj, 849, 130

\bibitem[{{Mordasini} {et~al.}(2016){Mordasini}, {van Boekel}, {Molli{\`e}re},
  {Henning}, \& {Benneke}}]{mordasini_imprint_2016}
{Mordasini}, C., {van Boekel}, R., {Molli{\`e}re}, P., {Henning}, T., \&
  {Benneke}, B. 2016, \apj, 832, 41

\bibitem[{{Morris} {et~al.}(2024){Morris}, {Wang}, {Hsu}, {Ruffio}, {Xuan},
  {Delorme}, {Hood}, {Bryan}, {Martin}, {Pezzato}, {Mawet}, {Skemer}, {Baker},
  {Bartos}, {Calvin}, {Cetre}, {Doppmann}, {Echeverri}, {Finnerty},
  {Fitzgerald}, {Jovanovic}, {Liberman}, {Lopez}, {Sappey}, {Schofield},
  {Wallace}, \& {Wang}}]{morris_kappa_2024}
{Morris}, E.~C., {Wang}, J.~J., {Hsu}, C.-C., {et~al.} 2024, \aj, 168, 144

\bibitem[{{Morris} {et~al.}(2020){Morris}, {Wang}, {Ruffio}, {Delorme},
  {Pezzato}, {Bond}, {Mawet}, \& {Skemer}}]{Morris2020}
{Morris}, E.~C., {Wang}, J.~J., {Ruffio}, J.-B., {et~al.} 2020, in Society of
  Photo-Optical Instrumentation Engineers (SPIE) Conference Series, Vol. 11447,
  Ground-based and Airborne Instrumentation for Astronomy VIII, ed. C.~J.
  {Evans}, J.~J. {Bryant}, \& K.~{Motohara}, 1144761

\bibitem[{{Nielsen} {et~al.}(2019){Nielsen}, {De Rosa}, {Macintosh}, {Wang},
  {Ruffio}, {Chiang}, {Marley}, {Saumon}, {Savransky}, {Ammons}, {Bailey},
  {Barman}, {Blain}, {Bulger}, {Burrows}, {Chilcote}, {Cotten}, {Czekala},
  {Doyon}, {Duch{\^e}ne}, {Esposito}, {Fabrycky}, {Fitzgerald}, {Follette},
  {Fortney}, {Gerard}, {Goodsell}, {Graham}, {Greenbaum}, {Hibon}, {Hinkley},
  {Hirsch}, {Hom}, {Hung}, {Dawson}, {Ingraham}, {Kalas}, {Konopacky},
  {Larkin}, {Lee}, {Lin}, {Maire}, {Marchis}, {Marois}, {Metchev},
  {Millar-Blanchaer}, {Morzinski}, {Oppenheimer}, {Palmer}, {Patience},
  {Perrin}, {Poyneer}, {Pueyo}, {Rafikov}, {Rajan}, {Rameau}, {Rantakyr{\"o}},
  {Ren}, {Schneider}, {Sivaramakrishnan}, {Song}, {Soummer}, {Tallis},
  {Thomas}, {Ward-Duong}, \& {Wolff}}]{Nielsen2019}
{Nielsen}, E.~L., {De Rosa}, R.~J., {Macintosh}, B., {et~al.} 2019, \aj, 158,
  13

\bibitem[{{Nortmann} {et~al.}(2024){Nortmann}, {Lesjak}, {Yan}, {Cont},
  {Czesla}, {Lavail}, {Rains}, {Nagel}, {Boldt-Christmas}, {Hatzes}, {Reiners},
  {Piskunov}, {Kochukhov}, {Heiter}, {Shulyak}, {Rengel}, \&
  {Seemann}}]{Nortmann2024}
{Nortmann}, L., {Lesjak}, F., {Yan}, F., {et~al.} 2024, arXiv e-prints,
  arXiv:2404.12363

\bibitem[{{{\"O}berg} {et~al.}(2011){{\"O}berg}, {Murray-Clay}, \&
  {Bergin}}]{oberg_effects_2011}
{{\"O}berg}, K.~I., {Murray-Clay}, R., \& {Bergin}, E.~A. 2011, \apjl, 743, L16

\bibitem[{{Palma-Bifani} {et~al.}(2023){Palma-Bifani}, {Chauvin}, {Bonnefoy},
  {Rojo}, {Petrus}, {Rodet}, {Langlois}, {Allard}, {Charnay}, {Desgrange},
  {Homeier}, {Lagrange}, {Beuzit}, {Baudoz}, {Boccaletti}, {Chomez}, {Delorme},
  {Desidera}, {Feldt}, {Ginski}, {Gratton}, {Maire}, {Meyer}, {Samland},
  {Snellen}, {Vigan}, \& {Zhang}}]{palma-bifani_peering_2023}
{Palma-Bifani}, P., {Chauvin}, G., {Bonnefoy}, M., {et~al.} 2023, \aap, 670,
  A90

\bibitem[{{Palma-Bifani} {et~al.}(2024){Palma-Bifani}, {Chauvin}, {Borja},
  {Bonnefoy}, {Petrus}, {Mesa}, {De Rosa}, {Gratton}, {Baudoz}, {Boccaletti},
  {Charnay}, {Desgrange}, {Tremblin}, \&
  {Vigan}}]{palma-bifani_atmospheric_2024}
{Palma-Bifani}, P., {Chauvin}, G., {Borja}, D., {et~al.} 2024, \aap, 683, A214

\bibitem[{{Pepe} {et~al.}(2021){Pepe}, {Cristiani}, {Rebolo}, {Santos},
  {Dekker}, {Cabral}, {Di Marcantonio}, {Figueira}, {Lo Curto}, {Lovis},
  {Mayor}, {M{\'e}gevand}, {Molaro}, {Riva}, {Zapatero Osorio}, {Amate},
  {Manescau}, {Pasquini}, {Zerbi}, {Adibekyan}, {Abreu}, {Affolter}, {Alibert},
  {Aliverti}, {Allart}, {Allende Prieto}, {{\'A}lvarez}, {Alves}, {Avila},
  {Baldini}, {Bandy}, {Barros}, {Benz}, {Bianco}, {Borsa}, {Bourrier},
  {Bouchy}, {Broeg}, {Calderone}, {Cirami}, {Coelho}, {Conconi}, {Coretti},
  {Cumani}, {Cupani}, {D'Odorico}, {Damasso}, {Deiries}, {Delabre},
  {Demangeon}, {Dumusque}, {Ehrenreich}, {Faria}, {Fragoso}, {Genolet},
  {Genoni}, {G{\'e}nova Santos}, {Gonz{\'a}lez Hern{\'a}ndez}, {Hughes},
  {Iwert}, {Kerber}, {Knudstrup}, {Landoni}, {Lavie}, {Lillo-Box}, {Lizon},
  {Maire}, {Martins}, {Mehner}, {Micela}, {Modigliani}, {Monteiro}, {Monteiro},
  {Moschetti}, {Murphy}, {Nunes}, {Oggioni}, {Oliveira}, {Oshagh}, {Pall{\'e}},
  {Pariani}, {Poretti}, {Rasilla}, {Rebord{\~a}o}, {Redaelli}, {Santana
  Tschudi}, {Santin}, {Santos}, {S{\'e}gransan}, {Schmidt}, {Segovia},
  {Sosnowska}, {Sozzetti}, {Sousa}, {Span{\`o}}, {Su{\'a}rez Mascare{\~n}o},
  {Tabernero}, {Tenegi}, {Udry}, \& {Zanutta}}]{pepe_espresso_2021}
{Pepe}, F., {Cristiani}, S., {Rebolo}, R., {et~al.} 2021, \aap, 645, A96

\bibitem[{{Perdelwitz} {et~al.}(2024){Perdelwitz}, {Trifonov}, {Teklu},
  {Sreenivas}, \& {Tal-Or}}]{perdelwitz_analysis_2024}
{Perdelwitz}, V., {Trifonov}, T., {Teklu}, J.~T., {Sreenivas}, K.~R., \&
  {Tal-Or}, L. 2024, \aap, 683, A125

\bibitem[{{Petrus} {et~al.}(2021){Petrus}, {Bonnefoy}, {Chauvin}, {Charnay},
  {Marleau}, {Gratton}, {Lagrange}, {Rameau}, {Mordasini}, {Nowak}, {Delorme},
  {Boccaletti}, {Carlotti}, {Houll{\'e}}, {Vigan}, {Allard}, {Desidera},
  {D'Orazi}, {Hoeijmakers}, {Wyttenbach}, \&
  {Lavie}}]{petrus_medium-resolution_2021}
{Petrus}, S., {Bonnefoy}, M., {Chauvin}, G., {et~al.} 2021, \aap, 648, A59

\bibitem[{{Petrus} {et~al.}(2023){Petrus}, {Chauvin}, {Bonnefoy}, {Tremblin},
  {Charnay}, {Delorme}, {Marleau}, {Bayo}, {Manjavacas}, {Lagrange},
  {Molli{\`e}re}, {Palma-Bifani}, {Biller}, {Jenkins}, {Goyal}, \&
  {Hoch}}]{petrus_x-shyne_2023}
{Petrus}, S., {Chauvin}, G., {Bonnefoy}, M., {et~al.} 2023, \aap, 670, L9

\bibitem[{{Petrus} {et~al.}(2024){Petrus}, {Whiteford}, {Patapis}, {Biller},
  {Skemer}, {Hinkley}, {Su{\'a}rez}, {Palma-Bifani}, {Morley}, {Tremblin},
  {Charnay}, {Vos}, {Wang}, {Stone}, {Bonnefoy}, {Chauvin}, {Miles}, {Carter},
  {Lueber}, {Helling}, {Sutlieff}, {Janson}, {Gonzales}, {Hoch}, {Absil},
  {Balmer}, {Boccaletti}, {Bonavita}, {Booth}, {Bowler}, {Briesemeister},
  {Bryan}, {Calissendorff}, {Cantalloube}, {Chen}, {Choquet}, {Christiaens},
  {Cugno}, {Currie}, {Danielski}, {De Furio}, {Dupuy}, {Factor}, {Faherty},
  {Fitzgerald}, {Fortney}, {Franson}, {Girard}, {Grady}, {Henning}, {Hines},
  {Hood}, {Howe}, {Kalas}, {Kammerer}, {Kennedy}, {Kenworthy}, {Kervella},
  {Kim}, {Kitzmann}, {Kraus}, {Kuzuhara}, {Lagage}, {Lagrange}, {Lawson},
  {Lazzoni}, {Leisenring}, {Lew}, {Liu}, {Liu}, {Llop-Sayson}, {Lloyd},
  {Macintosh}, {M{\^a}lin}, {Manjavacas}, {Marino}, {Marley}, {Marois},
  {Martinez}, {Matthews}, {Matthews}, {Mawet}, {Mazoyer}, {McElwain},
  {Metchev}, {Meyer}, {Millar-Blanchaer}, {Molli{\`e}re}, {Moran}, {Mukherjee},
  {Pantin}, {Perrin}, {Pueyo}, {Quanz}, {Quirrenbach}, {Ray}, {Rebollido},
  {Adams Redai}, {Ren}, {Rickman}, {Sallum}, {Samland}, {Sargent}, {Schlieder},
  {Stapelfeldt}, {Tamura}, {Tan}, {Theissen}, {Uyama}, {Vasist}, {Vigan},
  {Wagner}, {Ward-Duong}, {Wolff}, {Worthen}, {Wyatt}, {Ygouf}, {Zurlo},
  {Zhang}, {Zhang}, {Zhang}, \& {Zhou}}]{petrus_jwst_2024}
{Petrus}, S., {Whiteford}, N., {Patapis}, P., {et~al.} 2024, \apjl, 966, L11

\bibitem[{{Piskunov} {et~al.}(2021){Piskunov}, {Wehrhahn}, \&
  {Marquart}}]{piskunov_optimal_2021}
{Piskunov}, N., {Wehrhahn}, A., \& {Marquart}, T. 2021, \aap, 646, A32

\bibitem[{{Rickman} {et~al.}(2024){Rickman}, {Ceva}, {Matthews},
  {S{\'e}gransan}, {Bowler}, {Forveille}, {Franson}, {Hagelberg}, {Udry}, \&
  {Vigan}}]{rickman_discovery_2024}
{Rickman}, E.~L., {Ceva}, W., {Matthews}, E.~C., {et~al.} 2024, \aap, 684, A88

\bibitem[{{Ruffio} {et~al.}(2023){Ruffio}, {Horstman}, {Mawet}, {Rosenthal},
  {Batygin}, {Wang}, {Millar-Blanchaer}, {Wang}, {Fulton}, {Konopacky},
  {Agrawal}, {Hirsch}, {Howard}, {Blunt}, {Nielsen}, {Baker}, {Bartos}, {Bond},
  {Calvin}, {Cetre}, {Delorme}, {Doppmann}, {Echeverri}, {Finnerty},
  {Fitzgerald}, {Jovanovic}, {L{\'o}pez}, {Martin}, {Morris}, {Pezzato},
  {Ruane}, {Sappey}, {Schofield}, {Skemer}, {Venenciano}, {Wallace}, {Wallack},
  {Wizinowich}, \& {Xuan}}]{Ruffio2023}
{Ruffio}, J.-B., {Horstman}, K., {Mawet}, D., {et~al.} 2023, \aj, 165, 113

\bibitem[{{Santos} {et~al.}(2017){Santos}, {Adibekyan}, {Figueira},
  {Andreasen}, {Barros}, {Delgado-Mena}, {Demangeon}, {Faria}, {Oshagh},
  {Sousa}, {Viana}, \& {Ferreira}}]{santos_observational_2017}
{Santos}, N.~C., {Adibekyan}, V., {Figueira}, P., {et~al.} 2017, \aap, 603, A30

\bibitem[{{Schlaufman}(2018)}]{schlaufman_evidence_2018}
{Schlaufman}, K.~C. 2018, \apj, 853, 37

\bibitem[{Skilling(2006)}]{Skilling2006}
Skilling, J. 2006, Bayesian Analysis, 1, 833

\bibitem[{{Tannock} {et~al.}(2022){Tannock}, {Metchev}, {Hood}, {Mace},
  {Fortney}, {Morley}, {Jaffe}, \& {Lupu}}]{tannock_146248_2022}
{Tannock}, M.~E., {Metchev}, S., {Hood}, C.~E., {et~al.} 2022, \mnras, 514,
  3160

\bibitem[{{Vanderburg} {et~al.}(2018){Vanderburg}, {Rappaport}, \&
  {Mayo}}]{vanderburg_detecting_2018}
{Vanderburg}, A., {Rappaport}, S.~A., \& {Mayo}, A.~W. 2018, \aj, 156, 184

\bibitem[{{Vanderburg} \& {Rodriguez}(2021)}]{vanderburg_first_2021}
{Vanderburg}, A. \& {Rodriguez}, J.~E. 2021, \apjl, 922, L2

\bibitem[{{Vigan} {et~al.}(2024){Vigan}, {El Morsy}, {Lopez}, {Otten},
  {Garcia}, {Costes}, {Muslimov}, {Viret}, {Charles}, {Zins}, {Murray},
  {Costille}, {Paufique}, {Seemann}, {Houll{\'e}}, {Anwand-Heerwart},
  {Phillips}, {Abinanti}, {Balard}, {Baraffe}, {Benedetti}, {Blanchard},
  {Blanco}, {Beuzit}, {Choquet}, {Cristofari}, {Desidera}, {Dohlen}, {Dorn},
  {Ely}, {Fuenteseca}, {Garcia}, {Jaquet}, {Jaubert}, {Kasper}, {Le Merrer},
  {Maire}, {N'Diaye}, {Pallanca}, {Popovic}, {Pourcelot}, {Reiners}, {Rochat},
  {Sehim}, {Schmutzer}, {Smette}, {Tchoubaklian}, {Tomlinson}, \& {Valenzuela
  Soto}}]{Vigan2024}
{Vigan}, A., {El Morsy}, M., {Lopez}, M., {et~al.} 2024, \aap, 682, A16

\bibitem[{{Vigan} {et~al.}(2021){Vigan}, {Fontanive}, {Meyer}, {Biller},
  {Bonavita}, {Feldt}, {Desidera}, {Marleau}, {Emsenhuber}, {Galicher}, {Rice},
  {Forgan}, {Mordasini}, {Gratton}, {Le Coroller}, {Maire}, {Cantalloube},
  {Chauvin}, {Cheetham}, {Hagelberg}, {Lagrange}, {Langlois}, {Bonnefoy},
  {Beuzit}, {Boccaletti}, {D'Orazi}, {Delorme}, {Dominik}, {Henning}, {Janson},
  {Lagadec}, {Lazzoni}, {Ligi}, {Menard}, {Mesa}, {Messina}, {Moutou},
  {M{\"u}ller}, {Perrot}, {Samland}, {Schmid}, {Schmidt}, {Sissa}, {Turatto},
  {Udry}, {Zurlo}, {Abe}, {Antichi}, {Asensio-Torres}, {Baruffolo}, {Baudoz},
  {Baudrand}, {Bazzon}, {Blanchard}, {Bohn}, {Brown Sevilla}, {Carbillet},
  {Carle}, {Cascone}, {Charton}, {Claudi}, {Costille}, {De Caprio},
  {Delboulb{\'e}}, {Dohlen}, {Engler}, {Fantinel}, {Feautrier}, {Fusco},
  {Gigan}, {Girard}, {Giro}, {Gisler}, {Gluck}, {Gry}, {Hubin}, {Hugot},
  {Jaquet}, {Kasper}, {Le Mignant}, {Llored}, {Madec}, {Magnard}, {Martinez},
  {Maurel}, {M{\"o}ller-Nilsson}, {Mouillet}, {Moulin}, {Orign{\'e}}, {Pavlov},
  {Perret}, {Petit}, {Pragt}, {Puget}, {Rabou}, {Ramos}, {Rickman}, {Rigal},
  {Rochat}, {Roelfsema}, {Rousset}, {Roux}, {Salasnich}, {Sauvage}, {Sevin},
  {Soenke}, {Stadler}, {Suarez}, {Wahhaj}, {Weber}, \& {Wildi}}]{Vigan2021}
{Vigan}, A., {Fontanive}, C., {Meyer}, M., {et~al.} 2021, \aap, 651, A72

\bibitem[{{Vigan} {et~al.}(2022){Vigan}, {Lopez}, {El Morsy}, {Muslimov},
  {Viret}, {Zins}, {Murray}, {Costille}, {Otten}, {Seemann}, {Anwand-Heerwart},
  {Dohlen}, {Blanchard}, {Garcia}, {Charles}, {Tchoubaklian}, {Ely},
  {Phillips}, {Paufique}, {Beuzit}, {Houll{\'e}}, {Costes}, {Pourcelot},
  {Baraffe}, {Dorn}, {Jaquet}, {Kasper}, {Reiners}, {Smette}, {Blanco},
  {Pallanca}, {Carlotti}, {Choquet}, {Mouillet}, \& {N'Diaye}}]{Vigan2022spie}
{Vigan}, A., {Lopez}, M., {El Morsy}, M., {et~al.} 2022, in Society of
  Photo-Optical Instrumentation Engineers (SPIE) Conference Series, Vol. 12185,
  Adaptive Optics Systems VIII, ed. L.~{Schreiber}, D.~{Schmidt}, \&
  E.~{Vernet}, 121850S

\bibitem[{{Visscher} {et~al.}(2010){Visscher}, {Lodders}, \&
  {Fegley}}]{visscher_atmospheric_2010}
{Visscher}, C., {Lodders}, K., \& {Fegley}, Jr., B. 2010, \apj, 716, 1060

\bibitem[{{Vousden} {et~al.}(2016){Vousden}, {Farr}, \&
  {Mandel}}]{vousden_dynamic_2016}
{Vousden}, W.~D., {Farr}, W.~M., \& {Mandel}, I. 2016, \mnras, 455, 1919

\bibitem[{{Wang} {et~al.}(2022){Wang}, {Gao}, {Chilcote}, {Lozi}, {Guyon},
  {Marois}, {De Rosa}, {Sahoo}, {Groff}, {Vievard}, {Jovanovic}, {Greenbaum},
  \& {Macintosh}}]{wang_atmospheric_2022}
{Wang}, J.~J., {Gao}, P., {Chilcote}, J., {et~al.} 2022, \aj, 164, 143

\bibitem[{{Wang} {et~al.}(2021){Wang}, {Ruffio}, {Morris}, {Delorme},
  {Jovanovic}, {Pezzato}, {Echeverri}, {Finnerty}, {Hood}, {Zanazzi}, {Bryan},
  {Bond}, {Cetre}, {Martin}, {Mawet}, {Skemer}, {Baker}, {Xuan}, {Wallace},
  {Wang}, {Bartos}, {Blake}, {Boden}, {Buzard}, {Calvin}, {Chun}, {Doppmann},
  {Dupuy}, {Duch{\^e}ne}, {Feng}, {Fitzgerald}, {Fortney}, {Freedman},
  {Knutson}, {Konopacky}, {Lilley}, {Liu}, {Lopez}, {Lupu}, {Marley},
  {Meshkat}, {Miles}, {Millar-Blanchaer}, {Ragland}, {Roy}, {Ruane}, {Sappey},
  {Schofield}, {Weiss}, {Wetherell}, {Wizinowich}, \& {Ygouf}}]{Wang2021}
{Wang}, J.~J., {Ruffio}, J.-B., {Morris}, E., {et~al.} 2021, \aj, 162, 148

\bibitem[{{Xuan} {et~al.}(2024){Xuan}, {Hsu}, {Finnerty}, {Wang}, {Ruffio},
  {Zhang}, {Knutson}, {Mawet}, {Mamajek}, {Inglis}, {Wallack}, {Bryan},
  {Blake}, {Molli{\`e}re}, {Hejazi}, {Baker}, {Bartos}, {Calvin}, {Cetre},
  {Delorme}, {Doppmann}, {Echeverri}, {Fitzgerald}, {Jovanovic}, {Liberman},
  {L{\'o}pez}, {Morris}, {Pezzato}, {Sappey}, {Schofield}, {Skemer}, {Wallace},
  {Wang}, {Agrawal}, \& {Horstman}}]{Xuan2024}
{Xuan}, J.~W., {Hsu}, C.-C., {Finnerty}, L., {et~al.} 2024, \apj, 970, 71

\bibitem[{{Xuan} {et~al.}(2022){Xuan}, {Wang}, {Ruffio}, {Knutson}, {Mawet},
  {Molli{\`e}re}, {Kolecki}, {Vigan}, {Mukherjee}, {Wallack}, {Wang}, {Baker},
  {Bartos}, {Blake}, {Bond}, {Bryan}, {Calvin}, {Cetre}, {Chun}, {Delorme},
  {Doppmann}, {Echeverri}, {Finnerty}, {Fitzgerald}, {Horstman}, {Inglis},
  {Jovanovic}, {L{\'o}pez}, {Martin}, {Morris}, {Pezzato}, {Ragland}, {Ren},
  {Ruane}, {Sappey}, {Schofield}, {Skemer}, {Venenciano}, {Wallace}, \&
  {Wizinowich}}]{xuan_clear_2022}
{Xuan}, J.~W., {Wang}, J., {Ruffio}, J.-B., {et~al.} 2022, \apj, 937, 54

\bibitem[{{Zechmeister} {et~al.}(2014){Zechmeister}, {Anglada-Escud{\'e}}, \&
  {Reiners}}]{zechmeister_flat-relative_2014}
{Zechmeister}, M., {Anglada-Escud{\'e}}, G., \& {Reiners}, A. 2014, \aap, 561,
  A59

\bibitem[{{Zhang} {et~al.}(2024){Zhang}, {Huber}, {Weiss}, {Xuan}, {Burt},
  {Dai}, {Saunders}, {Petigura}, {Rubenzahl}, {Winn}, {Wang}, {Van Zandt},
  {Brodheim}, {Claytor}, {Crossfield}, {Deich}, {Fulton}, {Gibson},
  {Halverson}, {Hill}, {Holden}, {Householder}, {Howard}, {Isaacson}, {Kaye},
  {Lanclos}, {Laher}, {Lubin}, {Payne}, {Roy}, {Schwab}, {Shaum}, {Walawender},
  {Wishnow}, \& {Yeh}}]{zhang__testbed_2024}
{Zhang}, J., {Huber}, D., {Weiss}, L.~M., {et~al.} 2024, \aj, 168, 295

\bibitem[{{Zhang}(2024)}]{Zhang2024}
{Zhang}, Z. 2024, Research Notes of the American Astronomical Society, 8, 114

\bibitem[{{Zhang} {et~al.}(2023){Zhang}, {Molli{\`e}re}, {Hawkins}, {Manea},
  {Fortney}, {Morley}, {Skemer}, {Marley}, {Bowler}, {Carter}, {Franson},
  {Maas}, \& {Sneden}}]{zhang_elemental_2023}
{Zhang}, Z., {Molli{\`e}re}, P., {Hawkins}, K., {et~al.} 2023, \aj, 166, 198

\bibitem[{{Zhou} {et~al.}(2016){Zhou}, {Apai}, {Schneider}, {Marley}, \&
  {Showman}}]{zhou_discovery_2016}
{Zhou}, Y., {Apai}, D., {Schneider}, G.~H., {Marley}, M.~S., \& {Showman},
  A.~P. 2016, \apj, 818, 176

\bibitem[{{Zhou} {et~al.}(2020){Zhou}, {Bowler}, {Morley}, {Apai}, {Kataria},
  {Bryan}, \& {Benneke}}]{zhou_spectral_2020}
{Zhou}, Y., {Bowler}, B.~P., {Morley}, C.~V., {et~al.} 2020, \aj, 160, 77

\bibitem[{{Z{\'u}{\~n}iga-Fern{\'a}ndez}
  {et~al.}(2021){Z{\'u}{\~n}iga-Fern{\'a}ndez}, {Bayo}, {Elliott}, {Zamora},
  {Corval{\'a}n}, {Haubois}, {Corral-Santana}, {Olofsson}, {Hu{\'e}lamo},
  {Sterzik}, {Torres}, {Quast}, \& {Melo}}]{zuniga-fernandez_search_2021}
{Z{\'u}{\~n}iga-Fern{\'a}ndez}, S., {Bayo}, A., {Elliott}, P., {et~al.} 2021,
  \aap, 645, A30

\bibitem[{{Zucker}(2003)}]{zucker_cross-correlation_2003}
{Zucker}, S. 2003, \mnras, 342, 1291

\end{thebibliography}

\begin{appendix}
\label{Appendix}
\section{Detector traces}
\label{sec:detector traces}

\begin{figure}[!ht]
    \centering
    \includegraphics[width=\linewidth]{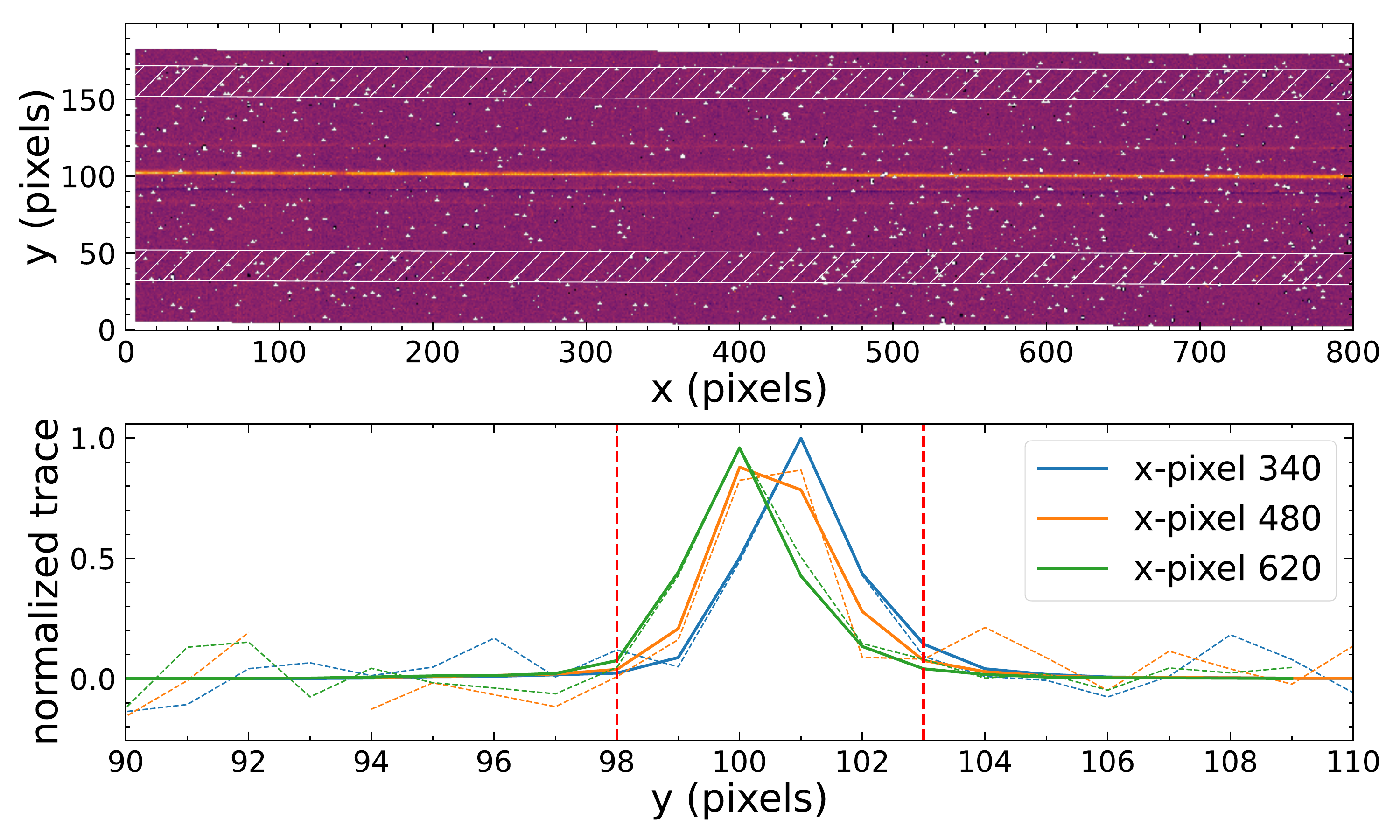}
    \caption{Trace of the CRIRES+ detector for a single order. Top panel: 2D detector image of the companion. The hatched area corresponds to the noise estimation region. The spectrum of the companion is located at the center of the image, around the y-pixel number 100. Around y-pixel 90 is the residual MACAO internal source fiber signal after background subtraction. Around y-pixel number 115 and y-pixel number 85, we can see the signal of 2 of the 3 reference fibers.  The bad pixels are depicted in white. Bottom panel: normalized profile between y-pixels 90 and 110 of the signal at x-pixels 340, 480 and 620 for the star (solid lines) and the companion (dashed lines). The extracted signal area is represented by the vertical red dashed lines.}
    \label{fig:trace}
\end{figure}

Figure~\ref{fig:trace} presents the trace of the companion data for a single order. The top panel depicts a 2D image of the data when the fiber is centered at the companion. In the 2D image of the companion, we can observe the residual MACAO internal source fiber signal around the y-pixel number 90. Secondary signals are also visible around y-pixel number 115 and y-pixel number 85, originating from 2 reference fibers. The signal from the last reference fiber is too faint to be seen on this image. The bottom panel shows the shape of the signal between y-pixels 90 and 110 at different wavelengths (x-pixels 340, 480 and 620). The dashed line is the profile of the companion data, whereas the solid line is the profile pf the star data. We can see that the profiles of the companion and star star are very similar across different wavelengths. Extracting the signal over 6 pixels is a reasonable choice to capture the entire signal at each wavelength without adding too much noise. By analyzing the trace at different positions on the x-axis, we see that the profiles of the companion and star data are globally very consistent.

\section{Impact of S/N ratio on RV and \vsini estimations}
\label{sec:S/N_ratio_impact}

\begin{figure}
    \centering
    \includegraphics[width=\linewidth]{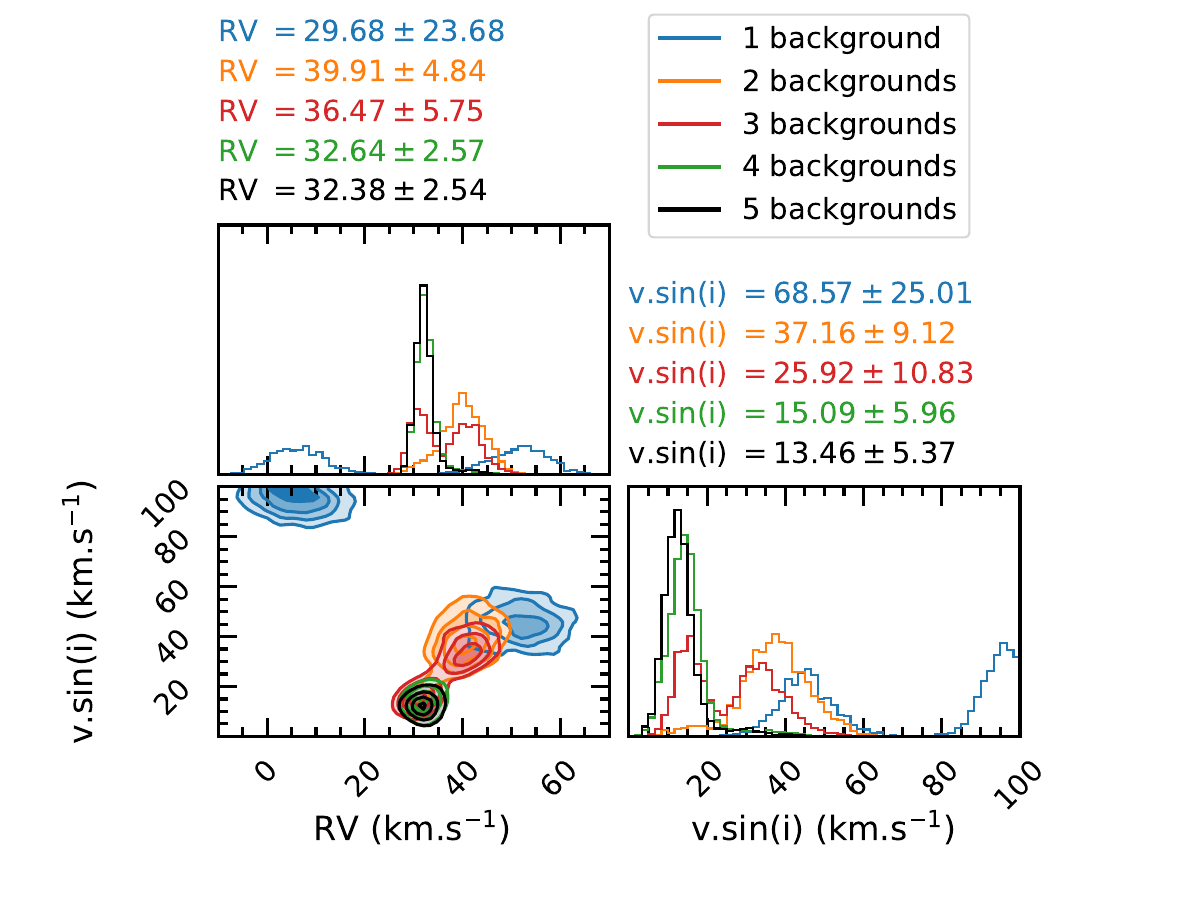}
    \caption{First night posterior distribution of RV and \vsini for different number of backgrounds used in the data reduction. The number of backgrounds impacts the final S/N of the reduced data, which has a direct impact on the valued derived for the RV and \vsini of the companion.}
    \label{fig:corner_first_night_bkgs}
\end{figure}

In Sect.~\ref{sec:atmosheric_models} we decided to exclude the data of the second offset of the second night from the analysis, mainly due to the lower number of available backgrounds for this second offset. To verify this hypothesis, we ran \formosa on the data of the first night with an increasing numbers of backgrounds used in the data reduction, going from one to five backgrounds. The backgrounds are mainly used to subtract the MACAO guide fiber leakage \citep{Vigan2024}. The removal of this signal leaves some residual noise which decreases as we increase the number of backgrounds used. The results of this analysis are shown in Fig.~\ref{fig:corner_first_night_bkgs}. We see that the more backgrounds we use to reduce the data, the more constrained the final RV and \vsini are. This effect is particularly strong when using a small number of backgrounds (less than three), where some distributions of RV and \vsini are bi-modal, and diminishes as the number of backgrounds increases. This is expected since we theoretically expect a reduction in the S/N of the reduced data of $\sqrt{N}$ with $N$ being the number of backgrounds used, up to a certain point. The differences between the results with four and five backgrounds become very small, which indicates that we are probably reaching a fundamental limit in our data, such as photon noise or a mixture of photon noise and instrumental systematics. We recall that the photon noise in our data comes mainly from the stellar contribution at the location of the companion, which is the dominant source in our data.

\FloatBarrier

\section{CCF analysis for the second night}
\label{sec:ccf_second_night}

\begin{figure*}
    \centering
    \includegraphics[width=\textwidth]{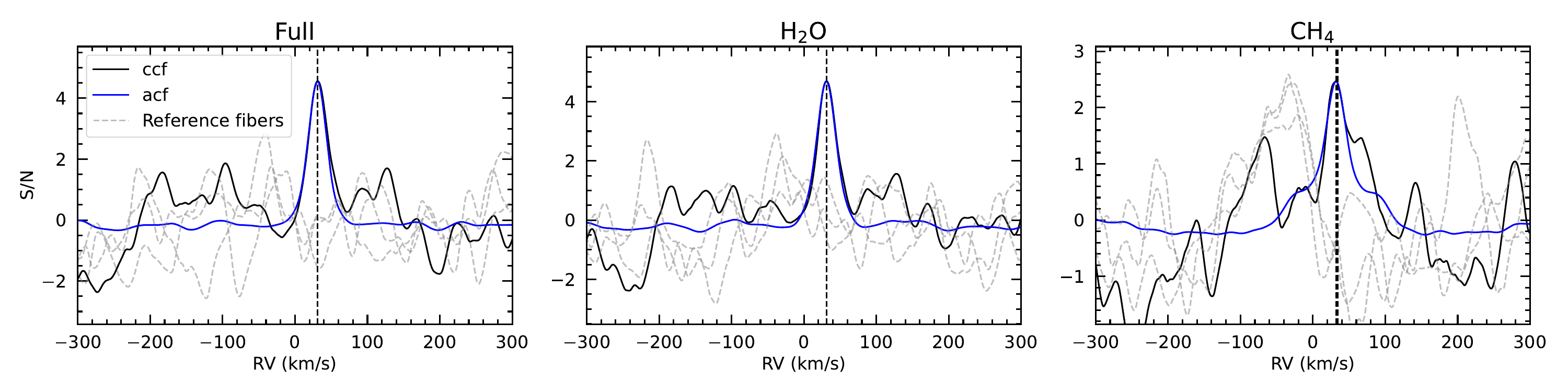}
    \caption{Cross-correlation functions for the first offset of the second night.}
    \label{fig:ccf_second_night_offset0}
\end{figure*}

\begin{figure*}
    \centering
    \includegraphics[width=\textwidth]{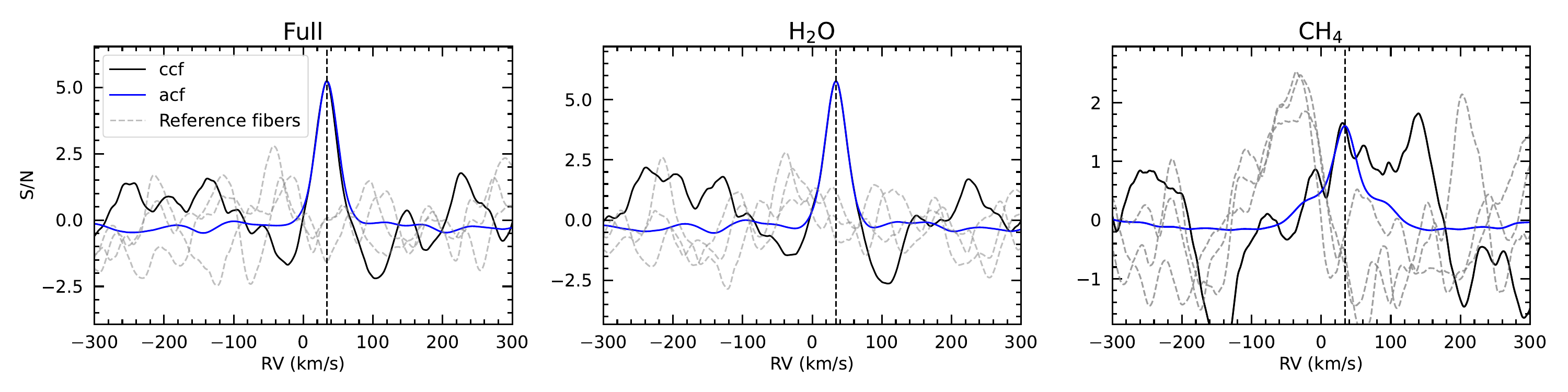}
    \caption{Cross-correlation functions for the second offset of the second night.}
    \label{fig:ccf_second_night_offset1}
\end{figure*}

Figures~\ref{fig:ccf_second_night_offset0} and \ref{fig:ccf_second_night_offset1} present the cross-correlation analysis for the first and second offset of the second night. Compared to the first night, the detections have lower S/N, confirming the better data quality of the first night. The estimated RV values are 31.9 \kms, 31.7 \kms and 31.8 \kms for the full model, H$_2$O and CH$_4$ with the first offset. These values are consistent with the values estimated by \formosa (see Table \ref{tab:formosa_results}). For the second offset, we estimate RVs of 33.1 \kms with the full model and 34.0 \kms with H$_2$O. The CCF with CF$_4$ does not show a convincing detection of CH$_4$, albeit it exhibits a small peak in the CCF around 34 \kms.

\FloatBarrier

\section{\texttt{Orvara} results}
\label{sec:orvara_results}

\begin{figure}[!ht]
    \centering
    \includegraphics[width=\linewidth]{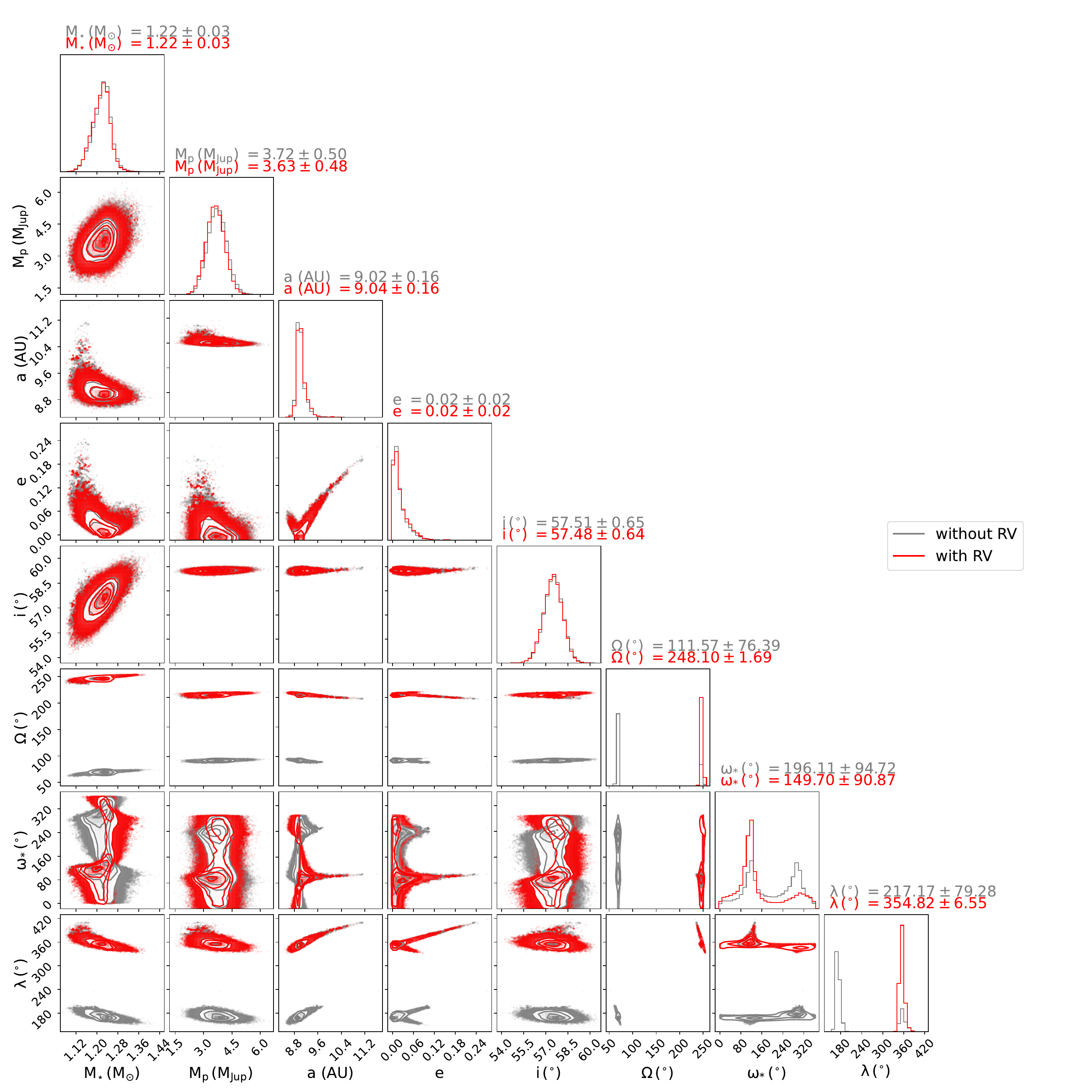}
    \caption{Posterior distribution for the orbit of AF\,Lep\,b with the astrometry from SPHERE, KECK/NIRC2 and GRAVITY (grey) and with our RV points included (red).}
    \label{fig:corner_plot_orvara}
\end{figure}

Figure~\ref{fig:corner_plot_orvara} presents the posterior distribution obtained with Orvara. 2 cases are depicted: without including our RV measurement (grey) and with our RV measurement (red). The inclusion of our RV measurement does not impact the estimation of the orbital parameters except for the argument of periastron $\omega$, the mean longitude at epoch 2010 $\lambda$, and the longitude of ascending node $\Omega$. The latter two present bimodal distributions when we do not include our radial velocity measurement.

\FloatBarrier

\section{\texttt{ForMoSA} results}
\label{sec:formosa_results}

\begin{figure}[!ht]
    \centering
    \includegraphics[width=\linewidth]{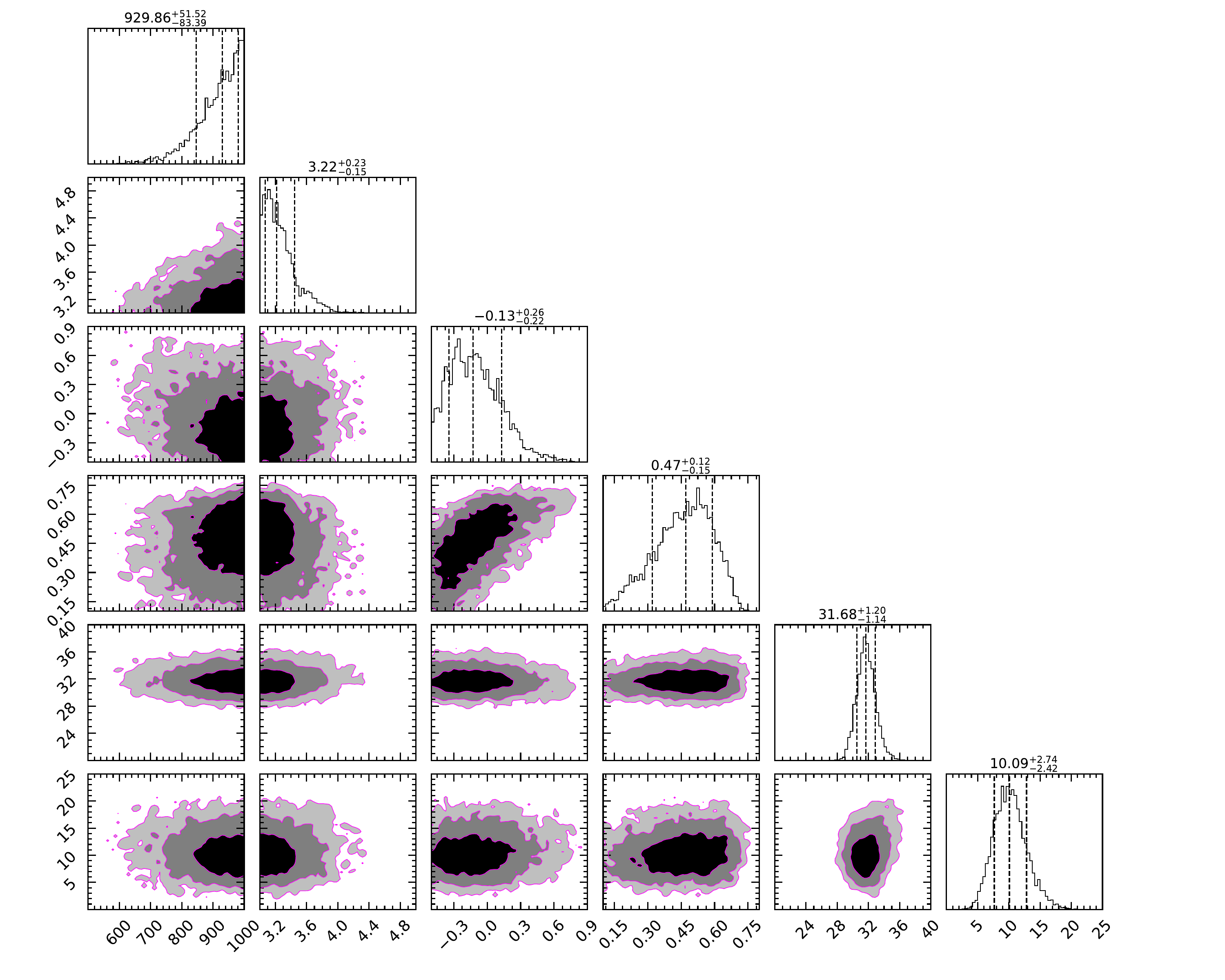}
    \caption{First night posterior distribution of atmospheric parameters for the Exo-REM/Exo\_k model in the case where all parameters are free}
    \label{fig:corner_first_night_free}
\end{figure}

\begin{figure}[!ht]
    \centering
    \includegraphics[width=\linewidth]{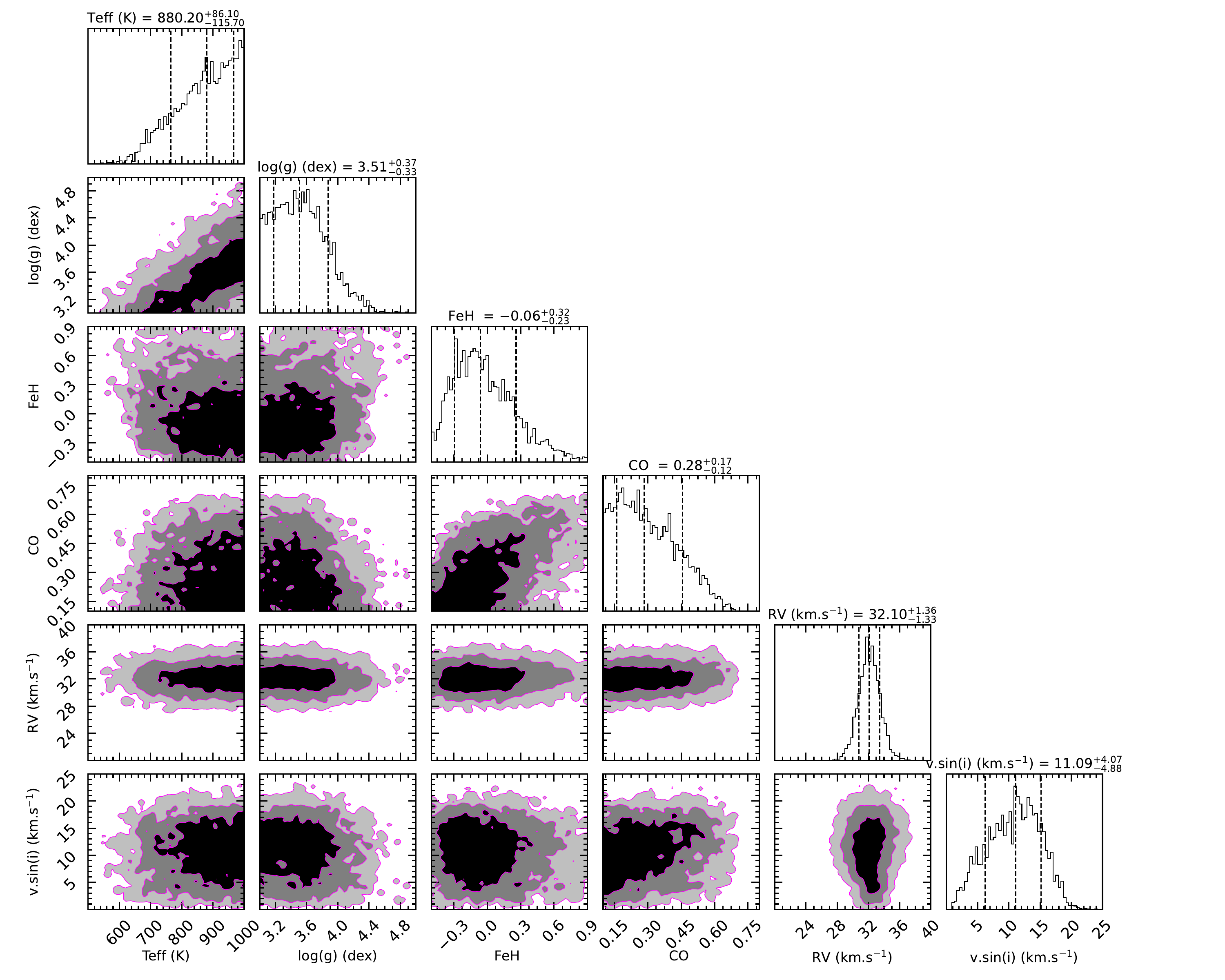}
    \caption{First offset second night posterior distribution of atmospheric parameters for the Exo-REM/Exo\_k model in the case where all parameters are free}
    \label{fig:corner_second_night_free}
\end{figure}

\begin{figure}[!ht]
    \centering
    \includegraphics[width=\linewidth]{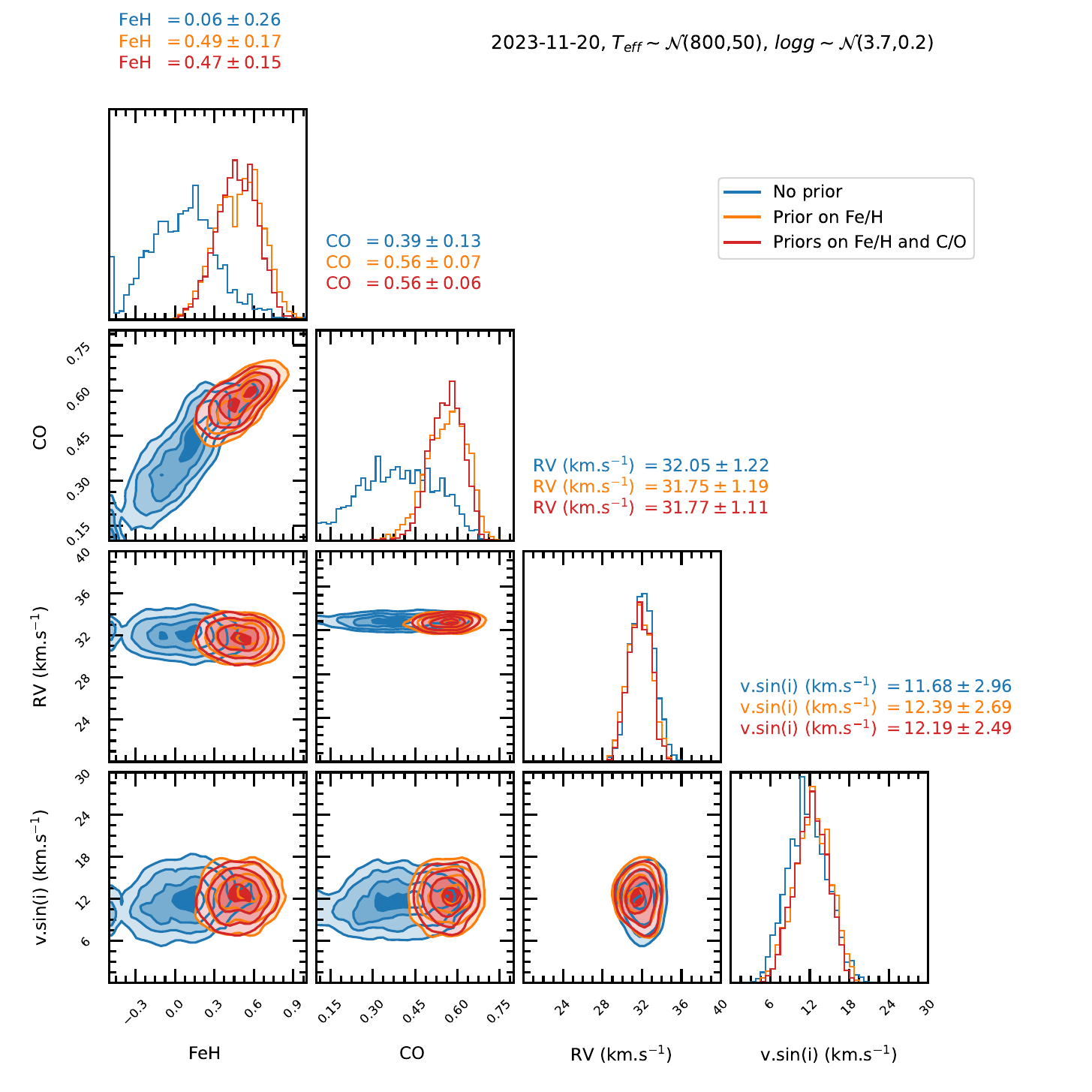}
    \caption{First night posterior distribution of atmospheric parameters for the Exo-REM/Exo\_k model with priors on \Teff and \logg. Three cases are considered : No prior (blue), Prior on \met (orange), Prior on \met and \co (red).}
    \label{fig:corner_first_night_priors}
\end{figure}

\begin{figure}[!ht]
    \centering
    \includegraphics[width=\linewidth]{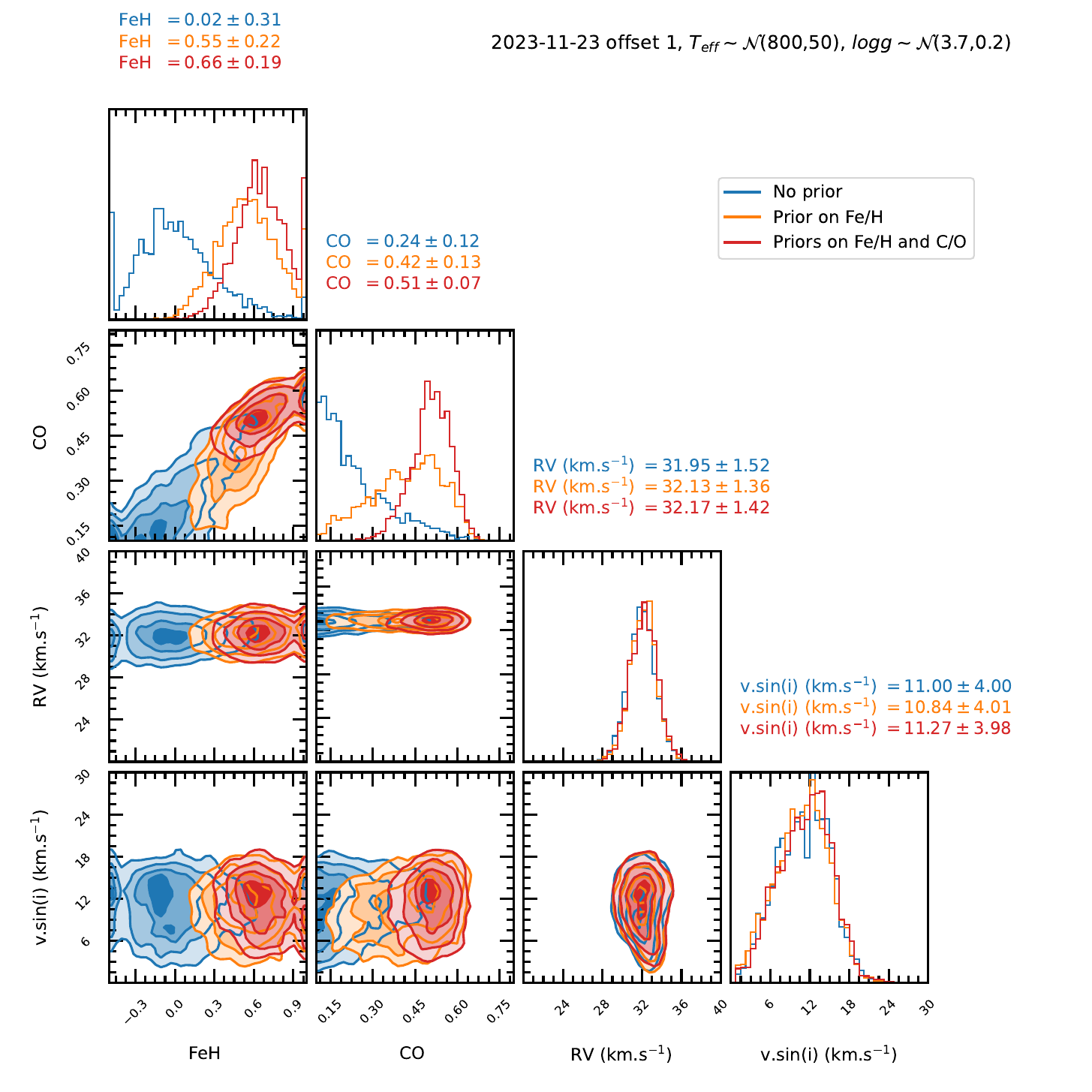}
    \caption{First offset of the second night posterior distributions of atmospheric parameters for the Exo-REM/Exo\_k model with priors on \Teff and \logg. Three cases are considered: No prior (blue), Prior on \met (orange), Prior on \met and \co (red). }
    \label{fig:corner_second_night_priors}
\end{figure}

This section presents some of the results obtained with ForMoSA. Figure~\ref{fig:corner_first_night_free} presents the posterior distribution for the first night without any prior. Figure ~\ref{fig:corner_first_night_priors} compares the posterior distributions for the first night with fixed \Teff and \logg under the different configurations of priors adopted (See Table \ref{tab:formosa_results}). Similarly, Figures~\ref{fig:corner_second_night_free} and \ref{fig:corner_second_night_priors} present the posterior distribution for the first offset of the second night under the different configurations adopted for the priors.

\end{appendix}
\end{document}